\documentclass[twocolumn,pre,aps,superscriptaddress,longbibliography]{revtex4-1}
\usepackage[T1]{fontenc}
\usepackage[latin9]{inputenc}

\usepackage{amsmath}
\usepackage{amssymb}
\usepackage{xcolor}
\usepackage{bbm}
\usepackage{braket}
\usepackage{mathrsfs}  
\usepackage{tikz-cd}
\usepackage{bbold}
\usepackage{comment}
\usepackage{centernot}
\usepackage{mathtools}
\usepackage{stmaryrd}

\makeatletter
\newcommand{\xMapsto}[2][]{\ext@arrow 0599{\Mapstofill@}{#1}{#2}}
\def\Mapstofill@{\arrowfill@{\Mapstochar\Relbar}\Relbar\Rightarrow}
\makeatother

\allowdisplaybreaks

\usepackage{graphicx}

\usepackage[colorlinks=true]{hyperref}  
\hypersetup{
    bookmarks=true,         
    unicode=false,          
    pdftoolbar=true,        
    pdfmenubar=true,        
    pdffitwindow=false,     
    pdfstartview={FitH},    
    pdfsubject={},   
    pdfcreator={},   
    pdfproducer={}, 
    pdfkeywords={} {} {}, 
    pdfnewwindow=true,      
    colorlinks=true,       
    linkcolor=blue, 
    citecolor=blue,        
    filecolor=magenta,      
    urlcolor=blue,           
	breaklinks=true
}

\definecolor{orange}{rgb}{1,0.5,0}

\newcommand{\sect}[1]{\vspace{0.3em}{\it #1.}---}

\newcommand{\be}{\begin{equation}}
\newcommand{\ee}{\end{equation}}
\newcommand{\bea}{\begin{eqnarray}}
\newcommand{\eea}{\end{eqnarray}}

\newcommand{\imi}{\,\text{i}}

\newcommand{\ie}{i.e.\;}
\newcommand{\eg}{e.g.\;}

\def\bs#1{\boldsymbol{#1}}

\def \H {\mathcal{H}}
\def \A {\mathcal{A}}
\def \F {\mathcal{F}}
\def \k {\mathbf{k}}

\def \K {\hat{\mathcal{K}}}
\def \Z {\mathbb{Z}}

\begin{document}
\title{Second Euler number in four dimensional synthetic matter}
\author{Adrien Bouhon}
\affiliation{TCM Group, Cavendish Laboratory, University of Cambridge, J. J. Thomson Avenue, Cambridge CB3 0HE, United Kingdom}
\affiliation{Nordita, Stockholm University and KTH Royal Institute of Technology, Hannes Alfv{\'e}ns v{\"a}g 12, SE-106 91 Stockholm, Sweden}
\author{Yan-Qing Zhu}
\affiliation{Guangdong-Hong Kong Joint Laboratory of Quantum Matter, Department of Physics, and HKU-UCAS Joint Institute for Theoretical and Computational Physics at Hong Kong,\\ The University of Hong Kong, Pokfulam Road, Hong Kong, China}
\author{Robert-Jan Slager}
\affiliation{TCM Group, Cavendish Laboratory, University of Cambridge, J. J. Thomson Avenue, Cambridge CB3 0HE, United Kingdom}
\author{Giandomenico Palumbo}
\affiliation{School of Theoretical Physics, Dublin Institute for Advanced Studies, 10 Burlington Road, Dublin 4, Ireland}

\date{\today}

\begin{abstract}
\noindent Two-dimensional Euler insulators are novel kind of systems that host multi-gap topological phases, quantified by a quantised first Euler number in their bulk. 
Recently, these phases have been experimentally realised in suitable two-dimensional synthetic matter setups.  Here we introduce the second Euler invariant, a familiar invariant in both differential topology (Chern-Gauss-Bonnet theorem) and in four-dimensional Euclidean gravity, whose existence has not been explored in condensed matter systems.  Specifically, we firstly define two specific novel models in four dimensions that support a non-zero second Euler number in the bulk together with peculiar gapless boundary states. Secondly, we discuss its robustness in general spacetime-inversion invariant phases and its role in the classification of topological degenerate real bands through real Grassmannians. In particular, we derive from homotopy arguments the minimal Bloch Hamiltonian form from which the tight-binding models of any second Euler phase can be generated. Considering more concretely the gapped Euler phase associated with the tangent bundle of the four-sphere, we show that the bulk band structure of the nontrivial 4D Euler phase necessarily exhibits triplets of linked nodal surfaces (where the three types of nodal surfaces are formed by the crossing of the three successive pairs of bands within one four-band subspace). Finally, we show how to engineer these new topological phases in a four-dimensional ultracold atom setup. Our results naturally generalize the second Chern and spin Chern numbers to the case of four-dimensional phases that are characterised by real Hamiltonians and open doors for implementing such unexplored higher-dimensional phases in artificial engineered systems, ranging from ultracold atoms to photonics and electric circuits.
\end{abstract}

\maketitle
\sect{Introduction} 
Topological phases of matter are largely recognized as one of the main pillars of modern physics. They embrace theoretical ideas and concepts that go much beyond the realm of condensed matter physics \cite{Wen}. As an interdisciplinary field, topological matter has been studied and explored by employing a myriad of methods and techniques ranging from different algebraic topology, quantum field theory, gravity to entanglement and band theory \cite{Senthil,Witten}. For instance, Chern numbers, originally formulated in the formal context of characteristic classes for complex vector bundles \cite{Nakahara_book,Eguchi}, are nowadays known to be also measurable quantities. They are, in fact, associated to well-defined physical observables lying at the core of quantum Hall states and Chern insulators in two and higher-even dimensional systems \cite{Zhang,Karabali,XLQi2008}. Importantly, some discrete symmetries such as time-reversal and particle-hole symmetries \cite{Altland} have been shown to give rise to novel topological invariants that characterise topological insulators and superconductors. This deeper understanding of the interplay between topology and symmetries has had far reaching consequences. Through the classification of free-fermion topological phases, different kinds of topological systems lie together within an elegant periodic table \cite{Kitaev,Ludwig}.
More recently, the addition of lattice discrete symmetries has enlarged this periodic table \cite{Morimoto_2013,Shiozaki14} by giving rise to novel important models and concepts, such as crystalline topological insulators~\cite{Fu,Slager_2013, Bradlyn_2017, Kruthoff_2017, mcom,Po_2017}, higher-order topological phases (HOTPs) \cite{Benalcazar,Slager_2015,Neupert,Slager_2014,Wieder_HOTI}, fragile topology~\cite{Po_2018, bouhon2018wilson,Bradlyn_fragile} to name a few. Within this framework, the combination of time-reversal symmetry and inversion symmetry $PT$ (parity-time reversal) or $C_2T$ (twofold rotations and time reversal) symmetry in spinless fermionic systems have unveiled the existence of novel topological phases characterized by the first Euler number and Stiefel-Whitney invariants~\cite{bouhon2019nonabelian,bouhon2020geometric, Ahn_2018,Ahn2019, BJY_nielsen, Bouhon2022braiding2,Palumbo, Wieder_axion}. These topological invariants, different from the Chern numbers, are associated to real vector bundles and their corresponding gauge connections, namely $\mathsf{SO}(2)$ Berry connections are related to degenerate bands of suitable real-valued Hamiltonians in two and three dimensions. \\
Although solid-state quantum materials represent the natural context for the experimental detection of topological phases, synthetic systems, ranging from photonic systems and metamaterials \cite{Ozawa3} to ultracold atoms \cite{Goldman,DWZhang2018,Cooper2019} have been largely employed to simulate quantum systems that could not exist in real quantum materials. 
For instance, a finite Euler class~\cite{bouhon2019nonabelian,bouhon2020geometric, Ahn_2018,Ahn2019, BJY_nielsen} results in physical signatures in out-of-equilibrium~\cite{Unal_quenched_Euler, slager2022floquet} as well as equilibrium contexts~\cite{Peng2022,Park2021, Peng2022b,Lange_2021, Chen_2022, Konye_2021}, including magnetic ones~\cite{Bouhon2020_mag}, that by now have been observed in experiments that range from metamaterials~\cite{Jiang2021euler,Guo1Dexp,jiang2022experimental} to trapped ion simulators~\cite{zhao2022observation}. 
On the other hand, synthetic matter is also the ideal playground to simulate higher-dimensional topological phases.
 The four-dimensional quantum Hall effect \cite{Zhang,Karabali,XLQi2008,Price} and the corresponding higher-dimensional Thouless pump \cite{Kraus} are among the most famous examples related to higher Chern numbers and have been experimentally realized in ultracold atoms and photonics \cite{Lohse,Zilberberg}. More recently, a four-dimensional tensor monopole \cite{Palumbo2,Palumbo3,Palumbo4, HTDing2020}, characterized by the first Dixmier-Douady (DD) invariant, has been theoretically investigated and experimentally realized in NV centers \cite{Cappellaro} and superconducting circuits \cite{Tan}. Furthermore, a theoretical proposal for the implementation of the second spin and valley Chern numbers in ultracold atoms has been presented in Ref. \cite{YQZhu2022}. We note that all these higher-dimensional invariants are related to the existence of complex vector Berry bundles. 
This opens the question of the identification and possible experimental implementation of novel four-dimensional systems that have a \emph{second Euler number}.\\
The main goal of this work is to fill this gap.
Although the second Euler number has been largely explored in the context of four-dimensional Euclidean gravity and differential geometry (Chern-Gauss-Bonnet theorem and Euler characteristic) \cite{Eguchi}, our work provides the first evidence of its importance for topological phases of matter. 

\begin{figure}[http]\centering
	\includegraphics[width=8.2cm]{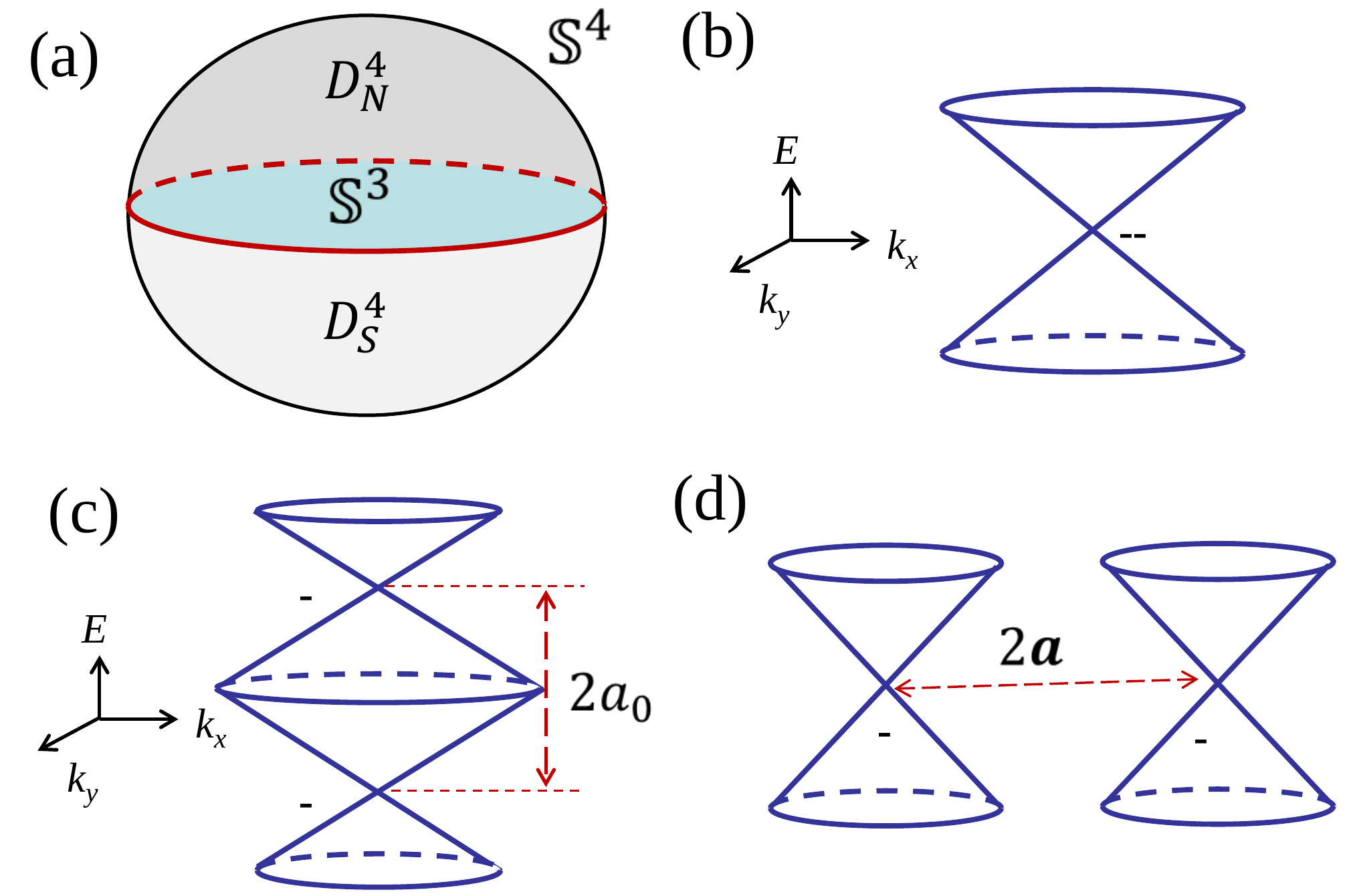}
	\caption{ (a) A base manifold $\mathbb{S}^4$ covered by the northern and southern hyper-hemispheres with the overlap being the equator $\mathbb{S}^3$. Figs. (b-d) denote the schematic of boundary energy spectra at $k_z=0$ plane near the origin when $2<m<4$. (b) A double Weyl point at $(E,\mathbf{k})=(a_0,{\boldsymbol a})=0$; (b) Nodal sphere structure formed by splitting a double Weyl point in energy when only $a_0\neq 0$; (c) Two Weyl points with the same chirality separated along $k_x$ axis when ${\boldsymbol a}=(a_1,0,0)$.}
	\label{BDmode}
\end{figure}

\emph{Bulk topology and edge states.--}
We start by considering a Dirac model of the 4D real topological insulator  which takes the form,
\begin{equation}\label{RCIHam}
	\mathcal{H}_0(k)=d_x\Gamma_1+d_y\Gamma_2+d_z\Gamma_3+d_w\Gamma_4+d_0\Gamma_0,
\end{equation}
 where $d_i=\sin k_i$, $d_0=m-\sum_{i}\cos k_i$ with $i=x,y,z,w$.
    The $8\times 8$ real Dirac matrices satisfying the Clifford algebra $\{\Gamma_{\alpha},\Gamma_{\beta}\}=2\delta_{\alpha\beta}$, represented by 	$\Gamma_1=G_{111}$, $\Gamma_2=G_{113}$,	$\Gamma_3=G_{130}$, $\Gamma_4=G_{232}$, and $\Gamma_0=G_{300}$. We label $G_{ijk}=\sigma_{i}\otimes\sigma_j\otimes\sigma_k$ hereafter.  Note that this model commutes with $\Gamma_{5}=G_{320}$, i.e., $[\Gamma_{5},\H_0]=0$. Model \eqref{RCIHam} preserves $PT$-symmetry, i.e., $[PT,\H_0]=0$, with $PT=\K$ satisfying $(PT)^2=+1$. $\K$ denotes complex conjugate.  It hosts two fourfold degenerate bands with  spectrum being $E=\pm \sqrt{\sum_{i}d_i^2+d_0^2}$. Below we discuss the gapped case where the occupied and unoccupied bands are fully separated. 
    Similar to the regular procedure \cite{Zhao_PT,BJY_linking}, we can assume the base manifold as $\mathbb{S}^4$ covered by two open sets $D_N^4$ and $D_S^4$ (the northern and southern hyper-hemispheres) with the overlap as the equator being $\mathbb{S}^3=D_N^4\cap D_S^4$; see Fig. \ref{BDmode}(a). The occupied eigenstates defined on these two sets satisfy the relation:  $|u_{N,n}\rangle=(g_{NS})_{nm}|u_{S,m}\rangle$, where the transition function $g_{NS}\in \mathsf{SO}(4)$ is defined on the equator $\mathbb{S}^3$. Say, $|u_{N/S,n}\rangle$ denote the occupied eigenstates on $D_{N/S}$ with $n=1,2,3,4$.
Thus, this system is purely real and its  structure group of the fourfold-degenerate occupied bands is $\mathsf{SO}(4)=SU(2)\times SU(2)$.
The homotopy group of transition functions $\mathbb{S}^3\rightarrow \mathsf{SO}(4)$ is given by 
\begin{equation}
\pi_3(\mathsf{SO}(4))=\pi_3(\mathbb{S}^3)\oplus\pi_3(\mathbb{S}^3)=\mathbb{Z}\oplus\mathbb{Z},
\end{equation}
where we used the relation: $SU(2)\sim \mathbb{S}^3$ \cite{Bott}. Based on the above discussion, therefore, the system $\H_0$ can be characterized by both the first Pontryagin $P_1$ and the second Euler numbers $\chi_2$ in the whole 4D Brillouin zone (BZ) $\mathbb{T}^4$ defined as follows \cite{Eguchi,Nakahara_book},
\begin{equation}
P_1=\frac{1}{32\pi^2}\int_{\mathbb{T}^4} d^4k\,\epsilon^{\mu\nu\alpha\beta}\mathcal{F}^{ab}_{\mu\nu}\mathcal{F}^{ab}_{\alpha\beta},
\end{equation}
and 
\begin{equation}
\label{eq_second_euler}
\chi_2=\frac{1}{128\pi^2}\int _{\mathbb{T}^4}d^4k\,\epsilon^{abcd}\epsilon^{\mu\nu\alpha\beta}\mathcal{F}^{ab}_{\mu\nu}\mathcal{F}^{cd}_{\alpha\beta}.
\end{equation}
Here the real non-Abelian Berry curvatures are the anti-symmetric matrices due to its $\mathsf{SO}(4)$ nature defined as $\F_{\alpha\beta}=\partial_{\alpha}\A_{\beta}-\partial_{\beta}\A_{\alpha}+[\A_{\alpha},\A_{\beta}]$ with associated non-Abelian Berry connection $(\A_{\alpha})^{mn}=\langle u^m(k)|\partial_{k_\alpha}|u^n(k)\rangle$ with $|u^n(k)\rangle$ denotes the eigenstate of $n$-th occupied band. The direct calculation shows that $\chi_2=P_1/2$ with  $P_1=6\,\text{sgn}(m)$ for $0<|m|<2$; $P_1=-2\,\text{sgn}(m)$ for $2<|m|<4$, and $P_1=0$ elsewhere. Although in our Dirac model, the second Euler number is connected to the first Pontryagin number, however, this situation is not always true, because in general $P_1$ and $\chi_2$ are two dinstinct topological invariants. This can be understood by noticing that the first Pontryagin class is related to both the second $w_2$ and fourth $w_4$ Stiefel-Whitney classes \cite{Aharony} while the second Euler class depends only on $w_4$ \cite{Flagga}. In the next section, we will present another model that hosts a non-trivial $\chi_2$ but $P_1$ is zero. 
On the other hand,  since there is a tight connection between the first Pontrygain and the second Chern classes \cite{Eguchi,Nakahara_book}, it implies that
this model can also be characterized by the second Chern number $C_2$.
For simplicity, by rotating $\Gamma_5$ into $G_{300}$ through a unitary transformation $U$, one can block diagonalize $\H_0$ as
$U\mathcal{H}_0U^{-1}=\mathcal{H}_+\oplus \mathcal{H}_-$ with $U=\exp\left[i\frac{\pi}{4}G_{200}\right]\exp\left[-i\frac{\pi}{4}G_{220}\right]$.
Now each block Hamiltonian is given by
\begin{equation}
\mathcal{H}_{\pm}(k)=\pm d_xG_{11}\pm d_yG_{13}\pm d_zG_{30}+ d_wG_{12}\pm d_0G_{20},
\end{equation}
with $G_{ij}=\sigma_i\otimes\sigma_j$. The total model $\mathcal{H}_+\oplus\mathcal{H}_-$ preserves the original $PT$-symmetry with $PT=G_{220}\K$ while $\mathcal{H}_{\pm}$ belongs to class AII being the 4D topological insulator (TI) \cite{XLQi2008} that can be characterized by the same second Chern number $C_2^{+}=C_2^-$. Therefore, the total second Chern number is $C_2=C_2^++C_2^-$ which hosts the same result as $P_1$. Notice that the occupied $SU(2)$-bundle of a 4D TI has a high-energy analog with the $SU(2)$ instanton in Yang-Mills theory \cite{Belavin,Avron,Shankar} while our 4D real topological insulator with a $\mathsf{SO}(4)$-bundle (for the occupied bands) is analogous to the $\mathsf{SO}(4)=SU(2)_+\times SU(2)_-$ gravitational instanton in 4D Euclidean gravity \cite{Eguchi,Hawking,Oh}. Thus this is the  reason why our model hosts an  invariant $P_1=2C_2^{\pm}$ with double-value compared to the $SU(2)$ TI.
In what follows we discuss its edge states on the 3D boundary when considering an open boundary condition along $w$-direction at $2<m<4$ with $\chi_2=P_1/2=-1$.
We find that  the effective boundary Hamiltonian around the origin is nothing but a double-Weyl cone \cite{ProHam}, which takes the form
\begin{equation}
	\mathcal{H}_{DW}(\k)=-k_xG_{23}-k_yG_{21}-k_zG_{02}.
\end{equation}
Note that this model preserves $CP$-symmetry, i.e.,
$\left\{CP,\mathcal{H}_{DW}\right\}=0$ with $CP=\K$ satisfying $(CP)^{2}=+1$. Here $C$ denotes the charge conjugate symmetry.
Thus, this model presents a double-Weyl monople with a  $2\mathbb{Z}$ classification and carry monopole charge $C_1=-2$ defined in terms of the first Chern number on the  2D sphere encloses it \cite{YXZhao2016}. If we further introduce some typical $PT$-symmetric perturbations into the bulk Hamiltonian \eqref{RCIHam} without gap closing, the total model shares the same topological bulk invariants and only the physical changes are the boundary gapless modes. Without loss of generality and for concreteness, we consider the perturbation as $\Delta=a_0G_{100}+a_1G_{331}+a_2G_{333}+a_3G_{310}$ in the following.
When only $a_0\neq 0$,  we have $\H_1=\H_0+a_0G_{100}$.
After projecting this term into the $w=0$ boundary \cite{ProHam}, we obtain the total boundary Hamiltonian
\begin{equation}\label{NSHam}
	\mathcal{H}_{NS}(\k)=\mathcal{H}_{DW}(\k)+a_0\gamma^5,
\end{equation}
with $\gamma^5=-\sigma_3\otimes\sigma_2$. Note that $\gamma^5$ commutes with $\mathcal{H}_{DW}$, i.e, $\left[\gamma^5,\mathcal{H}_{DW}\right]=0$. This model represents a 3D Weyl nodal surface and its spectrum  $E=\pm|\k|\pm|a_0|$,
exhibits a band degeneracy at the Fermi level $E=0$
on a sphere defined by $|\k|=|a_0|$ with $|\k|=\sqrt{k_x^2+k_y^2+k_z^2}$.
This model preserves the combined $CP$ symmetry and should be characterized by a $\mathbb{Z}_2$ invariant \cite{YXZhao2016}. 
Say that $\gamma^5$ term splits the double-Weyl point into two single Weyl points in the opposite directions in energy with the same chirality, i.e., model \eqref{NSHam} denotes an inflated double Weyl point \cite{Turker2018,Salerno}.
On the other hand, when ${\boldsymbol a}=(a_1,a_2,a_3)$ is non-zero, we obtain the effective boundary Hamiltonian
\begin{equation}
\H_{DWs}(\k)=\H_{DW}(\k)+a_1G_{31}+a_2G_{33}+a_3G_{10}\,.
\end{equation}
The spectrum is now given by $E=\pm\sqrt{(k_x\pm a_1)^2+(k_y\pm a_2)^2+(k_z\pm a_3)^2}$,
which represents two Weyl points with the same chirality $C_1=-1$ are separated at $(\pm a_1,\pm a_2,\pm a_3)$.
In summary, from the viewpoint of boundary physics, a 4D real topological insulator is quite robust while a $PT$-preserved perturbation will not gap out the boundary double-Weyl mode but at most will split it into two single-Weyl points in opposite directions with the same chirality in energy  or in momentum, as shown in Fig. \ref{BDmode}(b-d). Moreover, this system hosts different numbers of boundary double-Weyl modes in different parameter regions \cite{BDmodes}. The 2D topological charge of the double-Weyl points is essential for the bulk boundary correspondence while the 2D topological charge implies that the boundary band structure with an odd number of double Weyl points cannot be realized by an independent 3D system, and therefore has to be connected to a higher dimensional bulk. 
Furthermore,  adding a $PT$-protected term $\Delta=\lambda G_{022}$ into $\H_0$, the model  goes through a phase transition from a $2\Z$ TI to a nontrivial real nodal-line semimetallic phase that is characterized by the first Euler number and then finally becomes a trivial insulator by further increasing $|\lambda|$. A real version of the 5D Yang monopole and the corresponding nodal structure has been also investigated in the Supplemental Material \cite{SM}.

\sect{Systematic homotopy modeling of second Euler phases}We now turn to a setup that allows us to unveil, for the first time, the intrinsic bulk characterization of 4D Euler topology. The central idea is to consider the simplest physical system that supports Euler topology, namely a five-band $PT^+$ symmetric system with four occupied and one unoccupied bands, thus realizing the 4D real Grassmannian $\mathsf{Gr}_{4,5}^{\mathbb{R}} = \mathsf{O}(5)/[\mathsf{O}(4)\times \mathsf{O}(1)]$, that is nothing but the projective hyperplane $\mathbb{R}P^4$. Focusing on strict 4D topological phases, we are seeking the systematic building of Bloch Hamiltonians representing the fourth homotopy classes 
\begin{equation}
\label{eq_homotopy_classification}
\begin{aligned}
 \pi_4[\mathbb{R}P^4] = \pi_4[\mathbb{S}^4] = \mathbb{Z}  
 \cong 2\mathbb{Z} \ni \chi_2[\mathcal{B}_I] \,,
\end{aligned}
\end{equation} 
where the Bloch bundle $\mathcal{B}_I = \mathcal{B}_1 \oplus \mathcal{B}_2 \oplus \mathcal{B}_3 \oplus \mathcal{B}_4 = \bigcup_{\bs{k}\in \mathbb{T}^4} \langle u_1(\bs{k}),\dots,u_4(\bs{k})\rangle$ is defined as the vector bundle spanned by the frame of Bloch eigenvectors $R_I(\bs{k})=(u_1(\bs{k})~u_2(\bs{k})~u_3(\bs{k})~u_4(\bs{k}))\in \mathbb{R}^{5\times 4}$ corresponding to the four occupied energy levels $E_{1}(\bs{k})\leq E_{2}(\bs{k})\leq E_{3}(\bs{k})\leq E_{4}(\bs{k})$. The factor two is explained below. 
We now seek the minimal $4+1$-gapped real Hamiltonian characterized by an eigenframe $R = (R_I~R_{II}) \in \mathsf{O}(5)$ with $R_{I}=(u_1\cdots u_4)$ and $R_{II} = u_5$, that generates all the second Euler phases. For this, we take advantage of the equivalence between the oriented Grassmannian and the four-sphere, \ie $\widetilde{\mathsf{Gr}}_{4,5}^{\mathbb{R}} = \mathsf{SO}(5)/\mathsf{SO}(4) \cong \mathbb{S}^4$. (The use of the oriented Grassmannian is motivated by $\pi_4[\widetilde{\mathsf{Gr}}_{4,5}^{\mathbb{R}}] = \pi_4[\mathsf{Gr}_{4,5}^{\mathbb{R}}]$.) Points of the Grassmannian are represented by left-cosets of oriented eigenframes $[R] = \{ R\cdot (G_I\oplus 1)\vert R\in \mathsf{SO}(5), G_I \in \mathsf{SO}(4)\}$, that can then be parameterized by the points of a four-sphere. Starting from the generic parametrization of an element $R= (u_1\cdots u_5)\in \mathsf{SO}(5)$, we use the following constraint that guarantees the $(4+1)$-Hamiltonian to correspond to a point of the oriented Grassmannian (derived from the Pl{\"u}cker embedding \cite{SM})
\begin{equation}
\label{eq_plucker}
\begin{aligned}
    [u_5]_{j} &= \sum\limits_{\sigma\in S_4}
    (-1)^{\sigma} [u_1]_{\sigma(\bs{i}_{j,1})}
    [u_2]_{\sigma(\bs{i}_{j,2})}
    [u_3]_{\sigma(\bs{i}_{j,3})}
    [u_4]_{\sigma(\bs{i}_{j,4})}\,,
\end{aligned}
\end{equation}
with $\{\bs{i}_j \}_{j=1}^{5} = \{(a,b,c,d)\vert 1\leq a < b < c< d \leq 5\}$ (\eg $\bs{i}_1 = (1,2,3,4)$), where $[u_n]_m$ is the $m$-th component of the $n$-th Bloch eigenvector, $S_{4}$ is the symmetric group of permutations of four elements and $(-1)^{\sigma}$ is the parity of the permutation $\sigma$. The expression of the vector $u_5$ is effectively obtained from the wedge product of the four occupied Bloch eigenvectors, \ie $u_1\wedge u_2\wedge u_3\wedge u_4 ``=" u_5$ \cite{SM}, such that any $\mathsf{SO}(4)$ rotation of the occupied $u_i$'s leaves the above definition of $u_5$ invariant. Setting $u_5$ as a generic unit vector living on the four-sphere, it is hence natural to consider the frame $(u_1\cdots u_4)$ as a basis of the tangent hyper-plane of the sphere at $u_5$. Picking the hyper-spherical coordinates, each point of the sphere is parametrized by
\begin{subequations}
\begin{equation}
\label{eq_hyperspherical_unit_vector}
    \begin{aligned}
    u_5 = \bs{e}_r(\phi,\theta,\psi,\rho) = \left(
            \begin{array}{l}
                \sin \rho \sin \psi \sin \theta \sin \phi \\
                \sin \rho \sin \psi \sin \theta \cos \phi \\
                \sin \rho \sin \psi \cos \theta  \\
                \sin \rho \cos \psi  \\
                \cos \rho
            \end{array}
        \right) \in \mathbb{S}^4 \,,
    \end{aligned}
\end{equation}
for $\phi\in[0,2\pi)$ and $\theta,\psi,\rho\in[0,\pi]$, and a basis of the tangent space at $\bs{e}_r(\phi,\theta,\psi,\rho)$ is given by
\begin{equation}
 (u_1,u_2,u_3,u_4) = \left(\bs{\partial}_{\phi},
 \bs{\partial}_{\theta},
 \bs{\partial}_{\psi},
 \bs{\partial}_{\rho}
 \right)\,,
\end{equation}
\end{subequations}
with $\bs{\partial}_{J} = \partial_{J} \bs{e}_r/ \vert \partial_{J} \bs{e}_r \vert$ for $J\in\{\phi,\theta,\psi,\rho\}$. Inserting the eigenframe $R=(\bs{\partial}_{\phi}~
 \bs{\partial}_{\theta}~
 \bs{\partial}_{\psi}~
 \bs{\partial}_{\rho}~\bs{e}_r)$ and the eigenvalues $E_1=E_2=E_3=E_4=-1$ and $E_1=1$ in the (4+1)-Hamiltonian form $H^{(4+1)} = R \cdot \text{diag}[E_1,E_2,E_3,E_4,E_5]\cdot R^T$, we arrive at the minimal form 
\begin{equation}
\label{eq_euler_H}
    H^{(4+1)}[\bs{e}_r(\phi,\theta,\psi,\rho)] = 2 \bs{e}_r \cdot \bs{e}_r^T - \mathbb{1}_5 \,.
\end{equation}
Mapping the Brillouin zone torus on the four-sphere, $f_{tts}:\mathbb{T}^4\rightarrow \mathbb{S}^4_0: \bs{k}\mapsto (\phi_0(\bs{k}),\theta_0(\bs{k}),\psi_0(\bs{k}),\rho_0(\bs{k}))$ so that $\mathbb{S}^4_0$ is wrapped one time whenever $\bs{k}$ is scanned through the Brillouin zone torus one time, see \eg \cite{SM}, and setting
\begin{equation}
\label{eq_ansatz}
    \bs{e}_r(\phi,\theta,\psi,\rho) = \bs{e}_r(W_4\phi_0,\theta_0,\psi_0,\rho_0)\,,
\end{equation}
where the number $W_4\in \mathbb{Z}$ fixes the number of times $\bs{e}_r$ wraps the target four-sphere (\ie $W_4$ is the degree of the map $\bs{e}_r:\mathbb{S}^4_0\rightarrow \mathbb{S}^4$), we finally obtain the Bloch Hamiltonian [combining Eq.\,(\ref{eq_euler_H}) and (\ref{eq_hyperspherical_unit_vector})]
\begin{equation}
    H_{W_4}(\bs{k}) = H^{(4+1)}[\bs{e}_r(W_4\phi_0(\bs{k}),\theta_0(\bs{k}),\psi_0(\bs{k}),\rho_0(\bs{k}))]  \,,
\end{equation}
from which we can represent all second Euler phases classified by Eq.\,(\ref{eq_homotopy_classification}). This can be shown through direct computation. Taking $\mathbb{S}^4_0$ as the base parameter space (since $f_{tts}(\bs{k})$ is surjective and almost one-to-one), we define the $\mathsf{SO}(4)$-connection $\mathcal{A}^{mn}_J $ and curvature $\mathcal{F}^{mn}_{IJ}$ for $I,J\in\{\phi_0,\theta_0,\psi_0,\rho_0\}$. We find the integrand of Eq.\,(\ref{eq_second_euler}) to be $96 \,W_4 \sin\theta (\sin\psi)^2 (\sin\rho)^3$, leading to the second Euler number
\begin{equation}
    \chi_{2}[\langle \bs{\partial}_{\phi},
 \bs{\partial}_{\theta},
 \bs{\partial}_{\psi},
 \bs{\partial}_{\rho} \rangle] = 2 W_4 \,.
\end{equation}
We conclude that the second Euler number of the Bloch Hamiltonian Eq.\,(\ref{eq_euler_H}) is simply determined by the winding number $W_4$ entering the ansatz Eq.\,(\ref{eq_ansatz}) of the unit vector $\bs{e}_r$. We can also trace the factor $2$ in Eq.\,(\ref{eq_homotopy_classification}) from the fact that the square of $\bs{e}_r$ enters the Bloch Hamiltonian Eq.\,(\ref{eq_euler_H}). In particular, when $W_4=1$ there is a formal equivalence between $\mathcal{B}_I$ and the tangent bundle of the four-sphere, i.e.\;$T\mathbb{S}^4$, and we recover the Chern-Gauss-Bonnet theorem in 4D, \ie   \begin{equation}
    \chi_2[\mathcal{B}^{(W_4=1)}_I]= \chi[\mathbb{S}^4] = 2\,,
\end{equation}
where $\chi$ is the Euler characteristic of the four-sphere. 
Tight-binding models with arbitrary fixed second Euler classes can be now readily generated by approximating every matrix element $[H_{W_4}(\bs{k})]_{\alpha\beta}$ by a truncated Fourier series $\sum_{\bs{m}\in\mathbb{Z}^4}^{\max (m_i)_i = N_{\alpha\beta}} t_{\alpha\beta}(\bs{m}) \exp(\imi \pi \bs{k}\cdot \bs{m}) $, where the Fourier coefficients represent the hopping amplitudes \cite{SM}.   
\begin{figure}[thb!]
\centering
\begin{tabular}{ll}
    $(a)\quad k_4=0$ & $(b)\quad k_3=0$ \\
\includegraphics[width=0.49\linewidth]{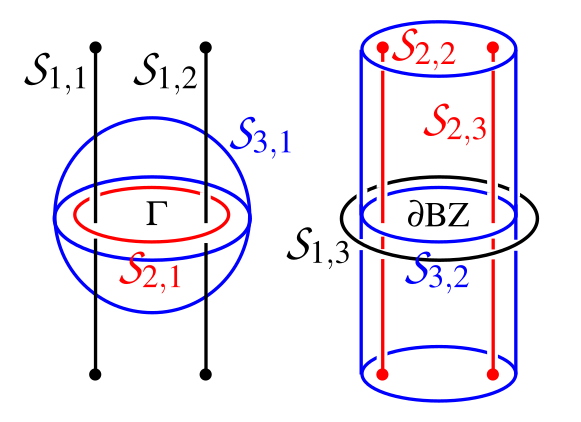} &
\includegraphics[width=0.49\linewidth]{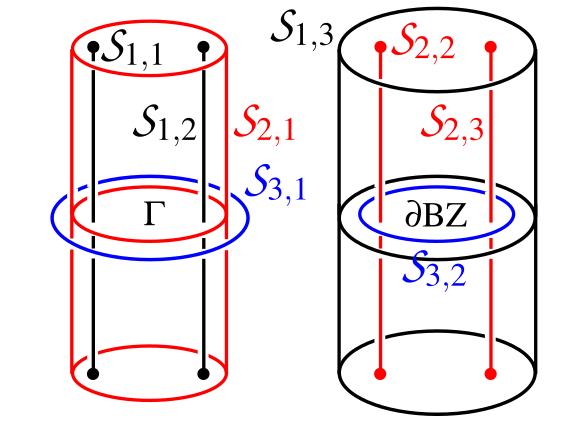} \\
    $(c)\quad k_3=k_4=0$, around $\Gamma$ & $(d)\quad k_3=k_4=0$, around  \\
    & $\quad\quad\quad\quad\quad\quad(k_1,k_2)=(1,1)$ \\
\includegraphics[width=0.48\linewidth]{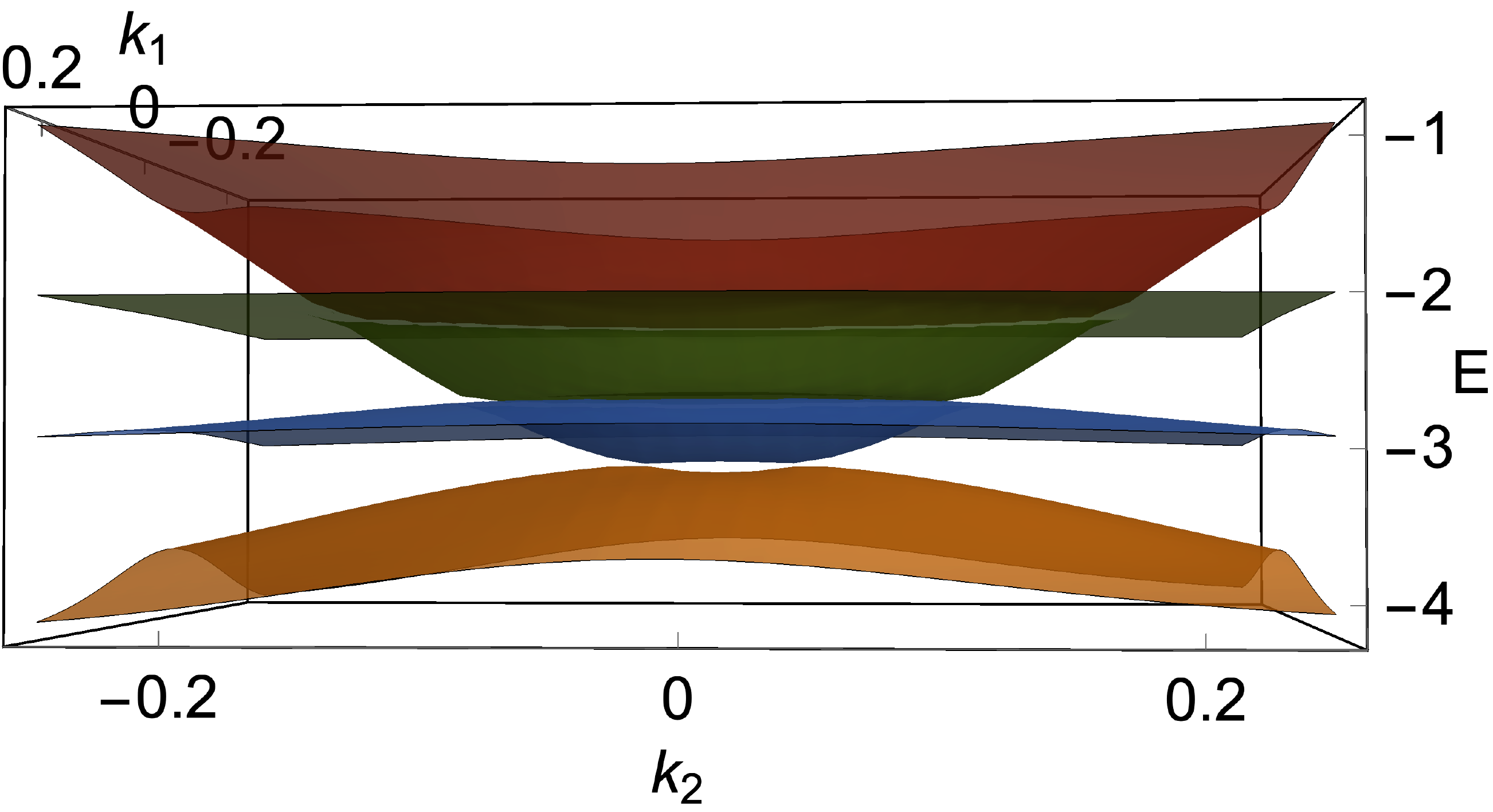} &
	\includegraphics[width=0.48\linewidth]{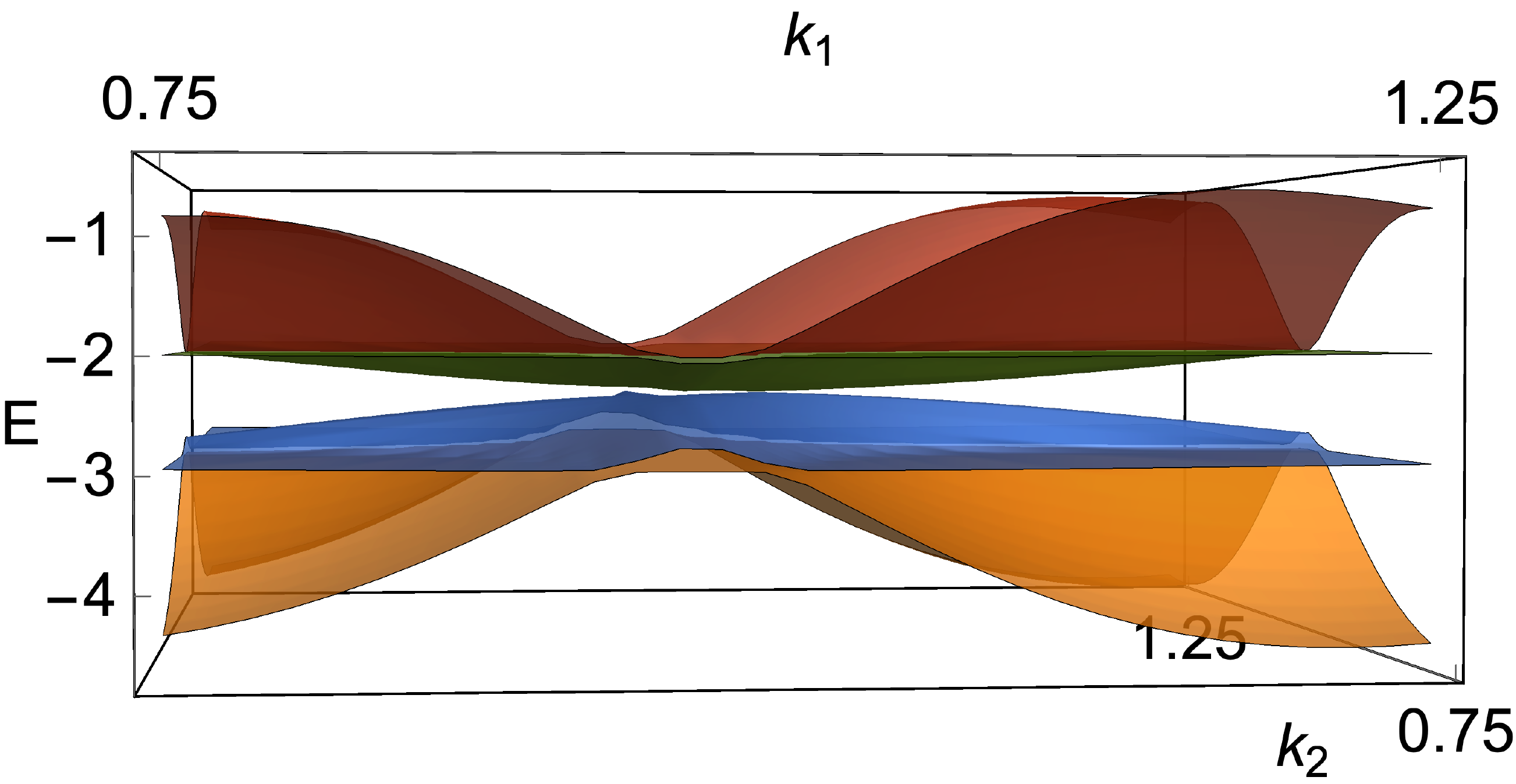}
\end{tabular}
	\caption{ Triplets of linked nodal surfaces (LNS) protected by the second Euler topology, seen in $(a)$ for the $(k_4=0)$-3D projection and in $(b)$ for the $(k_3=0)$-3D projection. The nodal surfaces between bands 1 and 2 ($\{\mathcal{S}_{1,1},\mathcal{S}_{1,2},\mathcal{S}_{1,3}\}$) are drawn in black, between bands 2 and 3 ($\{\mathcal{S}_{2,1},\mathcal{S}_{2,2},\mathcal{S}_{2,3}\}$) in red, and between bands 3 and 4 ($\{\mathcal{S}_{3,1},\mathcal{S}_{3,2}\}$) in blue. On the left-hand side of each panel, there is the triplet of LNS around the $\Gamma$ point, and, on the left-hand side, the triplet of LNS close to the Brillouin zone boundary $\partial \text{BZ}$. $(c,d)$ Band structure of the four occupied states over the $(k_3=k_4=0)$-2D projection $(c)$ around $\Gamma$ ($\bs{k}=\bs{0}$) and $(d)$ around the corner $(k_1,k_2)=(1,1)$. We find that every pair of nodal surfaces of a same gap is locally stable against their merging (as indicated by a nonzero first Euler number over a two-dimensional patch intersecting the nodal surfaces perpendicularly, see text).}
	\label{fig_LNS}
\end{figure}

\sect{Bulk nodal structure protected by the second Euler topology}We are now equipped with a generic homotopy-based tight-binding Bloch Hamiltonian which, by construction, only supports the second Euler topology. We show that any four-band subspace with a nontrivial second Euler number necessarily exhibits triplets of linked nodal surfaces connecting the four energy bands together, where each nodal surface is formed by the crossing of a pair of successive bands. More precisely, labeling the occupied bands from 1 to 4 with increasing energies, \ie $E_1(\bs{k})\leq E_2(\bs{k}) \leq E_3(\bs{k}) \leq E_4(\bs{k})$, the gap of every successive pair of bands, $\{E_n(\bs{k}),E_{n+1}(\bs{k})\}_{n=1,2,3}$, is closed along two-dimensional surfaces, noted $\mathcal{S}_{n,j_n} \in E_n(\bs{k}) \cap E_{n+1}(\bs{k})$ with $j_n\in J_n=\{1,2,\dots\}$, such that the nodal surfaces of the first energy gap $\{\mathcal{S}_{1,j_1}\}_{j_1\in J_1}$ and the third energy gap $\{\mathcal{S}_{3,j_3}\}_{j_3\in J_3}$ are linked with the nodal surfaces of the common adjacent energy gap $\{\mathcal{S}_{2,j_2}\}_{j_2\in J_2}$. Decomposing the total set of linked nodal surfaces into separated groups of linked nodal surfaces (LNS), which we label by $m=1,2,\dots,M$, each group thus forms a triplet of nodal surfaces $\{\mathcal{S}_{1,j^{m}_1},\mathcal{S}_{2,j^{m}_2},\mathcal{S}_{3,j^{m}_3}\}_{(j^m_1,j^m_2,j^m_3)\in J_1^m\times J_2^m \times J_3^m}$ with $\cup_{m} J^m_n = J_n$ for each $n=1,2,3$. We have observed that the number $M$ of separate LNS is fixed by the second Euler class, \ie in our case $M = \vert \chi_2[\mathcal{B}^{(W_4)}_{I}]\vert = 2 W_4 = 2$.   

Fig.\,\ref{fig_LNS} represents the two groups of LNS realized by the tight-binding model obtained for $W_4=1$ with a maximum hopping distance of $N_{\alpha\beta}=3$ in each direction, and setting the energy eigenvalues of the $(4+1)$-Hamiltonian form to $(E_1,E_2,E_3,E_4,E_5)=(-4,-3,-2,-1,2)$. Panel $(a)$ shows the $LNS$ in the $(k_4=0)$-3D projection, and $(b)$ in the $(k_3=0)$-3D projection. There are two separate $LNS$, one in the vicinity of the $\Gamma$ point ($\bs{k}=\bs{0}$) illustrated on the left-hand side of each panel, and one around the Brillouin zone boundary illustrated on the right-hand side. The panels $(c)$ and $(d)$ show the $(k_3=k_4=0)$-2D cut of the band structure of the occupied states, around $\Gamma$ in $(c)$ and around the corner $(k_1,k_2)=(1,1)$ in $(d)$. We have verified the local stability of every pair of nodal surfaces {\it from the same gap}, i.e.\;every pair in $\{\mathcal{S}_{n,j_n}\}_{j_n\in J_n}$ for a fixed $n$, through the evaluation of the first Euler number $\chi[\mathcal{B}_{n},\mathcal{B}_{n+1};\mathcal{D}_{j_n j_n'}]$ \cite{SM} computed over a two-dimensional patch $\mathcal{D}_{j_nj_n'}$ that crosses the pair of nodal surfaces perpendicularly. We conclude that the two triplets of LNS in Fig.\,\ref{fig_LNS}(a,b), \ie one around $\Gamma$ and one around the Brillouin zone boundary, are protected by the second Euler topology of the phase $\chi_2=2$, such that they cannot be annihilated as long as the main energy gap remains open, namely $\Delta_4 =  E_5(\bs{k})-E_4(\bs{k}) > 0$ for all $\bs{k}$. 

\sect{Experimental realizations} Finally, we turn to possible experimental realizations of 4D topological phases that are characterized by a second Euler number. In particular, the Dirac model proposed in our work represents the simplest theoretical model for 4D Euler insulators that can be realized in synthetic matter. Similarly, to the recent experimental proposal for the simulation of a 4D model presented in Ref.~\cite{YQZhu2022} by two of the authors, our Euler model can be also implemented in a similar cold-atom setup due to the same number of degrees of freedom. For instance, the fourth space-like dimensional can be emulated by employing synthetic dimension \cite{Celi,Ozawa2}, periodic driving (Floquet states) \cite{Demler,Peng}, quantum quench \cite{Chang,fluxesart,Ezawa2, Unal2019} or generalized Thouless pumping \cite{Kraus}. For these reasons, we envisage that 4D Euler insulators can be also engineered in several artificial systems ranging from ultracold atoms \cite{Lohse,Price,Spielman,Nascimbene}, photonics \cite{Zilberberg,Ozawa,Lu,DiColandrea} and meta-materials \cite{Prodan,Ma} to acoustic systems \cite{Chen,Chen2}, trapped-ion simulators \cite{Zhao2022}, superconducting systems \cite{Rastelli,Chan} and electric circuits \cite{Wang,Yu,Ezawa}.  

\sect{Conclusions and outlook} In conclusion, we have introduced two spacetime-inversion symmetric models exhibiting a quantized second Euler number. We have shown that the first model, given by a Dirac-like Hamiltonian, supports a rich number of boundary states, ranging from 3D double-Weyl monopoles to inflated Weyl points.  Moreover, we have presented a general approach to understand topological Euler phases through real Grassmanians for a generic number of Bloch bands. Importantly, our theory can directly be implemented in various synthetic settings that range from trapped ion insulators to ultracold atoms. There are several open questions in these novel topological phases that still need to be addressed, such as the developing of a semiclassical approach for wavepackets and an effective-field-theory description to study quantum transport, quantum anomalies and possible interacting phases. 
Notice that besides 4D synthetic Bloch bands, the second Chern number has been shown to play an important role also in the 4D phase-space of chiral magnets \cite{Rosch} and inhomogeneous crystals \cite{Niu}. Due to the natural link between the second Chern and Euler invariants shown in our paper, it would be then natural to consider $\chi_2$ as a possible topological number in the 4D phase-space of suitable quantum materials with defects and inhomogeneous order. This will open the door to the exploration of the second Euler invariant also in real quantum materials.
All these relevant points will be analyzed in future work.

\sect{Acknowledgments}
A.~B. was funded by a Marie-Curie fellowship, grant no. 101025315. 
R.~J.~S acknowledges funding from a New Investigator Award, EPSRC grant EP/W00187X/1, as well as Trinity college, Cambridge. 
\clearpage
\appendix
\section{Topological phase transitions:Insulator-metal-Insulator}\label{sec1}

\begin{figure}[http]\centering
	\includegraphics[width=8.6cm]{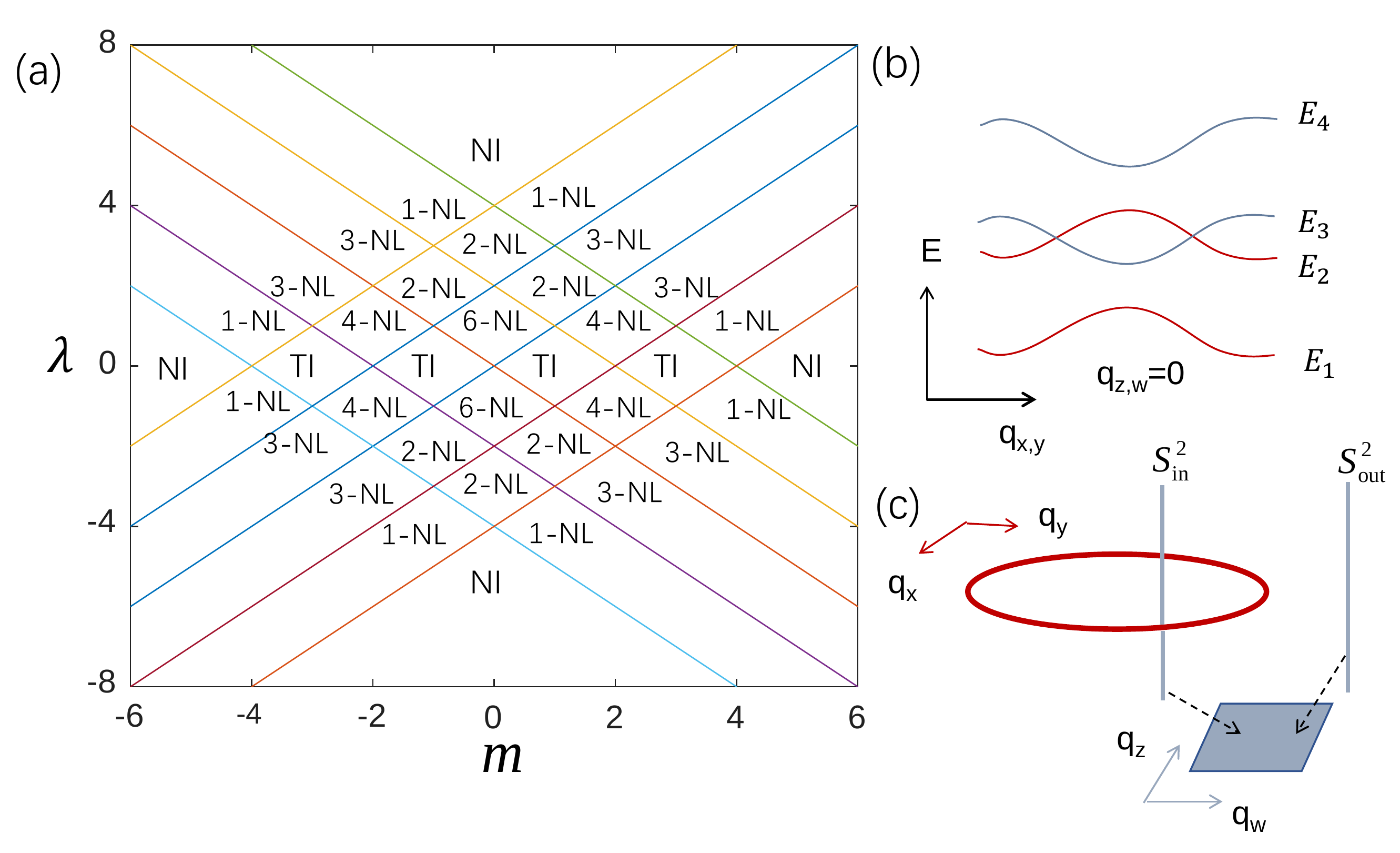}
	\caption{(a)Schematic of the phase diagram of $\H_1$ with parameters $(m,\lambda)$. The straight color lines denote the phase transition boundaries where the bulk gap closes at $E=0$.  Note that "TI" denotes real Euler TIs, while "NI" denotes normal insulators. "i-NL" denotes that the system hosts $i$ real nodal lines in the middle two degenerate bands, (c) Schematic diagram of a real nodal line. }
	\label{PD}
\end{figure}

By introducing a term $\Delta=\lambda G_{022}$ into $\H_0$, i.e., $\H_1=\H_0+\Delta$, the spectrum of the total system is given by,
\begin{equation}
E=\pm\sqrt{d_z^2+d_w^2+(\sqrt{d_x^2+d_y^2+d_0^2}\pm|\lambda|)^2},
\end{equation}
As we increase $|\lambda|$, the system will still be in the real insulating phase until the bulk gap closes and then the band crosses and forms nodal-line structure for the two middle double-degenerate bands satifying the condition,
\begin{equation}
	\begin{aligned}
		&d_z=d_w=0\rightarrow k_{z,w}=0,\pi,\\
		&d_x^2+d_y^2+d_0^2=\lambda^2\rightarrow \\
		&\sin^2 k_x+\sin^2k_y+(m-\sum_i \cos k_i)^2=\lambda^2.
	\end{aligned}
\end{equation}
The complete phase diagram is shown in Fig. \ref{PD}.

Note that $\Delta$ breaks $C_{2,xy}=G_{310}$ but keeps $C_{2,zw}=G_{302}$ and thus lifts the band degeneracy. The above $C_2$-symmetry is defined as 
\begin{equation}
	\begin{aligned}
	C_{2,xy}\H_0(k_x,k_y,k_z,k_w)C_{2,xy}^{-1}=\H_0(-k_x,-k_y,k_z,k_w),\\
	C_{2,zw}\H_0(k_x,k_y,k_z,k_w)C_{2,zw}^{-1}=\H_0(k_x,k_y,-k_z,-k_w),\\
	\end{aligned}
\end{equation}

Without loss of generality and for instance, we discuss a concrete example when $m=3$ and $5<\lambda<7$. The system supports one real nodal line expanded around $k_c= (\pi,\pi,\pi,\pi)$ satisfying the equation: $q^2_x + q^2_y = \rho^2_0$ with the radius $\rho_0 = \sqrt{\lambda^2-m^2}$ when $k_{z,w} =\pi$. The low-energy Hamiltonian of $H_1$ around $(\pi,\pi,\pi,\pi)$ is given by
\begin{equation}
\begin{aligned}
	H_{eff}&=H_0+V,\\
	H_0&=q_x\Gamma_1+q_y\Gamma_2+\tilde{m}\Gamma_0-i\lambda\Gamma_5\Gamma_4\Gamma_0,\\
	V&=-q_z\Gamma_3-q_w\Gamma_4.
	\end{aligned}
\end{equation}
where $\tilde{m}=m+\frac{q^2_{||}}{2}$ and $q_{||}=\sqrt{q_z^2+q_w^2}$.  For $H_{eff}$, there is a nodal-line structure formed by the middle two bands for each subsystem, thus we use the degenerate perturbation theory to calculate the four-level effective Hamiltonian by considering $V$ as a perturbation.

Its spectrum $E=\pm \lambda \pm\sqrt{q_x^2+q_z^2+\tilde{m}^2}$ as shown in Fig. \ref{PD}(b) when $q_z=q_w=0$. We label these bands as $E_i$ with $E_i<E_j$ for $i=1,2,3,4$.  Now we apply the degenerate perturbation theory for the middle two-degenerate bands, i.e., $(E_2,E_3)$, with the eigenstates $(|\psi_{2,1}\rangle,|\psi_{2,2}\rangle,|\psi_{3,1}\rangle,|\psi_{3,2}\rangle)$. After straightforward calculation into the first-order, i.e., $(\H_{RNL})_{ij}=\langle\psi_i|(H_{0}+V)|\psi_j\rangle$, we obtain the effective Hamiltonian $\H_{RNL}$ for this real nodal line is given by,
\begin{equation}
	\H_{RNL}=(\lambda-\sqrt{q_{\perp}^2+\tilde{m}^2})G_{03}-q_zG_{31}-q_wG_{11},
\end{equation} 
where $\tilde{m}=m+q^2_{||}/2$ with $q_{\perp}=\sqrt{q_x^2+q_y^2}$, and $q_{||}=\sqrt{q_z^2+q_w^2}$. 
Since the codimension of this nodal line is $d_c=d-d_{FS}-1=2$ and it is protected by $PT$-symmetry, we could introduce the first Euler number to characterize it.

One can define the topological invariant as 
\begin{equation}
	\nu_1=\chi_1^{\text{in}}-\chi_1^{\text{out}}~ \text{mod}~2,
\end{equation}
where $\chi_1^{\text{in}(\text{out})}$ denotes the first Euler number for the gapped subsystem $\H_{q_{\perp}}(q_z,q_w)$ inside (outside) the nodal line, i.e, $q_{\perp}<\rho_0(q_{\perp}>\rho_0)$, which is defined as 
\begin{equation}
	\begin{aligned}
	\chi_1&=\frac{1}{2\pi}\int^{q_{\perp}=q_0}_{\mathbb{R}^2} dq_zdq_w \F_{zw}^{12}\\
&	=\left\{\begin{array}{c}
	   1,~q_{0}<\rho_0,\\
	   0,~q_{0}>\rho_0.
	\end{array}\right.
\end{aligned}
\end{equation}

\section{5D Euler semimetals}
\subsection{Continuum models}
We start by considering a 5D real Weyl monopole with the continuum minimal model given by
\begin{equation}
\mathcal{H}_{RY}(k)=\sum_{i=1}^4 k_i\Gamma_i+k_5\Gamma_0,
\end{equation}
where $\Gamma_i$ takes the same form. These matrices satisfy a real Clifford algebra,
\begin{equation}
\left\{\Gamma_i,\Gamma_j\right\}=2\delta_{ij},~\Gamma_i^2=1.
\end{equation}
This monopole charge can be calculated by the first Pontryagin or second Euler number defined on the $S^4$ enclosing it with $\chi_2=P_1/2=1$.

\begin{figure}[http]\centering
\includegraphics[width=8.6cm]{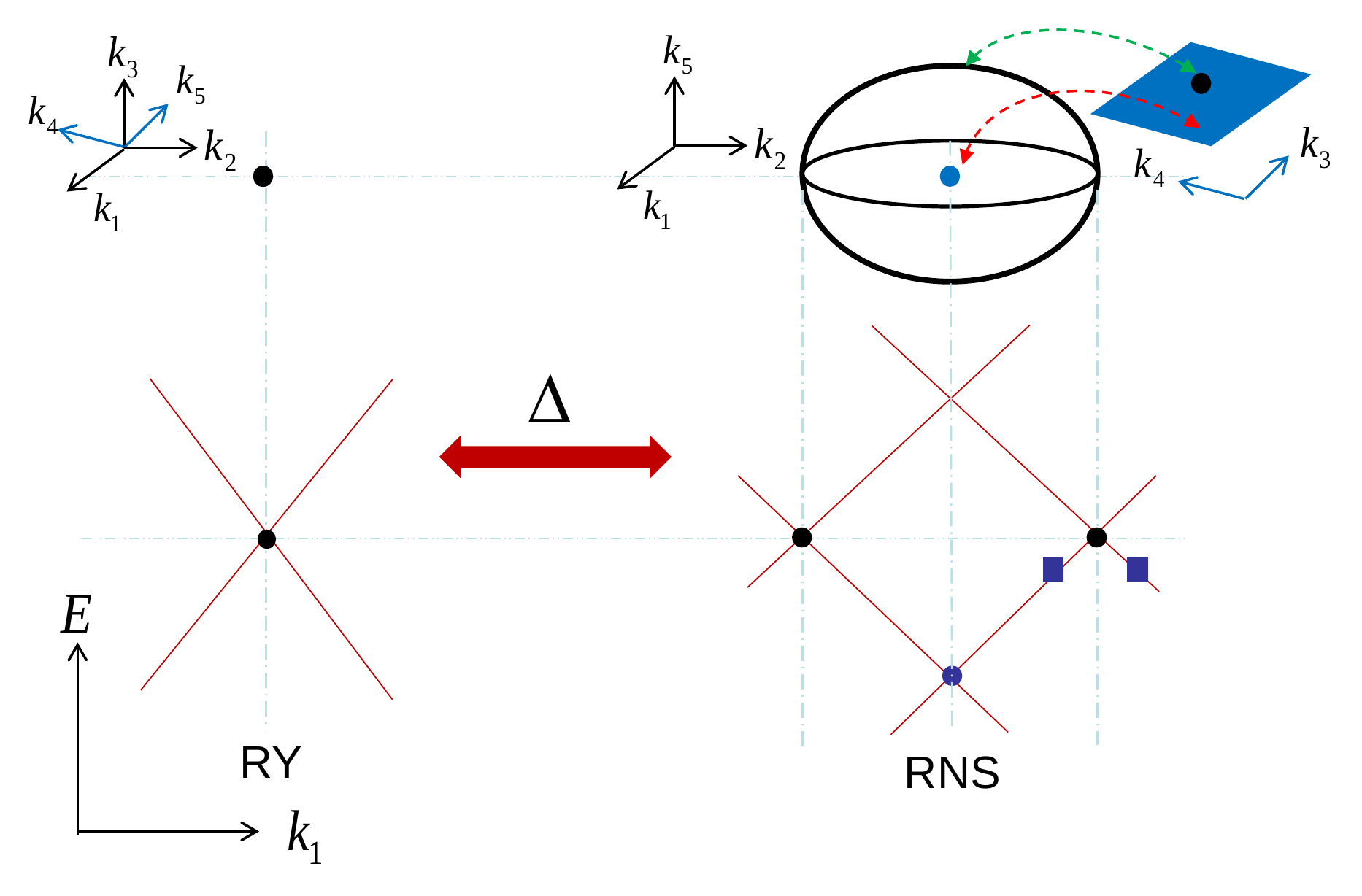}
\caption{ Schematic of energy spectra of a real Weyl point (left panel) and a real nodal sphere(right panel). (a) A 5D real Weyl point; (b) Nodal sphere structure (black) formed by splitting a double Weyl point in energy at $|\k|=|\lambda|$ when $k_{3,4}=0$. The lower two band crossing points are located on a 2D plane (blue)  $\k_p=(k_3,k_4)$ when $k_{1,2,5}=0$.}
\label{5DNS}
\end{figure}

Moreover, if we introduce an extra real perturbation to this monopole, e.g., $\Delta=\lambda G_{022}$, we have
\begin{equation}\label{RNS}
\H_{RNS}=\H_{RY}+\Delta.
\end{equation}
Since $G_{022}$ commutes with $\Gamma_{1,2,0}$ and anti-commutes with $\Gamma_{3,4}$, then we have the energy spectrum
\begin{equation}
E=\pm\sqrt{k_3^2+k_4^2+(\sqrt{k_1^2+k_2^2+k_5^2}\pm |\lambda|)^2}.
\end{equation}
It represents a real nodal sphere (RNS) band crossing with 4-fold degeneracy at $E=0$ on a sphere at $|\k|=|\lambda|$ with $\k=
\sqrt{k_1^2+k_2^2+k^2_5}$ when $k_{3,4}=0$. In addition, the two lower (upper) bands with two-fold degeneracy touch on an infinite 2D plane $\k_p=(k_3,k_4)$ when $k_{1,2,5}=0$, see Fig. \ref{5DNS}.

This nodal-surface monopole carries two topological invariants: one is the original second Euler number $\chi_2=1$ (or the first Pontryagin number $P_1=2$) defined for the lower two bands on a $\mathbb{S}^4$ enclosing it, the other one should be a 2D invariant since $d_c=d-d_{FS}-1=2$ with $d=5$ and $d_{FS}=2$\cite{YXZhao2016}. One can define the first Euler number for the higher occupied band with two-fold degeneracy anywhere inside and outside the sphere with varying $k_3$ and $k_4$, i.e., $|\k|<|\lambda|$ and $|\k|>|\lambda|$, respectively (labeled by blue squares as in Fig. \ref{5DNS}). For instance, we define the first Euler number as
\begin{equation}
\chi_1=\frac{1}{2\pi}\int_{|\k|=|\k_0|} dk_3dk_4 \mathcal{F}^{12}_{k_3k_4}.
\end{equation}
Thus the total invariant of this nodal surface can be expressed as
\begin{equation}
\nu_2=\chi_1^{\text{in}}-\chi_1^{\text{out}}~ \text{mod}~2.
\end{equation}
Therefore, a nodal surface described by Eq. \eqref{RNS} hosts two topological invariants $(\chi_2,\nu_2)=(1,1)$.

\subsection{Lattice models}
A lattice model of the corresponding 5D semimetal $\H_{ESM}$ can be given by
\begin{equation}
\H_{ESM}(k)=\sum_{a=1}^4d_a\Gamma_a+d_5\Gamma_0,
\end{equation}
with the Bloch vector
\begin{equation}
d_a=\sin k_a,~d_5=m-\sum_{i=1}^5\cos k_i, 
\end{equation}
where $a=1,2,3,4$. Without loss of generality,  we only consider the case when $3<m<5$ which hosts a pair of 5D real Weyl points along $k_5$ direction at $k_{\pm}=(0,0,0,0,\pm\arccos(m-4))$. For simplicity, we consider the case when $m=4$ where a pair of Weyl nodes located at $(0,0,0,0,\pm \pi/2)$. The low-energy effective Hamiltonian is given by
\begin{equation}
\H_{RY}^{\pm}=\sum_{a} k_a\Gamma_a\pm k_5\Gamma_0.
\end{equation}
Each monopole carry a topological charge $\chi_2^{\pm}=\pm1$.  Note that each 4D subsystem $\H_{ESM,k_5}(k_1,k_2,k_3,k_4)$ with a fixed $k_5$ beyond two monopoles is the 4D real Chern insulator as we discussed above. The 4D sub systems are nontrivial  when $k_5\in (-\pi/2,\pi/2)$ while the sub systems are trivial elsewhere.

After introducing a perturbation $\Delta$, i.e.,
\begin{equation}\label{ENSM}
 \H_{ENSM}=\H_{ESM}+\Delta.
 \end{equation}
 Each monopole inflates into a nodal surface and carries double charges $(\chi_2^{\pm},\nu_2^{\pm})=(\pm1,\pm1)$.  The gapped 4D subsystems between two nodal surfaces are also the nontrivial phases. The energy spectrum is given by
 \begin{equation}\label{ENSMSpec}
E=\pm\sqrt{d_3^2+d_4^2+(\sqrt{d_1^2+d_2^2+d_5^2}\pm |\lambda|)^2}.
\end{equation}

\begin{figure*}[htbp]
\centering
\includegraphics[width=16cm]{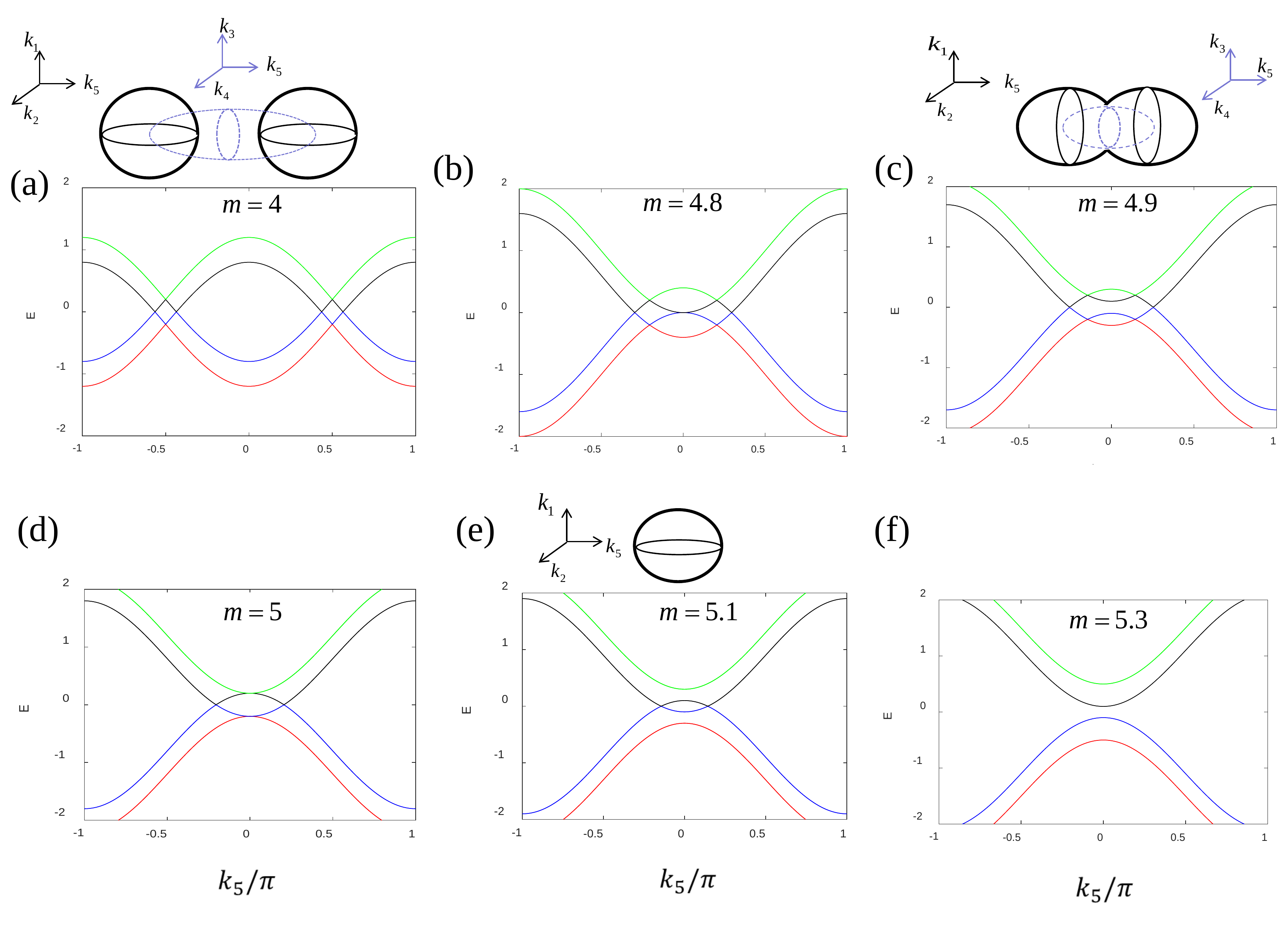}
\caption{Energy spectra evolution of a pair of RNSs  for Eq. (\ref{ENSMSpec})
with $k_{2,3,4,5}=0$. (a)The system hosts a pair of RNSs centered at  $k_{\pm}$. (b) Two RNSs are emerging together. (c) The centers of the two spheres are coincident. (d) Two RNSs vanish but the two middle bands (colored in black and blue) touch to form an ordinary NS. We choose $\lambda=0.2$ in the above pictures. } \label{Evolution}
\end{figure*}

For concreteness, we would like to work in a parameter region, e.g., $3+|\lambda|< m \leq 5+|\lambda|$ with $|\lambda|<1$, where we can study the merging process of a single pair of RNSs. The effective Hamiltonian of this merging process can be obtained by expanding the model Hamiltonian near the origin $k=0$, which  takes the form
\begin{equation}\label{EVNS}
\mathcal{H}_{m}(k)=\sum_ak_a\Gamma_a+(m-5+\frac{k^2}{2})\Gamma_0+\Delta,
\end{equation}
where $k=\sqrt{\sum_{i=1}^5k_i^2}$. Fig. \ref{Evolution} shows the evolution of the energy spectra with double band inversion\cite{BJY_linking}. As we increase $m$ from $m=3+|\lambda|$, two RNSs carrying opposite charge are created with ${k}_{\pm}$ as the center when $k_{3,4}=0$ [Fig. \ref{Evolution}(a)] and move together along $k_5$ axis, then touch ($m=5-|\lambda|$) [Fig. \ref{Evolution}(b)] and merge together [Fig. \ref{Evolution}(c)]  until $m=5$ [Fig. \ref{Evolution}(d)]. Subsequently, the RNSs totally vanish as the upper and lower two bands respectively open the gap  [Fig. \ref{Evolution}(e)]. However, the middle two bands still touch and become an ordinary NS characterized by $\nu_2$ and later shrinks to a point when $m=5+|\lambda|$ and finally disappears with gap opening when $m>5+|\lambda|$ [Fig. \ref{Evolution}(f)]. Note that the lower two bands touch with the condition: $k_1^2+k_2^2+(m-5+k^2/2)^2=0$, which leads
\begin{equation}
\begin{aligned}
k_{1,2}=0,~k_3^2+k_4^2+k_5^2=2(5-m).
\end{aligned}
\end{equation}
This forms a nodal sphere when $m<5$. The total nodal structure of this merging process is presented above the subfigures.

\section{Homotopy based modeling of {\it \textbf{ real}} 4D phases} 

We want to explore the properties of 4D topological phases that host a nontrivial second Euler number. (See the definition of the 4D reciprocal Bravais lattice and Brillouin zone in SM \ref{sec_bloch_periodicity}.) Every gapped 4D $PT^+$-symmetric phase corresponds to a homotopy class in $[\mathbb{T}^4,\mathsf{Gr}_{p,N}^{\mathbb{R}}]$, with as the base space the four-dimensional Brillouin zone $\mathbb{T}^4$ and as the target space the real non-orientable Grassmannian
\begin{equation}
    \mathsf{Gr}_{p,N}^{\mathbb{R}} = \mathsf{O}(N)/[\mathsf{O}(p)\times \mathsf{O}(N-p)]\,.
\end{equation}
Similarly to the first Euler number that is supported only by (2D) {\it orientable two-band} subspaces isolated from other energy bands \cite{BJY_linking,BJY_nielsen,bouhon2019nonabelian,bouhon2020geometric, Bouhon_geo2} (see also the earlier works that identified the $\mathbb{Z}$ winding of Wilson loop of two-band subspaces without invoking the notion of Euler class \cite{Wi1,BzduSigristRobust,bouhon2018wilson}), 
the second Euler number is supported only by (4D) {\it orientable four-band} subspaces. We discuss below the consequence on the homotopy classification of Hamiltonians of the fact that band-subspaces are at best {\it orientable} and not {\it oriented}, see also Refs.\,\cite{bouhon2020geometric} and \cite{Bouhon_geo2}. 

One remark here on the semantics. When we talk about a $d$-dimensional system, we refer to the dimension of its parameter (momentum) space, \ie compatible with a Brillouin zone $\mathbb{T}^d$. We then refer to the dimensionality of band subspaces as their {\it rank}. In other words, nontrivial first Euler number is supported by rank-2 band subspaces of two-dimensional systems, while nontrivial second Euler number is supported by rank-4 band subspaces of four-dimensional systems.

The condition of orientability concretely means that we exclude $\pi$-Berry phases over any non-contractible path of the 4D Brillouin zone torus \cite{bouhon2020geometric}. However, there remains a caveat from the fact that the vector bundle associated with the occupied Bloch eigenvectors is only {\it orientable} and not {\it oriented} \cite{bouhon2020geometric}. This follows from the fact that the orientation of the eigenframe can be changed via a gauge transformation, thus leaving the Bloch Hamiltonian unchanged, see below. 



Formally, we are thus considering $[\mathbb{T}^4,\widetilde{\mathsf{Gr}}_{p,N}^{\mathbb{R}}]$ with, as the target space, the {\it orientable} Grassmannian 
\begin{equation}
\label{eq_orgrass}
    \widetilde{\mathsf{Gr}}_{p,N}^{\mathbb{R}} = \mathsf{SO}(N)/[\mathsf{SO}(p)\times \mathsf{SO}(N-p)]\,.
\end{equation}
For simplicity we also exclude any $2D$ (first) Euler phase on the two-dimensional tori of the Brillouin zone. It turns out that all sub-dimensional topological phases are effectively discarded by considering the fourth homotopy group of the {\it non-orientable} Grassmannian, \ie $\pi_4(\mathsf{Gr}_{p,N}^\mathbb{R})$, since the CW structure of the four-sphere only contains a four-dimensional cell and a base point (also given that $\pi_0(\mathsf{Gr}_{p,N}^{\mathbb{R}}) = \{e\}$). In particular, we have $\pi_4(\widetilde{\mathsf{Gr}}_{p,N}^{\mathbb{R}})=\pi_4(\mathsf{Gr}_{p,N}^{\mathbb{R}})$. 

We emphasize that by avoiding any non-trivial winding over the sub-dimensional cells of the Brillouin zone, we are effectively restricting ourselves to periodic Bloch Hamiltonians, \ie 
\begin{equation}
    H(\bs{k}+\bs{K}) = H(\bs{k})\;\text{for\;all}\;\bs{K}\in \Lambda^*\,.
\end{equation}
This justifies {\it a posteriori} that we chose the Brillouin zone torus as the base space (see Section \ref{sec_bloch_periodicity}). Furthermore, by avoiding sub-dimensional topologies over non-contractible cells of the Brillouin zone, the image of the Bloch Hamiltonian restricted on the boundary of the Brillouin zone, \ie $H(\partial \text{BZ})\subset \mathsf{Gr}_{p,N}^{\mathbb{R}}$, is null-homotopic within the Grassmannian. In other words, there exists an adiabatic deformation of the Bloch Hamiltonian shrinking the image of the Brillouin zone boundary, $H(\partial \text{BZ})$, to a point. We will use this in the construction of tight-binding models below.

Addressing now the restriction to rank-4 subspaces, we thus focus on the homotopy classification of real Hamiltonian with at least one four-band subspace, \ie for $p=4$ or $N-p=4$. Since $\mathsf{Gr}_{p,N}^{\mathbb{R}} = \mathsf{Gr}_{N-p,N}^{\mathbb{R}}$, we hence consider $\mathsf{Gr}_{4,N}^{\mathbb{R}}$ without loss of generality. 

The long exact sequence of homotopy groups of fiber bundles $F\xhookrightarrow{i} E \xrightarrow{p} B$, with the total space $E=\mathsf{O}(N)$, the base space $B=\mathsf{Gr}_{4,N}^{\mathbb{R}}$ and the fiber $F=\mathsf{O}(4)\times \mathsf{O}(N-4)$, gives
\begin{equation}
    \pi_4(\mathsf{Gr}_{4,N}^{\mathbb{R}}) = \mathbb{Z}^{(\chi_{2,I})} \oplus \mathbb{Z}^{(\text{P})}
    \,,\;\text{for~all}\; N\in \{3\}\cup \mathbb{N}_{\geq5 } \,,
\end{equation}
while $N=4$ gives
\begin{equation}
    \pi_4(\mathsf{Gr}_{4,8}^{\mathbb{R}}) = \mathbb{Z}^{(\chi_{2,I})} \oplus
    \mathbb{Z}^{(\chi_{2,II})} \oplus
    \mathbb{Z}^{(\text{P})}\,,
\end{equation}
and 
\begin{equation}
    \pi_4(\mathsf{Gr}_{4,N}^{\mathbb{R}}) = \mathbb{Z}^{(\chi_{2,I})} \,,
    \;\text{for}\;N\in \{1,2\}\,.
\end{equation}
For all these cases, the $\mathbb{Z}^{(\chi_{2,I})}$ invariant stands for the second Euler number of one rank-4 band subspace. Then, there is an extra $\mathbb{Z}^{(\chi_{2,II})}$ invariant when also the second band subspace has four bands, \ie when $p=N-p=4$. Indeed, similarly to the first Euler number that can be hosted independently by every two-band subspace of a (two-dimensional) multi-band system, every four-band subspace of a (four-dimensional) multi-band system can host a second Euler number independently. (There is however a restriction on the parity of the sum of all Euler classes in order to guarantee a global trivial Stiefel-Whitney class [see below for more details].) Finally, the third $\mathbb{Z}^{(\text{P})}$ invariant, found when both $p\geq 3$ and $N-p\geq 3$ hold, comes from the existence of a complementary stable invariant in 4D, namely the Pontryagin number. The detailed discussion of the Pontryagin topology goes beyond the scope of this work and will be addressed elsewhere in detail. 

\subsection{Classification in terms of transition functions}\label{ap_transition_fct}

Embedding $\mathcal{H}_0$ in a Bloch Hamiltonian with an infinite number of unoccupied bands, $\mathcal{H}_0^{(\infty)}$, we remind the bijection between the group of homotopy classes of clutching functions $\pi_3( \mathsf{SO}(4))$ and the group of homotopy classes of Bloch Hamiltonian $\pi_4(\mathsf{Gr}_{4,\infty}^{\mathbb{R}})$ . Below we address the geometric modelling of tight-binding models with Euler topology based on the structure of the real Grassmannians.

 The homotopy classification of all Euler phases in that context is obtained from
\begin{multline}
    [\mathbb{T}^4,\mathsf{Gr}_{4,5}^{\mathbb{R}}] \subset \pi_4(\mathsf{Gr}_{4,5}^{\mathbb{R}}) \oplus \left\{\pi_2^{(e^2_i)}(\mathsf{Gr}_{4,5}^{\mathbb{R}})\right\}_{i=1}^{6} \\
    \oplus \left\{\pi_1^{(e^1_i)}(\mathsf{Gr}_{4,5}^{\mathbb{R}})\right\}_{i=1}^{4} \,,    
\end{multline}
where $e_i^{d}$ is the $i$-th $d$-dimensional cell of the CW structure of the Grassmannian (the number of $d$-th homotopy groups matches with the number of $e_i^{d}$ cells). The first homotopy classes are indicated by the first Stiefel-Whitney class (equivalent to the Berry phase factor) $w_{1,i}=w_1[e^1_i] = e^{\text{i}\gamma_B[e^1_i]} \in \mathbb{Z}_2$ and the second homotopy classes by the first Euler class $\chi_{1,i}=\chi_1[e^2_i]\in \mathbb{Z}$ \cite{bouhon2020geometric}, while there is no contribution from the 3D cells since $\pi_3(\mathsf{Gr}_{4,5}^{\mathbb{R}})=0$. All Berry phases must vanish since the Euler classes are only defined for orientable phases \cite{bouhon2020geometric} (see also \cite{SM}).

\subsection{{Orientability and reduced homotopy class.}}

Strictly speaking the topological classification of gapped Bloch Hamiltonians is obtained from the set of {\it free} (\ie without fixed base point) homotopy classes over the Brillouin zone torus, \ie in our case $[\mathbb{T}^4,\mathbb{R}P^4]$, which also contains the 1D non-orientable topology indicated by the first Stiefel-Whintey class (equivalently, the Berry phase factor) over the four loop-cycles of the Brillouin zone torus $\{l_i\}_{i=1}^{4}$, \ie $w_1[l_i] = e^{\text{i} \gamma_B[l_i]} \in \pi_1[\mathbb{R}P^4] = \mathbb{Z}_2$. (There is no other sub-dimensional topologies since $\pi_2[\mathbb{R}P^4]=\pi_3[\mathbb{R}P^4]=0$.) The Euler class is only defined for {\it oriented} vector bundles, such that we restrict ourselves to {\it orientable} Bloch bundles $\mathcal{B}_I$ characterized by trivial Berry phase factors $e^{\text{i} \gamma_B[l_i]}=0$ (we say orientable because the $\mathcal{B}_I$'s orientation can be flipped by a change of gauge). A remaining caveat is the non-trivial action of the generator $[l^{[1]}]\in\pi_1[\mathbb{R}P^4]$ on the elements of the fourth homotopy group $\beta\in \pi_4[\mathbb{R}P] = \mathbb{Z}$ via $\beta\rightarrow \beta^{-1}=-\beta$, which reverses the orientation of the Bloch bundle and with it the sign of the second Euler class \cite{bouhon2020geometric}. (The gauge freedom in the choice of an orientation for the Bloch eigenframe is manifested by the absence of favored base point in the definition of the homotopy classes of Bloch Hamiltonians.) As a consequence the topological Euler phases are classified up to homotopy by the unsigned second Euler class $\vert \chi_2\vert \in 2\mathbb{N}$.

\subsection{Tight-binding model for $\mathsf{Gr}_{4,5}^{\mathbb{R}}$ with a fixed second Euler class}

In the following we will focus on the systematic derivation of $4D$ tight-binding models of the Bloch Hamiltonians representing the homotopy classes of $\pi_4(\mathsf{Gr}_{4,5}^{\mathbb{R}})=\mathbb{Z}^{(\chi_{2,I})}$. The final result must thus be a five-band Hamiltonian with a four-band subspace hosting an arbitrarily fixed second Euler number. 

For this, we first remind the relation with the four-sphere
\begin{equation}
    \mathsf{Gr}_{4,5}^{\mathbb{R}} = \mathbb{R}P^4 = \mathbb{S}^4/\sim\,,
\end{equation}
with $x_1 \sim x_2$ whenever $x_1$ and $x_2$ are two antipodal points of the four-sphere. More precisely, there is a two-to-one (surjective) universal covering map 
\begin{equation}
    q_4: \mathbb{S}^4 \rightarrow \mathbb{R}P^4\,,
\end{equation}
such that if we wrap the four-sphere a number $W_4$ of times via the map
\begin{multline}
    f_{W_4}:\mathbb{S}^4_0\rightarrow \mathbb{S}^4 : \\
    (\phi_0,\theta_0,\psi_0,\rho_0) \mapsto 
    (\phi,\theta,\psi,\rho)=(W_4\phi_0,\theta_0,\psi_0,\rho_0)\,,
\end{multline}
where we have used the hyper-spherical coordinates of the four-sphere of radius $r>0$, \ie
\begin{equation}
\label{eq_hyperspherical}
    \begin{aligned}
        \bs{x} = \left(
            \begin{array}{l}
                r\sin \rho \sin \psi \sin \theta \sin \phi \\
                r\sin \rho \sin \psi \sin \theta \cos \phi \\
                r\sin \rho \sin \psi \cos \theta  \\
                r\sin \rho \cos \psi  \\
                r\cos \rho
            \end{array}
        \right) \in \mathbb{R}^5-\{ \bs{0}\} \,,
    \end{aligned}
\end{equation}
with $\phi\in[0,2\pi)$ and $\theta,\psi,\rho\in[0,\pi]$, the composition with $q_4$ leads to the wrapping of $\mathbb{R}P^4$ a number $2W_4$ of times. In the following we will refer to $W_4$ as the four-dimensional winding number. It then only remains to map the Brillouin zone torus on the base sphere $\mathbb{S}^4_0$, which we simply do through 
\begin{equation}
    \begin{aligned}
    f_{tts} : \mathbb{T}^4 &\rightarrow \mathbb{S}^4_0 : 
    \bs{k}=(k_1,k_2,k_3,k_4) \mapsto \\
    \phi_0 &= \arg (k_1+\imi k_2)    \,,\\
    \theta_0 &= \arccos \left(k_3/\sqrt{k_1^2+k_2^2+k_3^2}\right) \,,\\
    \psi_0&= \arccos \left(k_4/\sqrt{k_1^2+k_2^2+k_3^2+k_4^2}\right)\,,\\
    \rho_0&= \pi \max\{ \vert k_1\vert, 
    \vert k_2\vert,
    \vert k_3\vert,
    \vert k_4\vert\} \,,
    \end{aligned}
\end{equation} 
with $-1\leq k_i \leq 1$ for $i=1,2,3,4$.

We recapitulate the above relations with the diagram,
\begin{equation}
\label{eq_diag}
\begin{tikzcd}[]
\mathbb{T}^4 \arrow[r, "f_{tts}", line width=0.5, black] & \mathbb{S}^4_0 
\arrow[d, "f_{W_4}"', line width=0.5, black] \arrow[rd, "q_4 \circ f_{W_4}", line width=0.5, black] & \\
& \mathbb{S}^4 \arrow[r, "q_4", line width=0.5, black] & \mathbb{R}P^4\,.
\end{tikzcd}
\end{equation}
Our goal is thus to obtain a generic Bloch Hamiltonian that realizes the map 
\begin{equation}
    H_{W_4}(\bs{k}) = (q_4 \circ f_{W_4}\circ f_{tts})(\bs{k}) \,,
\end{equation}
for any fixed winding number $W_4\in \mathbb{Z}$. It is also clear from above that the second Euler class simply matches with $W_4$, \ie 
\begin{equation}
    \chi_2[H_{W_4}(\bs{k})] = W_4 \in \pi_4(\mathbb{S}^4) = \pi_4( \mathsf{Gr}_{4,5}^{\mathbb{R}}) =  \mathbb{Z}^{(\chi_{2,I})}\,.
\end{equation}
We note that by construction the Bloch Hamiltonian $H_{W_4}(\bs{k})$ is periodic over the first Brillouin zone and, in particular, it is orientable, \ie every non-contractible path of the Brillouin zone has a zero Berry phase, which is a necessary condition to define a nonzero second Euler class \cite{Hatcher_2}. Reminding the homotopy equivalence $H_A(\bs{k}) \equiv H_B(\bs{k})$ whenever $ \chi_2[H_B(\bs{k})] = -\chi_2[H_A(\bs{k})]$, we conclude 

The generic form of the flat (\ie with a flat spectrum) Bloch Hamiltonian is 
\begin{equation}
\begin{aligned}
    \mathcal{H}^{(\text{flat})} &=
    \sum\limits_{\bs{k}\in \mathbb{T}^4} 
    \vert \bs{\phi},\bs{k} \rangle 
    H^{(\text{flat})}(\bs{k}) \langle \bs{\phi},\bs{k} \vert \,,\\
    &= \sum\limits_{\bs{k}\in \mathbb{T}^4} 
    \vert \bs{\phi},\bs{k} \rangle 
    R(\bs{k})  
    \left[\begin{array}{cc}
        -\mathbb{1}_4 & 0 \\
        0 & 1
    \end{array}\right]  
    R(\bs{k})^T \langle \bs{\phi},\bs{k} \vert \,,
\end{aligned}
\end{equation}
with the eigenframe 
\begin{equation}
    R(\bs{k}) = \left(u_1(\bs{k})~
    u_2(\bs{k})~
    u_3(\bs{k})~
    u_4(\bs{k})~
    u_5(\bs{k})\right) \in \mathsf{O}(5) \,. 
\end{equation}
Our task is thus to derive a analytical parametrization of the eigenframe $R(\bs{k})$ such that we capture the nontrivial winding associated to the Euler topology. First of all, we can always take the eigenframe in $\mathsf{SO}(5)$, by substituting $R(\bs{k})\rightarrow R(\bs{k})\,\text{diag}[1,1,1,1,\det( R(\bs{k}))]$. Actually, the flat Hamiltonian is explicitly invariant under every gauge transformation of the form
\begin{equation}
    R(\bs{k}) \rightarrow R(\bs{k})\cdot \left[\begin{array}{cc}
        G_I & 0 \\
        0 & s_{II}
    \end{array}
    \right]\,,
\end{equation}
with $s_{II}\in \{+1,-1\}$ and $G_I\in\mathsf{O}(4)$. The above gauge structure of the Bloch Hamiltonian is summarized in the quotient form of the Grassmannian 
\begin{equation}
    \mathbb{R}P^4= \mathsf{Gr}_{4,5}^{\mathbb{R}} =  \mathsf{O}(5)/[\mathsf{O}(4)\times \mathsf{O}(1)] \,,
\end{equation}
such that each Hamiltonian is represented by a right coset $[R]=R\cdot [\mathsf{O}(4)\oplus \mathsf{O}(1)]$. Since we have seen that there is no difference between $\mathbb{R}P^4$ and $\mathbb{S}^4$ from the fourth homotopy group's viewpoint, we can thus represent the Bloch Hamiltonian by an element of the oriented Grassmannian
\begin{equation}
     \mathbb{S}^4= \widetilde{\mathsf{Gr}}_{4,5}^{\mathbb{R}} = \mathsf{SO}(5)/\mathsf{SO}(4)  \ni [\widetilde{R}] = R\cdot [\mathsf{SO}(4)\oplus 1] \,.
\end{equation}
We conclude that the $10$ angles parametrizing a generic element $R=(u_1~u_2~u_3~u_4~u_5)\in \mathsf{SO}(5)$ can be reduced to only the four angles, say the hyper-spherical angles $(\phi,\theta,\psi,\rho)$, that parametrize the points of $\mathbb{S}^4$. 

This reduction is naturally obtained through the Pl{\"u}cker embedding, \ie the exterior product of the occupied eigenvectors gives \cite{bouhon2020geometric}
\begin{equation}
    u_1 \wedge u_2 \wedge u_3 \wedge u_4  \in  \mathbb{S}^4 \subset \bigwedge^4(\mathbb{R}^5) \,,
\end{equation}
while the Hodge dual is
\begin{equation}
    *(u_1 \wedge u_2 \wedge u_3 \wedge u_4) = u_5 \in \mathbb{S}^4 \subset \mathbb{R}^5\,,
\end{equation}
with the explicit expression of the fourth-order wedge product given by
\begin{equation}
\begin{aligned}
    u_1\wedge & u_2 \wedge u_3 \wedge u_4 =  
    \sum\limits_{j=1}^5 v_j \check{e}_{\bs{i}_{j}} \in \mathbb{S}^4 \,,\\
    v_{j} &= \sum\limits_{\sigma\in S_4}
    (-1)^{\sigma} [u_1]_{\sigma(\bs{i}_{j,1})}
    [u_2]_{\sigma(\bs{i}_{j,2})}
    [u_3]_{\sigma(\bs{i}_{j,3})}
    [u_4]_{\sigma(\bs{i}_{j,4})}\,,\\
    \check{e}_{\bs{i}_j} &= e_{\bs{i}_{j,1}}\wedge
    e_{\bs{i}_{j,2}}\wedge
    e_{\bs{i}_{j,3}}\wedge
    e_{\bs{i}_{j,4}}\,,
\end{aligned}
\end{equation}
with $\{\bs{i}_j \}_{j=1}^{5} = \{(a,b,c,d)\vert 1\leq a < b < c< d \leq 5\}$ (\eg $\bs{i}_1 = (1,2,3,4)$), where $[u_n]_m$ is the $m$-th component of the $n$-th Bloch eigenvector, $S_{4}$ is the symmetric group of permutations of four elements and $(-1)^{\sigma}$ is the parity of the permutation $\sigma$, and $\check{e}_{\bs{i}_j}$ is the exterior (wedge) product of four unit vectors from the Cartesian basis $(e_1,e_2,e_3,e_4,e_5)$ of $\mathbb{R}^5$.

Noting the isomorphism $\bigwedge^4(\mathbb{R}^5) \cong \mathbb{R}^5$, we then set 
\begin{equation}
    u_1 \wedge u_2 \wedge u_3 \wedge u_4 ``=" u_5 =\bs{e}_r \in \mathbb{S}^4\,.
\end{equation}
A geometric interpretation of the above constraint is to consider the partial frame $(u_1~u_2~u_3~u_4)$ as a basis of the tangent hyper-plane of the four-sphere at the point $u_5$, \ie $T_{u_5}\mathbb{S}^4 = \text{span}\langle u_1,u_2,u_3,u_4\rangle$, since any $\mathsf{SO}(4)$ rotation $(u_1~u_2~u_3~u_4)\cdot S[G_I]$ leaves the tangent plane invariant. Since we want to cover the whole four-sphere as we scan one time through the Brillouin zone, we actually seek a section of the whole tangent bundle $T\mathbb{S}^4$. Such a section is readily obtained from the frame of unit vectors in the hyper-spherical coordinates of $\mathbb{S}^4$, namely from $\bs{x}$ defined in Eq.\,(\ref{eq_hyperspherical}), we obtain 
\begin{equation}
    \begin{aligned}
        u_5(\phi,\theta,\psi,\rho) &= \bs{e}_r = \bs{x}_{r=1}(\phi,\theta,\psi,\rho) \,,\\
        u_1(\phi,\theta,\psi,\rho) &= \bs{\partial}_{\phi} = \dfrac{\partial_{\phi} \bs{e}_r}{\vert \partial_{\phi} \bs{e}_r \vert} = (c\phi,-s\phi,0,0,0) \,, \\
        u_2(\phi,\theta,\psi,\rho) &= \bs{\partial}_{\theta} = \dfrac{\partial_{\theta} \bs{e}_r}{\vert \partial_{\theta} \bs{e}_r \vert} = 
        (s \phi \,c\theta ,c \phi\, c\theta,- s\theta,0,0) \,, \\
        u_3(\phi,\theta,\psi,\rho) &= \bs{\partial}_{\psi} = \dfrac{\partial_{\psi} \bs{e}_r}{\vert \partial_{\psi} \bs{e}_r \vert}\\
        &= (s \phi \,s\theta\, c\psi ,c \phi \,s\theta \,c\psi, c\theta \,c\psi,-s\psi,0) \,, \\
        u_4(\phi,\theta,\psi,\rho) &= \bs{\partial}_{\rho} = \dfrac{\partial_{\rho} \bs{e}_r}{\vert \partial_{\rho} \bs{e}_r \vert} \\
        &= (s \phi s\theta s\psi c\rho,c\phi s\theta s\psi c\rho, c\theta s\psi c\rho,c\psi c\rho,-s\rho),
    \end{aligned}
\end{equation}
with the short notation $c\phi=\cos\phi$ and $s\phi=\sin\phi$ and similarly for $\{\theta,\psi,\rho\}$, and we set
\begin{equation}
    \widetilde{R}(\phi,\theta,\psi,\rho) = \left(
        \begin{array}{ccccc}
            \bs{\partial}_{\phi} &
            \bs{\partial}_{\theta} &
            \bs{\partial}_{\psi} &
            \bs{\partial}_{\rho} &
            \bs{e}_r
        \end{array}
    \right)\,. 
\end{equation}

Pulling back the coordinates of the Grassmannian to a point $\bs{k}$ of the Brillouin zone four-torus, we finally define
\begin{equation}
    H_{W_4}(\bs{k}) = \widetilde{R}_{W_4}(\bs{k})\cdot \text{diag}\left[\begin{array}{c}
    E_1\\
    E_2 \\ 
    E_3 \\ E_4 \\ E_5
    \end{array}\right]\cdot \widetilde{R}_{W_4}(\bs{k})^T\,,
\end{equation}
with
\begin{equation}
    \widetilde{R}_{W_4}(\bs{k}) = 
    \widetilde{R}(W_4\phi_0(\bs{k}),\theta_0(\bs{k}),\psi_0(\bs{k}),\rho_0(\bs{k}))\,.
\end{equation}

We finally get a tight-binding model with a finite hopping range by approximating every element of the matrix $H_{W_4}(\bs{k})$ by a finite Fourier series, \ie
\begin{multline}
    [H_{W_4}(\bs{k})]_{\alpha\beta} = \epsilon_{\alpha\beta}   \\
    +\sum\limits_{\{n_1,\dots,n_4\}=1}^{N_{\alpha\beta}} \Big[t_{c,\alpha\beta}(\vert\bs{R}_{\bs{n}} \vert) \cos (\bs{k}\cdot\bs{R}_{\bs{n}}) \\
     +t_{s,\alpha\beta}(\vert\bs{R}_{\bs{n}} \vert) \sin (\bs{k}\cdot\bs{R}_{\bs{n}})
    \Big] \,,
\end{multline}
with $\bs{R}_{\bs{n}}= \sum_{i=1}^{4} n_i \bs{a}_i$ the position of the $\bs{n}$-th unit cell. The bulk nodal structure presented in the main text has been obtained for the tight-binding model with $N_{\alpha\beta} = 3$ (for all $\alpha$ and $\beta$) and setting the energy eigenvalues $(E_1,E_2,E_3,E_4,E_5) = (-4,-3,-2,-1,2)$.

\section{4D reciprocal Bravais lattice and Brillouin zone}\label{sec_bloch_periodicity}

We define the Bravais lattice 
\begin{equation}
    \Lambda = \bigcup\limits_{\bs{n}\in \mathbb{Z}^4} \sum\limits_{i=1}^4 n_i \bs{a}_i  \,,    
\end{equation}
spanned by the primitive vectors $\{\bs{a}_i\}_{i=1,\dots,4}$, with $\bs{n}=(n_1,n_2,n_3,n_4)$. Defining the reciprocal basis vectors
\begin{equation}
\begin{aligned}
    \bs{b}_1 = 2\pi \dfrac{\bs{a}_2\wedge \bs{a}_3\wedge\bs{a}_4}{\vert \bs{a}_1\wedge\bs{a}_2\wedge \bs{a}_3\wedge\bs{a}_4\vert},
    \bs{b}_2 = 2\pi \dfrac{\bs{a}_3\wedge \bs{a}_4\wedge\bs{a}_1}{\vert \bs{a}_1\wedge\bs{a}_2\wedge \bs{a}_3\wedge\bs{a}_4\vert},\\
    \bs{b}_3 = 2\pi \dfrac{\bs{a}_4\wedge \bs{a}_1\wedge\bs{a}_2}{\vert \bs{a}_1\wedge\bs{a}_2\wedge \bs{a}_3\wedge\bs{a}_4\vert},
    \bs{b}_4 = 2\pi \dfrac{\bs{a}_1\wedge \bs{a}_2\wedge\bs{a}_3}{\vert \bs{a}_1\wedge\bs{a}_2\wedge \bs{a}_3\wedge\bs{a}_4\vert},
\end{aligned}
\end{equation}
we define the lattice of reciprocal vectors 
\begin{equation}
    \Lambda^* = \bigcup\limits_{\bs{m}\in \mathbb{Z}^4} \sum\limits_{i=1}^4 m_i \bs{b}_i \,,  
\end{equation}
with $\bs{m} = (m_1,\dots, m_4)$. The Brillouin zone torus is then formally defined as $\text{BZ}=\mathbb{R}^4/\Lambda^*\cong \mathbb{T}^4$.


\begin{thebibliography}{110}%
\makeatletter
\providecommand \@ifxundefined [1]{%
 \@ifx{#1\undefined}
}%
\providecommand \@ifnum [1]{%
 \ifnum #1\expandafter \@firstoftwo
 \else \expandafter \@secondoftwo
 \fi
}%
\providecommand \@ifx [1]{%
 \ifx #1\expandafter \@firstoftwo
 \else \expandafter \@secondoftwo
 \fi
}%
\providecommand \natexlab [1]{#1}%
\providecommand \enquote  [1]{``#1''}%
\providecommand \bibnamefont  [1]{#1}%
\providecommand \bibfnamefont [1]{#1}%
\providecommand \citenamefont [1]{#1}%
\providecommand \href@noop [0]{\@secondoftwo}%
\providecommand \href [0]{\begingroup \@sanitize@url \@href}%
\providecommand \@href[1]{\@@startlink{#1}\@@href}%
\providecommand \@@href[1]{\endgroup#1\@@endlink}%
\providecommand \@sanitize@url [0]{\catcode `\\12\catcode `\$12\catcode
  `\&12\catcode `\#12\catcode `\^12\catcode `\_12\catcode `\%12\relax}%
\providecommand \@@startlink[1]{}%
\providecommand \@@endlink[0]{}%
\providecommand \url  [0]{\begingroup\@sanitize@url \@url }%
\providecommand \@url [1]{\endgroup\@href {#1}{\urlprefix }}%
\providecommand \urlprefix  [0]{URL }%
\providecommand \Eprint [0]{\href }%
\providecommand \doibase [0]{http://dx.doi.org/}%
\providecommand \selectlanguage [0]{\@gobble}%
\providecommand \bibinfo  [0]{\@secondoftwo}%
\providecommand \bibfield  [0]{\@secondoftwo}%
\providecommand \translation [1]{[#1]}%
\providecommand \BibitemOpen [0]{}%
\providecommand \bibitemStop [0]{}%
\providecommand \bibitemNoStop [0]{.\EOS\space}%
\providecommand \EOS [0]{\spacefactor3000\relax}%
\providecommand \BibitemShut  [1]{\csname bibitem#1\endcsname}%
\let\auto@bib@innerbib\@empty
\bibitem [{\citenamefont {Wen}(2017)}]{Wen}%
  \BibitemOpen
  \bibfield  {author} {\bibinfo {author} {\bibfnamefont {Xiao-Gang}\
  \bibnamefont {Wen}},\ }\bibfield  {title} {\enquote {\bibinfo {title}
  {Colloquium: Zoo of quantum-topological phases of matter},}\ }\href {\doibase
  10.1103/RevModPhys.89.041004} {\bibfield  {journal} {\bibinfo  {journal}
  {Rev. Mod. Phys.}\ }\textbf {\bibinfo {volume} {89}},\ \bibinfo {pages}
  {041004} (\bibinfo {year} {2017})}\BibitemShut {NoStop}%
\bibitem [{\citenamefont {Senthil}(2015)}]{Senthil}%
  \BibitemOpen
  \bibfield  {author} {\bibinfo {author} {\bibfnamefont {T.}~\bibnamefont
  {Senthil}},\ }\bibfield  {title} {\enquote {\bibinfo {title}
  {Symmetry-protected topological phases of quantum matter},}\ }\href {\doibase
  10.1146/annurev-conmatphys-031214-014740} {\bibfield  {journal} {\bibinfo
  {journal} {Annual Review of Condensed Matter Physics}\ }\textbf {\bibinfo
  {volume} {6}},\ \bibinfo {pages} {299--324} (\bibinfo {year} {2015})},\
  \Eprint
  {http://arxiv.org/abs/https://doi.org/10.1146/annurev-conmatphys-031214-014740}
  {https://doi.org/10.1146/annurev-conmatphys-031214-014740} \BibitemShut
  {NoStop}%
\bibitem [{\citenamefont {Witten}(2016)}]{Witten}%
  \BibitemOpen
  \bibfield  {author} {\bibinfo {author} {\bibfnamefont {Edward}\ \bibnamefont
  {Witten}},\ }\bibfield  {title} {\enquote {\bibinfo {title} {Fermion path
  integrals and topological phases},}\ }\href {\doibase
  10.1103/RevModPhys.88.035001} {\bibfield  {journal} {\bibinfo  {journal}
  {Rev. Mod. Phys.}\ }\textbf {\bibinfo {volume} {88}},\ \bibinfo {pages}
  {035001} (\bibinfo {year} {2016})}\BibitemShut {NoStop}%
\bibitem [{\citenamefont {Nakahara}(1990)}]{Nakahara_book}%
  \BibitemOpen
  \bibfield  {author} {\bibinfo {author} {\bibfnamefont {Mikio}\ \bibnamefont
  {Nakahara}},\ }\href {https://cds.cern.ch/record/206619} {\emph {\bibinfo
  {title} {{Geometry, topology and physics}}}},\ Graduate student series in
  physics\ (\bibinfo  {publisher} {Hilger},\ \bibinfo {address} {Bristol},\
  \bibinfo {year} {1990})\BibitemShut {NoStop}%
\bibitem [{\citenamefont {Eguchi}\ \emph {et~al.}(1980)\citenamefont {Eguchi},
  \citenamefont {Gilkey},\ and\ \citenamefont {Hanson}}]{Eguchi}%
  \BibitemOpen
  \bibfield  {author} {\bibinfo {author} {\bibfnamefont {Tohru}\ \bibnamefont
  {Eguchi}}, \bibinfo {author} {\bibfnamefont {Peter~B.}\ \bibnamefont
  {Gilkey}}, \ and\ \bibinfo {author} {\bibfnamefont {Andrew~J.}\ \bibnamefont
  {Hanson}},\ }\bibfield  {title} {\enquote {\bibinfo {title} {Gravitation,
  gauge theories and differential geometry},}\ }\href {\doibase
  https://doi.org/10.1016/0370-1573(80)90130-1} {\bibfield  {journal} {\bibinfo
   {journal} {Physics Reports}\ }\textbf {\bibinfo {volume} {66}},\ \bibinfo
  {pages} {213--393} (\bibinfo {year} {1980})}\BibitemShut {NoStop}%
\bibitem [{\citenamefont {Zhang}\ and\ \citenamefont {Hu}(2001)}]{Zhang}%
  \BibitemOpen
  \bibfield  {author} {\bibinfo {author} {\bibfnamefont {Shou-Cheng}\
  \bibnamefont {Zhang}}\ and\ \bibinfo {author} {\bibfnamefont {Jiangping}\
  \bibnamefont {Hu}},\ }\bibfield  {title} {\enquote {\bibinfo {title} {A
  four-dimensional generalization of the quantum hall effect},}\ }\href
  {\doibase 10.1126/science.294.5543.823} {\bibfield  {journal} {\bibinfo
  {journal} {Science}\ }\textbf {\bibinfo {volume} {294}},\ \bibinfo {pages}
  {823--828} (\bibinfo {year} {2001})}\BibitemShut {NoStop}%
\bibitem [{\citenamefont {Karabali}\ and\ \citenamefont
  {Nair}(2002)}]{Karabali}%
  \BibitemOpen
  \bibfield  {author} {\bibinfo {author} {\bibfnamefont {Dimitra}\ \bibnamefont
  {Karabali}}\ and\ \bibinfo {author} {\bibfnamefont {V.P.}\ \bibnamefont
  {Nair}},\ }\bibfield  {title} {\enquote {\bibinfo {title} {Quantum hall
  effect in higher dimensions},}\ }\href {\doibase
  https://doi.org/10.1016/S0550-3213(02)00634-X} {\bibfield  {journal}
  {\bibinfo  {journal} {Nuclear Physics B}\ }\textbf {\bibinfo {volume}
  {641}},\ \bibinfo {pages} {533--546} (\bibinfo {year} {2002})}\BibitemShut
  {NoStop}%
\bibitem [{\citenamefont {Qi}\ \emph {et~al.}(2008)\citenamefont {Qi},
  \citenamefont {Hughes},\ and\ \citenamefont {Zhang}}]{XLQi2008}%
  \BibitemOpen
  \bibfield  {author} {\bibinfo {author} {\bibfnamefont {Xiao-Liang}\
  \bibnamefont {Qi}}, \bibinfo {author} {\bibfnamefont {Taylor~L.}\
  \bibnamefont {Hughes}}, \ and\ \bibinfo {author} {\bibfnamefont {Shou-Cheng}\
  \bibnamefont {Zhang}},\ }\bibfield  {title} {\enquote {\bibinfo {title}
  {Topological field theory of time-reversal invariant insulators},}\ }\href
  {\doibase 10.1103/PhysRevB.78.195424} {\bibfield  {journal} {\bibinfo
  {journal} {Phys. Rev. B}\ }\textbf {\bibinfo {volume} {78}},\ \bibinfo
  {pages} {195424} (\bibinfo {year} {2008})}\BibitemShut {NoStop}%
\bibitem [{\citenamefont {Altland}\ and\ \citenamefont
  {Zirnbauer}(1997)}]{Altland}%
  \BibitemOpen
  \bibfield  {author} {\bibinfo {author} {\bibfnamefont {Alexander}\
  \bibnamefont {Altland}}\ and\ \bibinfo {author} {\bibfnamefont {Martin~R.}\
  \bibnamefont {Zirnbauer}},\ }\bibfield  {title} {\enquote {\bibinfo {title}
  {Nonstandard symmetry classes in mesoscopic normal-superconducting hybrid
  structures},}\ }\href {\doibase 10.1103/PhysRevB.55.1142} {\bibfield
  {journal} {\bibinfo  {journal} {Phys. Rev. B}\ }\textbf {\bibinfo {volume}
  {55}},\ \bibinfo {pages} {1142--1161} (\bibinfo {year} {1997})}\BibitemShut
  {NoStop}%
\bibitem [{\citenamefont {Kitaev}(2009)}]{Kitaev}%
  \BibitemOpen
  \bibfield  {author} {\bibinfo {author} {\bibfnamefont {Alexei}\ \bibnamefont
  {Kitaev}},\ }\bibfield  {title} {\enquote {\bibinfo {title} {Periodic table
  for topological insulators and superconductors},}\ }\href {\doibase
  10.1063/1.3149495} {\bibfield  {journal} {\bibinfo  {journal} {AIP Conference
  Proceedings}\ }\textbf {\bibinfo {volume} {1134}},\ \bibinfo {pages} {22--30}
  (\bibinfo {year} {2009})}\BibitemShut {NoStop}%
\bibitem [{\citenamefont {Schnyder}\ \emph {et~al.}(2008)\citenamefont
  {Schnyder}, \citenamefont {Ryu}, \citenamefont {Furusaki},\ and\
  \citenamefont {Ludwig}}]{Ludwig}%
  \BibitemOpen
  \bibfield  {author} {\bibinfo {author} {\bibfnamefont {Andreas~P.}\
  \bibnamefont {Schnyder}}, \bibinfo {author} {\bibfnamefont {Shinsei}\
  \bibnamefont {Ryu}}, \bibinfo {author} {\bibfnamefont {Akira}\ \bibnamefont
  {Furusaki}}, \ and\ \bibinfo {author} {\bibfnamefont {Andreas W.~W.}\
  \bibnamefont {Ludwig}},\ }\bibfield  {title} {\enquote {\bibinfo {title}
  {Classification of topological insulators and superconductors in three
  spatial dimensions},}\ }\href {\doibase 10.1103/PhysRevB.78.195125}
  {\bibfield  {journal} {\bibinfo  {journal} {Phys. Rev. B}\ }\textbf {\bibinfo
  {volume} {78}},\ \bibinfo {pages} {195125} (\bibinfo {year}
  {2008})}\BibitemShut {NoStop}%
\bibitem [{\citenamefont {Morimoto}\ and\ \citenamefont
  {Furusaki}(2013)}]{Morimoto_2013}%
  \BibitemOpen
  \bibfield  {author} {\bibinfo {author} {\bibfnamefont {Takahiro}\
  \bibnamefont {Morimoto}}\ and\ \bibinfo {author} {\bibfnamefont {Akira}\
  \bibnamefont {Furusaki}},\ }\bibfield  {title} {\enquote {\bibinfo {title}
  {Topological classification with additional symmetries from {C}lifford
  algebras},}\ }\href {\doibase 10.1103/PhysRevB.88.125129} {\bibfield
  {journal} {\bibinfo  {journal} {Phys. Rev. B}\ }\textbf {\bibinfo {volume}
  {88}},\ \bibinfo {pages} {125129} (\bibinfo {year} {2013})}\BibitemShut
  {NoStop}%
\bibitem [{\citenamefont {Shiozaki}\ and\ \citenamefont
  {Sato}(2014)}]{Shiozaki14}%
  \BibitemOpen
  \bibfield  {author} {\bibinfo {author} {\bibfnamefont {Ken}\ \bibnamefont
  {Shiozaki}}\ and\ \bibinfo {author} {\bibfnamefont {Masatoshi}\ \bibnamefont
  {Sato}},\ }\bibfield  {title} {\enquote {\bibinfo {title} {Topology of
  crystalline insulators and superconductors},}\ }\href {\doibase
  10.1103/PhysRevB.90.165114} {\bibfield  {journal} {\bibinfo  {journal} {Phys.
  Rev. B}\ }\textbf {\bibinfo {volume} {90}},\ \bibinfo {pages} {165114}
  (\bibinfo {year} {2014})}\BibitemShut {NoStop}%
\bibitem [{\citenamefont {Fu}(2011)}]{Fu}%
  \BibitemOpen
  \bibfield  {author} {\bibinfo {author} {\bibfnamefont {Liang}\ \bibnamefont
  {Fu}},\ }\bibfield  {title} {\enquote {\bibinfo {title} {Topological
  crystalline insulators},}\ }\href {\doibase 10.1103/PhysRevLett.106.106802}
  {\bibfield  {journal} {\bibinfo  {journal} {Phys. Rev. Lett.}\ }\textbf
  {\bibinfo {volume} {106}},\ \bibinfo {pages} {106802} (\bibinfo {year}
  {2011})}\BibitemShut {NoStop}%
\bibitem [{\citenamefont {Slager}\ \emph {et~al.}(2012)\citenamefont {Slager},
  \citenamefont {Mesaros}, \citenamefont {Juri{\v c}i{\'c}},\ and\
  \citenamefont {Zaanen}}]{Slager_2013}%
  \BibitemOpen
  \bibfield  {author} {\bibinfo {author} {\bibfnamefont {Robert-Jan}\
  \bibnamefont {Slager}}, \bibinfo {author} {\bibfnamefont {Andrej}\
  \bibnamefont {Mesaros}}, \bibinfo {author} {\bibfnamefont {Vladimir}\
  \bibnamefont {Juri{\v c}i{\'c}}}, \ and\ \bibinfo {author} {\bibfnamefont
  {Jan}\ \bibnamefont {Zaanen}},\ }\bibfield  {title} {\enquote {\bibinfo
  {title} {The space group classification of topological band-insulators},}\
  }\href {http://dx.doi.org/10.1038/nphys2513} {\bibfield  {journal} {\bibinfo
  {journal} {Nat. Phys.}\ }\textbf {\bibinfo {volume} {9}},\ \bibinfo {pages}
  {98} (\bibinfo {year} {2013})}\BibitemShut {NoStop}%
\bibitem [{\citenamefont {Bradlyn}\ \emph {et~al.}(2017)\citenamefont
  {Bradlyn}, \citenamefont {Elcoro}, \citenamefont {Cano}, \citenamefont
  {Vergniory}, \citenamefont {Wang}, \citenamefont {Felser}, \citenamefont
  {Aroyo},\ and\ \citenamefont {Bernevig}}]{Bradlyn_2017}%
  \BibitemOpen
  \bibfield  {author} {\bibinfo {author} {\bibfnamefont {Barry}\ \bibnamefont
  {Bradlyn}}, \bibinfo {author} {\bibfnamefont {L.}~\bibnamefont {Elcoro}},
  \bibinfo {author} {\bibfnamefont {Jennifer}\ \bibnamefont {Cano}}, \bibinfo
  {author} {\bibfnamefont {M.~G.}\ \bibnamefont {Vergniory}}, \bibinfo {author}
  {\bibfnamefont {Zhijun}\ \bibnamefont {Wang}}, \bibinfo {author}
  {\bibfnamefont {C.}~\bibnamefont {Felser}}, \bibinfo {author} {\bibfnamefont
  {M.~I.}\ \bibnamefont {Aroyo}}, \ and\ \bibinfo {author} {\bibfnamefont
  {B.~Andrei}\ \bibnamefont {Bernevig}},\ }\bibfield  {title} {\enquote
  {\bibinfo {title} {Topological quantum chemistry},}\ }\href
  {http://dx.doi.org/10.1038/nature23268} {\bibfield  {journal} {\bibinfo
  {journal} {Nature}\ }\textbf {\bibinfo {volume} {547}},\ \bibinfo {pages}
  {298} (\bibinfo {year} {2017})}\BibitemShut {NoStop}%
\bibitem [{\citenamefont {Kruthoff}\ \emph {et~al.}(2017)\citenamefont
  {Kruthoff}, \citenamefont {de~Boer}, \citenamefont {van Wezel}, \citenamefont
  {Kane},\ and\ \citenamefont {Slager}}]{Kruthoff_2017}%
  \BibitemOpen
  \bibfield  {author} {\bibinfo {author} {\bibfnamefont {Jorrit}\ \bibnamefont
  {Kruthoff}}, \bibinfo {author} {\bibfnamefont {Jan}\ \bibnamefont {de~Boer}},
  \bibinfo {author} {\bibfnamefont {Jasper}\ \bibnamefont {van Wezel}},
  \bibinfo {author} {\bibfnamefont {Charles~L.}\ \bibnamefont {Kane}}, \ and\
  \bibinfo {author} {\bibfnamefont {Robert-Jan}\ \bibnamefont {Slager}},\
  }\bibfield  {title} {\enquote {\bibinfo {title} {Topological classification
  of crystalline insulators through band structure combinatorics},}\ }\href
  {\doibase 10.1103/PhysRevX.7.041069} {\bibfield  {journal} {\bibinfo
  {journal} {Phys. Rev. X}\ }\textbf {\bibinfo {volume} {7}},\ \bibinfo {pages}
  {041069} (\bibinfo {year} {2017})}\BibitemShut {NoStop}%
\bibitem [{\citenamefont {Scheurer}\ and\ \citenamefont {Slager}(2020)}]{mcom}%
  \BibitemOpen
  \bibfield  {author} {\bibinfo {author} {\bibfnamefont {Mathias~S.}\
  \bibnamefont {Scheurer}}\ and\ \bibinfo {author} {\bibfnamefont {Robert-Jan}\
  \bibnamefont {Slager}},\ }\bibfield  {title} {\enquote {\bibinfo {title}
  {Unsupervised machine learning and band topology},}\ }\href {\doibase
  10.1103/PhysRevLett.124.226401} {\bibfield  {journal} {\bibinfo  {journal}
  {Phys. Rev. Lett.}\ }\textbf {\bibinfo {volume} {124}},\ \bibinfo {pages}
  {226401} (\bibinfo {year} {2020})}\BibitemShut {NoStop}%
\bibitem [{\citenamefont {Po}\ \emph {et~al.}(2017)\citenamefont {Po},
  \citenamefont {Vishwanath},\ and\ \citenamefont {Watanabe}}]{Po_2017}%
  \BibitemOpen
  \bibfield  {author} {\bibinfo {author} {\bibfnamefont {Hoi~Chun}\
  \bibnamefont {Po}}, \bibinfo {author} {\bibfnamefont {Ashvin}\ \bibnamefont
  {Vishwanath}}, \ and\ \bibinfo {author} {\bibfnamefont {Haruki}\ \bibnamefont
  {Watanabe}},\ }\bibfield  {title} {\enquote {\bibinfo {title} {Symmetry-based
  indicators of band topology in the 230 space groups},}\ }\href {\doibase
  10.1038/s41467-017-00133-2} {\bibfield  {journal} {\bibinfo  {journal} {Nat.
  Commun.}\ }\textbf {\bibinfo {volume} {8}},\ \bibinfo {pages} {50} (\bibinfo
  {year} {2017})}\BibitemShut {NoStop}%
\bibitem [{\citenamefont {Benalcazar}\ \emph {et~al.}(2017)\citenamefont
  {Benalcazar}, \citenamefont {Bernevig},\ and\ \citenamefont
  {Hughes}}]{Benalcazar}%
  \BibitemOpen
  \bibfield  {author} {\bibinfo {author} {\bibfnamefont {Wladimir~A.}\
  \bibnamefont {Benalcazar}}, \bibinfo {author} {\bibfnamefont {B.~Andrei}\
  \bibnamefont {Bernevig}}, \ and\ \bibinfo {author} {\bibfnamefont
  {Taylor~L.}\ \bibnamefont {Hughes}},\ }\bibfield  {title} {\enquote {\bibinfo
  {title} {Quantized electric multipole insulators},}\ }\href {\doibase
  10.1126/science.aah6442} {\bibfield  {journal} {\bibinfo  {journal}
  {Science}\ }\textbf {\bibinfo {volume} {357}},\ \bibinfo {pages} {61--66}
  (\bibinfo {year} {2017})},\ \Eprint
  {http://arxiv.org/abs/https://www.science.org/doi/pdf/10.1126/science.aah6442}
  {https://www.science.org/doi/pdf/10.1126/science.aah6442} \BibitemShut
  {NoStop}%
\bibitem [{\citenamefont {Slager}\ \emph {et~al.}(2015)\citenamefont {Slager},
  \citenamefont {Rademaker}, \citenamefont {Zaanen},\ and\ \citenamefont
  {Balents}}]{Slager_2015}%
  \BibitemOpen
  \bibfield  {author} {\bibinfo {author} {\bibfnamefont {Robert-Jan}\
  \bibnamefont {Slager}}, \bibinfo {author} {\bibfnamefont {Louk}\ \bibnamefont
  {Rademaker}}, \bibinfo {author} {\bibfnamefont {Jan}\ \bibnamefont {Zaanen}},
  \ and\ \bibinfo {author} {\bibfnamefont {Leon}\ \bibnamefont {Balents}},\
  }\bibfield  {title} {\enquote {\bibinfo {title} {Impurity-bound states and
  green's function zeros as local signatures of topology},}\ }\href {\doibase
  10.1103/PhysRevB.92.085126} {\bibfield  {journal} {\bibinfo  {journal} {Phys.
  Rev. B}\ }\textbf {\bibinfo {volume} {92}},\ \bibinfo {pages} {085126}
  (\bibinfo {year} {2015})}\BibitemShut {NoStop}%
\bibitem [{\citenamefont {Schindler}\ \emph {et~al.}(2018)\citenamefont
  {Schindler}, \citenamefont {Cook}, \citenamefont {Vergniory}, \citenamefont
  {Wang}, \citenamefont {Parkin}, \citenamefont {Bernevig},\ and\ \citenamefont
  {Neupert}}]{Neupert}%
  \BibitemOpen
  \bibfield  {author} {\bibinfo {author} {\bibfnamefont {Frank}\ \bibnamefont
  {Schindler}}, \bibinfo {author} {\bibfnamefont {Ashley~M.}\ \bibnamefont
  {Cook}}, \bibinfo {author} {\bibfnamefont {Maia~G.}\ \bibnamefont
  {Vergniory}}, \bibinfo {author} {\bibfnamefont {Zhijun}\ \bibnamefont
  {Wang}}, \bibinfo {author} {\bibfnamefont {Stuart S.~P.}\ \bibnamefont
  {Parkin}}, \bibinfo {author} {\bibfnamefont {B.~Andrei}\ \bibnamefont
  {Bernevig}}, \ and\ \bibinfo {author} {\bibfnamefont {Titus}\ \bibnamefont
  {Neupert}},\ }\bibfield  {title} {\enquote {\bibinfo {title} {Higher-order
  topological insulators},}\ }\href {\doibase 10.1126/sciadv.aat0346}
  {\bibfield  {journal} {\bibinfo  {journal} {Science Advances}\ }\textbf
  {\bibinfo {volume} {4}},\ \bibinfo {pages} {eaat0346} (\bibinfo {year}
  {2018})},\ \Eprint
  {http://arxiv.org/abs/https://www.science.org/doi/pdf/10.1126/sciadv.aat0346}
  {https://www.science.org/doi/pdf/10.1126/sciadv.aat0346} \BibitemShut
  {NoStop}%
\bibitem [{\citenamefont {Slager}\ \emph {et~al.}(2014)\citenamefont {Slager},
  \citenamefont {Mesaros}, \citenamefont {Juri\ifmmode \check{c}\else
  \v{c}\fi{}i\ifmmode~\acute{c}\else \'{c}\fi{}},\ and\ \citenamefont
  {Zaanen}}]{Slager_2014}%
  \BibitemOpen
  \bibfield  {author} {\bibinfo {author} {\bibfnamefont {Robert-Jan}\
  \bibnamefont {Slager}}, \bibinfo {author} {\bibfnamefont {Andrej}\
  \bibnamefont {Mesaros}}, \bibinfo {author} {\bibfnamefont {Vladimir}\
  \bibnamefont {Juri\ifmmode \check{c}\else \v{c}\fi{}i\ifmmode~\acute{c}\else
  \'{c}\fi{}}}, \ and\ \bibinfo {author} {\bibfnamefont {Jan}\ \bibnamefont
  {Zaanen}},\ }\bibfield  {title} {\enquote {\bibinfo {title} {Interplay
  between electronic topology and crystal symmetry: Dislocation-line modes in
  topological band insulators},}\ }\href {\doibase 10.1103/PhysRevB.90.241403}
  {\bibfield  {journal} {\bibinfo  {journal} {Phys. Rev. B}\ }\textbf {\bibinfo
  {volume} {90}},\ \bibinfo {pages} {241403} (\bibinfo {year}
  {2014})}\BibitemShut {NoStop}%
\bibitem [{\citenamefont {Wang}\ \emph {et~al.}(2019)\citenamefont {Wang},
  \citenamefont {Wieder}, \citenamefont {Li}, \citenamefont {Yan},\ and\
  \citenamefont {Bernevig}}]{Wieder_HOTI}%
  \BibitemOpen
  \bibfield  {author} {\bibinfo {author} {\bibfnamefont {Zhijun}\ \bibnamefont
  {Wang}}, \bibinfo {author} {\bibfnamefont {Benjamin~J.}\ \bibnamefont
  {Wieder}}, \bibinfo {author} {\bibfnamefont {Jian}\ \bibnamefont {Li}},
  \bibinfo {author} {\bibfnamefont {Binghai}\ \bibnamefont {Yan}}, \ and\
  \bibinfo {author} {\bibfnamefont {B.~Andrei}\ \bibnamefont {Bernevig}},\
  }\bibfield  {title} {\enquote {\bibinfo {title} {Higher-order topology,
  monopole nodal lines, and the origin of large {F}ermi arcs in transition
  metal dichalcogenides $x{\mathrm{te}}_{2}$ ($x=\mathrm{Mo},\mathrm{W}$)},}\
  }\href {\doibase 10.1103/PhysRevLett.123.186401} {\bibfield  {journal}
  {\bibinfo  {journal} {Phys. Rev. Lett.}\ }\textbf {\bibinfo {volume} {123}},\
  \bibinfo {pages} {186401} (\bibinfo {year} {2019})}\BibitemShut {NoStop}%
\bibitem [{\citenamefont {Po}\ \emph {et~al.}(2018)\citenamefont {Po},
  \citenamefont {Watanabe},\ and\ \citenamefont {Vishwanath}}]{Po_2018}%
  \BibitemOpen
  \bibfield  {author} {\bibinfo {author} {\bibfnamefont {Hoi~Chun}\
  \bibnamefont {Po}}, \bibinfo {author} {\bibfnamefont {Haruki}\ \bibnamefont
  {Watanabe}}, \ and\ \bibinfo {author} {\bibfnamefont {Ashvin}\ \bibnamefont
  {Vishwanath}},\ }\bibfield  {title} {\enquote {\bibinfo {title} {Fragile
  topology and wannier obstructions},}\ }\href {\doibase
  10.1103/PhysRevLett.121.126402} {\bibfield  {journal} {\bibinfo  {journal}
  {Phys. Rev. Lett.}\ }\textbf {\bibinfo {volume} {121}},\ \bibinfo {pages}
  {126402} (\bibinfo {year} {2018})}\BibitemShut {NoStop}%
\bibitem [{\citenamefont {Bouhon}\ \emph {et~al.}(2019)\citenamefont {Bouhon},
  \citenamefont {Black-Schaffer},\ and\ \citenamefont
  {Slager}}]{bouhon2018wilson}%
  \BibitemOpen
  \bibfield  {author} {\bibinfo {author} {\bibfnamefont {Adrien}\ \bibnamefont
  {Bouhon}}, \bibinfo {author} {\bibfnamefont {Annica~M.}\ \bibnamefont
  {Black-Schaffer}}, \ and\ \bibinfo {author} {\bibfnamefont {Robert-Jan}\
  \bibnamefont {Slager}},\ }\bibfield  {title} {\enquote {\bibinfo {title}
  {Wilson loop approach to fragile topology of split elementary band
  representations and topological crystalline insulators with time-reversal
  symmetry},}\ }\href {\doibase 10.1103/PhysRevB.100.195135} {\bibfield
  {journal} {\bibinfo  {journal} {Phys. Rev. B}\ }\textbf {\bibinfo {volume}
  {100}},\ \bibinfo {pages} {195135} (\bibinfo {year} {2019})}\BibitemShut
  {NoStop}%
\bibitem [{\citenamefont {Bradlyn}\ \emph {et~al.}(2019)\citenamefont
  {Bradlyn}, \citenamefont {Wang}, \citenamefont {Cano},\ and\ \citenamefont
  {Bernevig}}]{Bradlyn_fragile}%
  \BibitemOpen
  \bibfield  {author} {\bibinfo {author} {\bibfnamefont {Barry}\ \bibnamefont
  {Bradlyn}}, \bibinfo {author} {\bibfnamefont {Zhijun}\ \bibnamefont {Wang}},
  \bibinfo {author} {\bibfnamefont {Jennifer}\ \bibnamefont {Cano}}, \ and\
  \bibinfo {author} {\bibfnamefont {B.~Andrei}\ \bibnamefont {Bernevig}},\
  }\bibfield  {title} {\enquote {\bibinfo {title} {Disconnected elementary band
  representations, fragile topology, and wilson loops as topological indices:
  {A}n example on the triangular lattice},}\ }\href {\doibase
  10.1103/PhysRevB.99.045140} {\bibfield  {journal} {\bibinfo  {journal} {Phys.
  Rev. B}\ }\textbf {\bibinfo {volume} {99}},\ \bibinfo {pages} {045140}
  (\bibinfo {year} {2019})}\BibitemShut {NoStop}%
\bibitem [{\citenamefont {Bouhon}\ \emph
  {et~al.}(2020{\natexlab{a}})\citenamefont {Bouhon}, \citenamefont {Wu},
  \citenamefont {Slager}, \citenamefont {Weng}, \citenamefont {Yazyev},\ and\
  \citenamefont {Bzdu{\v s}ek}}]{bouhon2019nonabelian}%
  \BibitemOpen
  \bibfield  {author} {\bibinfo {author} {\bibfnamefont {Adrien}\ \bibnamefont
  {Bouhon}}, \bibinfo {author} {\bibfnamefont {QuanSheng}\ \bibnamefont {Wu}},
  \bibinfo {author} {\bibfnamefont {Robert-Jan}\ \bibnamefont {Slager}},
  \bibinfo {author} {\bibfnamefont {Hongming}\ \bibnamefont {Weng}}, \bibinfo
  {author} {\bibfnamefont {Oleg~V.}\ \bibnamefont {Yazyev}}, \ and\ \bibinfo
  {author} {\bibfnamefont {Tom{\'a}{\v s}}\ \bibnamefont {Bzdu{\v s}ek}},\
  }\bibfield  {title} {\enquote {\bibinfo {title} {Non-abelian reciprocal
  braiding of weyl points and its manifestation in zrte},}\ }\href {\doibase
  10.1038/s41567-020-0967-9} {\bibfield  {journal} {\bibinfo  {journal} {Nature
  Physics}\ } (\bibinfo {year} {2020}{\natexlab{a}}),\
  10.1038/s41567-020-0967-9}\BibitemShut {NoStop}%
\bibitem [{\citenamefont {Bouhon}\ \emph
  {et~al.}(2020{\natexlab{b}})\citenamefont {Bouhon}, \citenamefont
  {Bzdu\v{s}ek},\ and\ \citenamefont {Slager}}]{bouhon2020geometric}%
  \BibitemOpen
  \bibfield  {author} {\bibinfo {author} {\bibfnamefont {Adrien}\ \bibnamefont
  {Bouhon}}, \bibinfo {author} {\bibfnamefont {Tomas}\ \bibnamefont
  {Bzdu\v{s}ek}}, \ and\ \bibinfo {author} {\bibfnamefont {Robert-Jan}\
  \bibnamefont {Slager}},\ }\bibfield  {title} {\enquote {\bibinfo {title}
  {Geometric approach to fragile topology beyond symmetry indicators},}\ }\href
  {\doibase 10.1103/PhysRevB.102.115135} {\bibfield  {journal} {\bibinfo
  {journal} {Phys. Rev. B}\ }\textbf {\bibinfo {volume} {102}},\ \bibinfo
  {pages} {115135} (\bibinfo {year} {2020}{\natexlab{b}})}\BibitemShut
  {NoStop}%
\bibitem [{\citenamefont {Ahn}\ \emph {et~al.}(2018{\natexlab{a}})\citenamefont
  {Ahn}, \citenamefont {Kim}, \citenamefont {Kim},\ and\ \citenamefont
  {Yang}}]{Ahn_2018}%
  \BibitemOpen
  \bibfield  {author} {\bibinfo {author} {\bibfnamefont {Junyeong}\
  \bibnamefont {Ahn}}, \bibinfo {author} {\bibfnamefont {Dongwook}\
  \bibnamefont {Kim}}, \bibinfo {author} {\bibfnamefont {Youngkuk}\
  \bibnamefont {Kim}}, \ and\ \bibinfo {author} {\bibfnamefont {Bohm-Jung}\
  \bibnamefont {Yang}},\ }\bibfield  {title} {\enquote {\bibinfo {title} {Band
  topology and linking structure of nodal line semimetals with ${Z}_{2}$
  monopole charges},}\ }\href {\doibase 10.1103/PhysRevLett.121.106403}
  {\bibfield  {journal} {\bibinfo  {journal} {Phys. Rev. Lett.}\ }\textbf
  {\bibinfo {volume} {121}},\ \bibinfo {pages} {106403} (\bibinfo {year}
  {2018}{\natexlab{a}})}\BibitemShut {NoStop}%
\bibitem [{\citenamefont {Ahn}\ and\ \citenamefont {Yang}(2019)}]{Ahn2019}%
  \BibitemOpen
  \bibfield  {author} {\bibinfo {author} {\bibfnamefont {Junyeong}\
  \bibnamefont {Ahn}}\ and\ \bibinfo {author} {\bibfnamefont {Bohm-Jung}\
  \bibnamefont {Yang}},\ }\bibfield  {title} {\enquote {\bibinfo {title}
  {Symmetry representation approach to topological invariants in
  ${C}_{2z}{T}$-symmetric systems},}\ }\href {\doibase
  10.1103/PhysRevB.99.235125} {\bibfield  {journal} {\bibinfo  {journal} {Phys.
  Rev. B}\ }\textbf {\bibinfo {volume} {99}},\ \bibinfo {pages} {235125}
  (\bibinfo {year} {2019})}\BibitemShut {NoStop}%
\bibitem [{\citenamefont {Ahn}\ \emph {et~al.}(2019)\citenamefont {Ahn},
  \citenamefont {Park},\ and\ \citenamefont {Yang}}]{BJY_nielsen}%
  \BibitemOpen
  \bibfield  {author} {\bibinfo {author} {\bibfnamefont {Junyeong}\
  \bibnamefont {Ahn}}, \bibinfo {author} {\bibfnamefont {Sungjoon}\
  \bibnamefont {Park}}, \ and\ \bibinfo {author} {\bibfnamefont {Bohm-Jung}\
  \bibnamefont {Yang}},\ }\bibfield  {title} {\enquote {\bibinfo {title}
  {Failure of {N}ielsen-{N}inomiya {T}heorem and {F}ragile {T}opology in
  {T}wo-{D}imensional {S}ystems with {S}pace-{T}ime {I}nversion {S}ymmetry:
  {A}pplication to {T}wisted {B}ilayer {G}raphene at {M}agic {A}ngle},}\ }\href
  {\doibase 10.1103/PhysRevX.9.021013} {\bibfield  {journal} {\bibinfo
  {journal} {Phys. Rev. X}\ }\textbf {\bibinfo {volume} {9}},\ \bibinfo {pages}
  {021013} (\bibinfo {year} {2019})}\BibitemShut {NoStop}%
\bibitem [{\citenamefont {Bouhon}\ and\ \citenamefont
  {Slager}(2022{\natexlab{a}})}]{Bouhon2022braiding2}%
  \BibitemOpen
  \bibfield  {author} {\bibinfo {author} {\bibfnamefont {Adrien}\ \bibnamefont
  {Bouhon}}\ and\ \bibinfo {author} {\bibfnamefont {Robert-Jan}\ \bibnamefont
  {Slager}},\ }\href {\doibase 10.48550/ARXIV.2203.16741} {\enquote {\bibinfo
  {title} {Multi-gap topological conversion of euler class via band-node
  braiding: minimal models, $pt$-linked nodal rings, and chiral heirs},}\ }
  (\bibinfo {year} {2022}{\natexlab{a}})\BibitemShut {NoStop}%
\bibitem [{\citenamefont {Palumbo}(2021)}]{Palumbo}%
  \BibitemOpen
  \bibfield  {author} {\bibinfo {author} {\bibfnamefont {Giandomenico}\
  \bibnamefont {Palumbo}},\ }\bibfield  {title} {\enquote {\bibinfo {title}
  {Non-abelian tensor berry connections in multiband topological systems},}\
  }\href {\doibase 10.1103/PhysRevLett.126.246801} {\bibfield  {journal}
  {\bibinfo  {journal} {Phys. Rev. Lett.}\ }\textbf {\bibinfo {volume} {126}},\
  \bibinfo {pages} {246801} (\bibinfo {year} {2021})}\BibitemShut {NoStop}%
\bibitem [{\citenamefont {Wieder}\ and\ \citenamefont
  {Bernevig}(2018)}]{Wieder_axion}%
  \BibitemOpen
  \bibfield  {author} {\bibinfo {author} {\bibfnamefont {Benjamin~J.}\
  \bibnamefont {Wieder}}\ and\ \bibinfo {author} {\bibfnamefont {B.~Andrei}\
  \bibnamefont {Bernevig}},\ }\href@noop {} {\enquote {\bibinfo {title} {The
  axion insulator as a pump of fragile topology},}\ } (\bibinfo {year}
  {2018}),\ \Eprint {http://arxiv.org/abs/arXiv:1810.02373} {arXiv:1810.02373}
  \BibitemShut {NoStop}%
\bibitem [{\citenamefont {Ozawa}\ \emph {et~al.}(2019)\citenamefont {Ozawa},
  \citenamefont {Price}, \citenamefont {Amo}, \citenamefont {Goldman},
  \citenamefont {Hafezi}, \citenamefont {Lu}, \citenamefont {Rechtsman},
  \citenamefont {Schuster}, \citenamefont {Simon}, \citenamefont {Zilberberg},\
  and\ \citenamefont {Carusotto}}]{Ozawa3}%
  \BibitemOpen
  \bibfield  {author} {\bibinfo {author} {\bibfnamefont {Tomoki}\ \bibnamefont
  {Ozawa}}, \bibinfo {author} {\bibfnamefont {Hannah~M.}\ \bibnamefont
  {Price}}, \bibinfo {author} {\bibfnamefont {Alberto}\ \bibnamefont {Amo}},
  \bibinfo {author} {\bibfnamefont {Nathan}\ \bibnamefont {Goldman}}, \bibinfo
  {author} {\bibfnamefont {Mohammad}\ \bibnamefont {Hafezi}}, \bibinfo {author}
  {\bibfnamefont {Ling}\ \bibnamefont {Lu}}, \bibinfo {author} {\bibfnamefont
  {Mikael~C.}\ \bibnamefont {Rechtsman}}, \bibinfo {author} {\bibfnamefont
  {David}\ \bibnamefont {Schuster}}, \bibinfo {author} {\bibfnamefont
  {Jonathan}\ \bibnamefont {Simon}}, \bibinfo {author} {\bibfnamefont {Oded}\
  \bibnamefont {Zilberberg}}, \ and\ \bibinfo {author} {\bibfnamefont {Iacopo}\
  \bibnamefont {Carusotto}},\ }\bibfield  {title} {\enquote {\bibinfo {title}
  {Topological photonics},}\ }\href {\doibase 10.1103/RevModPhys.91.015006}
  {\bibfield  {journal} {\bibinfo  {journal} {Rev. Mod. Phys.}\ }\textbf
  {\bibinfo {volume} {91}},\ \bibinfo {pages} {015006} (\bibinfo {year}
  {2019})}\BibitemShut {NoStop}%
\bibitem [{\citenamefont {Goldman}\ \emph {et~al.}(2016)\citenamefont
  {Goldman}, \citenamefont {Budich},\ and\ \citenamefont {Zoller}}]{Goldman}%
  \BibitemOpen
  \bibfield  {author} {\bibinfo {author} {\bibfnamefont {N.}~\bibnamefont
  {Goldman}}, \bibinfo {author} {\bibfnamefont {J.~C.}\ \bibnamefont {Budich}},
  \ and\ \bibinfo {author} {\bibfnamefont {P.}~\bibnamefont {Zoller}},\
  }\bibfield  {title} {\enquote {\bibinfo {title} {Topological quantum matter
  with ultracold gases in optical lattices},}\ }\href {\doibase
  10.1038/nphys3803} {\bibfield  {journal} {\bibinfo  {journal} {Nature
  Physics}\ }\textbf {\bibinfo {volume} {12}},\ \bibinfo {pages} {639--645}
  (\bibinfo {year} {2016})}\BibitemShut {NoStop}%
\bibitem [{\citenamefont {Zhang}\ \emph {et~al.}(2018)\citenamefont {Zhang},
  \citenamefont {Zhu}, \citenamefont {Zhao}, \citenamefont {Yan},\ and\
  \citenamefont {Zhu}}]{DWZhang2018}%
  \BibitemOpen
  \bibfield  {author} {\bibinfo {author} {\bibfnamefont {D.-W.}\ \bibnamefont
  {Zhang}}, \bibinfo {author} {\bibfnamefont {Y.-Q.}\ \bibnamefont {Zhu}},
  \bibinfo {author} {\bibfnamefont {Y.~X.}\ \bibnamefont {Zhao}}, \bibinfo
  {author} {\bibfnamefont {H.}~\bibnamefont {Yan}}, \ and\ \bibinfo {author}
  {\bibfnamefont {S-.L.}\ \bibnamefont {Zhu}},\ }\bibfield  {title} {\enquote
  {\bibinfo {title} {Topological quantum matter with cold atoms},}\ }\href
  {\doibase 10.1080/00018732.2019.1594094} {\bibfield  {journal} {\bibinfo
  {journal} {Adv. Phys.}\ }\textbf {\bibinfo {volume} {67}},\ \bibinfo {pages}
  {253--402} (\bibinfo {year} {2018})}\BibitemShut {NoStop}%
\bibitem [{\citenamefont {Cooper}\ \emph {et~al.}(2019)\citenamefont {Cooper},
  \citenamefont {Dalibard},\ and\ \citenamefont {Spielman}}]{Cooper2019}%
  \BibitemOpen
  \bibfield  {author} {\bibinfo {author} {\bibfnamefont {N.~R.}\ \bibnamefont
  {Cooper}}, \bibinfo {author} {\bibfnamefont {J.}~\bibnamefont {Dalibard}}, \
  and\ \bibinfo {author} {\bibfnamefont {I.~B.}\ \bibnamefont {Spielman}},\
  }\bibfield  {title} {\enquote {\bibinfo {title} {Topological bands for
  ultracold atoms},}\ }\href {\doibase 10.1103/RevModPhys.91.015005} {\bibfield
   {journal} {\bibinfo  {journal} {Rev. Mod. Phys.}\ }\textbf {\bibinfo
  {volume} {91}},\ \bibinfo {pages} {015005} (\bibinfo {year}
  {2019})}\BibitemShut {NoStop}%
\bibitem [{\citenamefont {\"Unal}\ \emph {et~al.}(2020)\citenamefont {\"Unal},
  \citenamefont {Bouhon},\ and\ \citenamefont {Slager}}]{Unal_quenched_Euler}%
  \BibitemOpen
  \bibfield  {author} {\bibinfo {author} {\bibfnamefont {F.~Nur}\ \bibnamefont
  {\"Unal}}, \bibinfo {author} {\bibfnamefont {Adrien}\ \bibnamefont {Bouhon}},
  \ and\ \bibinfo {author} {\bibfnamefont {Robert-Jan}\ \bibnamefont
  {Slager}},\ }\bibfield  {title} {\enquote {\bibinfo {title} {Topological
  euler class as a dynamical observable in optical lattices},}\ }\href
  {\doibase 10.1103/PhysRevLett.125.053601} {\bibfield  {journal} {\bibinfo
  {journal} {Phys. Rev. Lett.}\ }\textbf {\bibinfo {volume} {125}},\ \bibinfo
  {pages} {053601} (\bibinfo {year} {2020})}\BibitemShut {NoStop}%
\bibitem [{\citenamefont {Slager}\ \emph {et~al.}(2022)\citenamefont {Slager},
  \citenamefont {Bouhon},\ and\ \citenamefont {{\"U}nal}}]{slager2022floquet}%
  \BibitemOpen
  \bibfield  {author} {\bibinfo {author} {\bibfnamefont {Robert-Jan}\
  \bibnamefont {Slager}}, \bibinfo {author} {\bibfnamefont {Adrien}\
  \bibnamefont {Bouhon}}, \ and\ \bibinfo {author} {\bibfnamefont {F~Nur}\
  \bibnamefont {{\"U}nal}},\ }\bibfield  {title} {\enquote {\bibinfo {title}
  {Floquet multi-gap topology: Non-abelian braiding and anomalous dirac string
  phase},}\ }\href {https://arxiv.org/abs/2208.12824} {\bibfield  {journal}
  {\bibinfo  {journal} {arXiv preprint arXiv:2208.12824}\ } (\bibinfo {year}
  {2022})}\BibitemShut {NoStop}%
\bibitem [{\citenamefont {Peng}\ \emph
  {et~al.}(2022{\natexlab{a}})\citenamefont {Peng}, \citenamefont {Bouhon},
  \citenamefont {Monserrat},\ and\ \citenamefont {Slager}}]{Peng2022}%
  \BibitemOpen
  \bibfield  {author} {\bibinfo {author} {\bibfnamefont {Bo}~\bibnamefont
  {Peng}}, \bibinfo {author} {\bibfnamefont {Adrien}\ \bibnamefont {Bouhon}},
  \bibinfo {author} {\bibfnamefont {Bartomeu}\ \bibnamefont {Monserrat}}, \
  and\ \bibinfo {author} {\bibfnamefont {Robert-Jan}\ \bibnamefont {Slager}},\
  }\bibfield  {title} {\enquote {\bibinfo {title} {Phonons as a platform for
  non-abelian braiding and its manifestation in layered silicates},}\ }\href
  {\doibase 10.1038/s41467-022-28046-9} {\bibfield  {journal} {\bibinfo
  {journal} {Nature Communications}\ }\textbf {\bibinfo {volume} {13}},\
  \bibinfo {pages} {423} (\bibinfo {year} {2022}{\natexlab{a}})}\BibitemShut
  {NoStop}%
\bibitem [{\citenamefont {Park}\ \emph {et~al.}(2021)\citenamefont {Park},
  \citenamefont {Hwang}, \citenamefont {Choi},\ and\ \citenamefont
  {Yang}}]{Park2021}%
  \BibitemOpen
  \bibfield  {author} {\bibinfo {author} {\bibfnamefont {Sungjoon}\
  \bibnamefont {Park}}, \bibinfo {author} {\bibfnamefont {Yoonseok}\
  \bibnamefont {Hwang}}, \bibinfo {author} {\bibfnamefont {Hong~Chul}\
  \bibnamefont {Choi}}, \ and\ \bibinfo {author} {\bibfnamefont {Bohm~Jung}\
  \bibnamefont {Yang}},\ }\bibfield  {title} {\enquote {\bibinfo {title}
  {{Topological acoustic triple point}},}\ }\href {\doibase
  10.1038/s41467-021-27158-y} {\bibfield  {journal} {\bibinfo  {journal}
  {Nature Communications}\ }\textbf {\bibinfo {volume} {12}},\ \bibinfo {pages}
  {1--9} (\bibinfo {year} {2021})}\BibitemShut {NoStop}%
\bibitem [{\citenamefont {Peng}\ \emph
  {et~al.}(2022{\natexlab{b}})\citenamefont {Peng}, \citenamefont {Bouhon},
  \citenamefont {Slager},\ and\ \citenamefont {Monserrat}}]{Peng2022b}%
  \BibitemOpen
  \bibfield  {author} {\bibinfo {author} {\bibfnamefont {Bo}~\bibnamefont
  {Peng}}, \bibinfo {author} {\bibfnamefont {Adrien}\ \bibnamefont {Bouhon}},
  \bibinfo {author} {\bibfnamefont {Robert-Jan}\ \bibnamefont {Slager}}, \ and\
  \bibinfo {author} {\bibfnamefont {Bartomeu}\ \bibnamefont {Monserrat}},\
  }\bibfield  {title} {\enquote {\bibinfo {title} {Multigap topology and
  non-abelian braiding of phonons from first principles},}\ }\href {\doibase
  10.1103/PhysRevB.105.085115} {\bibfield  {journal} {\bibinfo  {journal}
  {Phys. Rev. B}\ }\textbf {\bibinfo {volume} {105}},\ \bibinfo {pages}
  {085115} (\bibinfo {year} {2022}{\natexlab{b}})}\BibitemShut {NoStop}%
\bibitem [{\citenamefont {Lange}\ \emph {et~al.}(2021)\citenamefont {Lange},
  \citenamefont {Bouhon},\ and\ \citenamefont {Slager}}]{Lange_2021}%
  \BibitemOpen
  \bibfield  {author} {\bibinfo {author} {\bibfnamefont {Gunnar~F.}\
  \bibnamefont {Lange}}, \bibinfo {author} {\bibfnamefont {Adrien}\
  \bibnamefont {Bouhon}}, \ and\ \bibinfo {author} {\bibfnamefont {Robert-Jan}\
  \bibnamefont {Slager}},\ }\bibfield  {title} {\enquote {\bibinfo {title}
  {Subdimensional topologies, indicators, and higher order boundary effects},}\
  }\href {\doibase 10.1103/PhysRevB.103.195145} {\bibfield  {journal} {\bibinfo
   {journal} {Phys. Rev. B}\ }\textbf {\bibinfo {volume} {103}},\ \bibinfo
  {pages} {195145} (\bibinfo {year} {2021})}\BibitemShut {NoStop}%
\bibitem [{\citenamefont {Chen}\ \emph
  {et~al.}(2022{\natexlab{a}})\citenamefont {Chen}, \citenamefont {Bouhon},
  \citenamefont {Slager},\ and\ \citenamefont {Monserrat}}]{Chen_2022}%
  \BibitemOpen
  \bibfield  {author} {\bibinfo {author} {\bibfnamefont {Siyu}\ \bibnamefont
  {Chen}}, \bibinfo {author} {\bibfnamefont {Adrien}\ \bibnamefont {Bouhon}},
  \bibinfo {author} {\bibfnamefont {Robert-Jan}\ \bibnamefont {Slager}}, \ and\
  \bibinfo {author} {\bibfnamefont {Bartomeu}\ \bibnamefont {Monserrat}},\
  }\bibfield  {title} {\enquote {\bibinfo {title} {Non-abelian braiding of weyl
  nodes via symmetry-constrained phase transitions},}\ }\href {\doibase
  10.1103/PhysRevB.105.L081117} {\bibfield  {journal} {\bibinfo  {journal}
  {Phys. Rev. B}\ }\textbf {\bibinfo {volume} {105}},\ \bibinfo {pages}
  {L081117} (\bibinfo {year} {2022}{\natexlab{a}})}\BibitemShut {NoStop}%
\bibitem [{\citenamefont {K\"onye}\ \emph {et~al.}(2021)\citenamefont
  {K\"onye}, \citenamefont {Bouhon}, \citenamefont {Fulga}, \citenamefont
  {Slager}, \citenamefont {van~den Brink},\ and\ \citenamefont
  {Facio}}]{Konye_2021}%
  \BibitemOpen
  \bibfield  {author} {\bibinfo {author} {\bibfnamefont {Viktor}\ \bibnamefont
  {K\"onye}}, \bibinfo {author} {\bibfnamefont {Adrien}\ \bibnamefont
  {Bouhon}}, \bibinfo {author} {\bibfnamefont {Ion~Cosma}\ \bibnamefont
  {Fulga}}, \bibinfo {author} {\bibfnamefont {Robert-Jan}\ \bibnamefont
  {Slager}}, \bibinfo {author} {\bibfnamefont {Jeroen}\ \bibnamefont {van~den
  Brink}}, \ and\ \bibinfo {author} {\bibfnamefont {Jorge~I.}\ \bibnamefont
  {Facio}},\ }\bibfield  {title} {\enquote {\bibinfo {title} {Chirality flip of
  weyl nodes and its manifestation in strained ${\mathrm{mote}}_{2}$},}\ }\href
  {\doibase 10.1103/PhysRevResearch.3.L042017} {\bibfield  {journal} {\bibinfo
  {journal} {Phys. Rev. Research}\ }\textbf {\bibinfo {volume} {3}},\ \bibinfo
  {pages} {L042017} (\bibinfo {year} {2021})}\BibitemShut {NoStop}%
\bibitem [{\citenamefont {Bouhon}\ \emph {et~al.}(2021)\citenamefont {Bouhon},
  \citenamefont {Lange},\ and\ \citenamefont {Slager}}]{Bouhon2020_mag}%
  \BibitemOpen
  \bibfield  {author} {\bibinfo {author} {\bibfnamefont {Adrien}\ \bibnamefont
  {Bouhon}}, \bibinfo {author} {\bibfnamefont {Gunnar~F.}\ \bibnamefont
  {Lange}}, \ and\ \bibinfo {author} {\bibfnamefont {Robert-Jan}\ \bibnamefont
  {Slager}},\ }\bibfield  {title} {\enquote {\bibinfo {title} {Topological
  correspondence between magnetic space group representations and
  subdimensions},}\ }\href {\doibase 10.1103/PhysRevB.103.245127} {\bibfield
  {journal} {\bibinfo  {journal} {Phys. Rev. B}\ }\textbf {\bibinfo {volume}
  {103}},\ \bibinfo {pages} {245127} (\bibinfo {year} {2021})}\BibitemShut
  {NoStop}%
\bibitem [{\citenamefont {Jiang}\ \emph {et~al.}(2021)\citenamefont {Jiang},
  \citenamefont {Bouhon}, \citenamefont {Lin}, \citenamefont {Zhou},
  \citenamefont {Hou}, \citenamefont {Li}, \citenamefont {Slager},\ and\
  \citenamefont {Jiang}}]{Jiang2021euler}%
  \BibitemOpen
  \bibfield  {author} {\bibinfo {author} {\bibfnamefont {Bin}\ \bibnamefont
  {Jiang}}, \bibinfo {author} {\bibfnamefont {Adrien}\ \bibnamefont {Bouhon}},
  \bibinfo {author} {\bibfnamefont {Zhi-Kang}\ \bibnamefont {Lin}}, \bibinfo
  {author} {\bibfnamefont {Xiaoxi}\ \bibnamefont {Zhou}}, \bibinfo {author}
  {\bibfnamefont {Bo}~\bibnamefont {Hou}}, \bibinfo {author} {\bibfnamefont
  {Feng}\ \bibnamefont {Li}}, \bibinfo {author} {\bibfnamefont {Robert-Jan}\
  \bibnamefont {Slager}}, \ and\ \bibinfo {author} {\bibfnamefont {Jian-Hua}\
  \bibnamefont {Jiang}},\ }\bibfield  {title} {\enquote {\bibinfo {title}
  {Experimental observation of non-abelian topological acoustic semimetals and
  their phase transitions},}\ }\href {\doibase 10.1038/s41567-021-01340-x}
  {\bibfield  {journal} {\bibinfo  {journal} {Nature Physics}\ }\textbf
  {\bibinfo {volume} {17}},\ \bibinfo {pages} {1239--1246} (\bibinfo {year}
  {2021})}\BibitemShut {NoStop}%
\bibitem [{\citenamefont {Guo}\ \emph {et~al.}(2021)\citenamefont {Guo},
  \citenamefont {Jiang}, \citenamefont {Zhang}, \citenamefont {Zhang},
  \citenamefont {Zhang}, \citenamefont {Yang}, \citenamefont {Zhang},\ and\
  \citenamefont {Chan}}]{Guo1Dexp}%
  \BibitemOpen
  \bibfield  {author} {\bibinfo {author} {\bibfnamefont {Qinghua}\ \bibnamefont
  {Guo}}, \bibinfo {author} {\bibfnamefont {Tianshu}\ \bibnamefont {Jiang}},
  \bibinfo {author} {\bibfnamefont {Ruo-Yang}\ \bibnamefont {Zhang}}, \bibinfo
  {author} {\bibfnamefont {Lei}\ \bibnamefont {Zhang}}, \bibinfo {author}
  {\bibfnamefont {Zhao-Qing}\ \bibnamefont {Zhang}}, \bibinfo {author}
  {\bibfnamefont {Biao}\ \bibnamefont {Yang}}, \bibinfo {author} {\bibfnamefont
  {Shuang}\ \bibnamefont {Zhang}}, \ and\ \bibinfo {author} {\bibfnamefont
  {C.~T.}\ \bibnamefont {Chan}},\ }\bibfield  {title} {\enquote {\bibinfo
  {title} {Experimental observation of non-abelian topological charges and edge
  states},}\ }\href {\doibase 10.1038/s41586-021-03521-3} {\bibfield  {journal}
  {\bibinfo  {journal} {Nature}\ }\textbf {\bibinfo {volume} {594}},\ \bibinfo
  {pages} {195--200} (\bibinfo {year} {2021})}\BibitemShut {NoStop}%
\bibitem [{\citenamefont {Jiang}\ \emph {et~al.}(2022)\citenamefont {Jiang},
  \citenamefont {Bouhon}, \citenamefont {Wu}, \citenamefont {Kong},
  \citenamefont {Lin}, \citenamefont {Slager},\ and\ \citenamefont
  {Jiang}}]{jiang2022experimental}%
  \BibitemOpen
  \bibfield  {author} {\bibinfo {author} {\bibfnamefont {Bin}\ \bibnamefont
  {Jiang}}, \bibinfo {author} {\bibfnamefont {Adrien}\ \bibnamefont {Bouhon}},
  \bibinfo {author} {\bibfnamefont {Shi-Qiao}\ \bibnamefont {Wu}}, \bibinfo
  {author} {\bibfnamefont {Ze-Lin}\ \bibnamefont {Kong}}, \bibinfo {author}
  {\bibfnamefont {Zhi-Kang}\ \bibnamefont {Lin}}, \bibinfo {author}
  {\bibfnamefont {Robert-Jan}\ \bibnamefont {Slager}}, \ and\ \bibinfo {author}
  {\bibfnamefont {Jian-Hua}\ \bibnamefont {Jiang}},\ }\bibfield  {title}
  {\enquote {\bibinfo {title} {Experimental observation of meronic topological
  acoustic euler insulators},}\ }\href {https://arxiv.org/abs/2205.03429}
  {\bibfield  {journal} {\bibinfo  {journal} {arXiv preprint arXiv:2205.03429}\
  } (\bibinfo {year} {2022})}\BibitemShut {NoStop}%
\bibitem [{\citenamefont {Zhao}\ \emph
  {et~al.}(2022{\natexlab{a}})\citenamefont {Zhao}, \citenamefont {Yang},
  \citenamefont {Jiang}, \citenamefont {Mao}, \citenamefont {Guo},
  \citenamefont {Qiu}, \citenamefont {Wang}, \citenamefont {Yao}, \citenamefont
  {He}, \citenamefont {Zhou}, \citenamefont {Xu},\ and\ \citenamefont
  {Duan}}]{zhao2022observation}%
  \BibitemOpen
  \bibfield  {author} {\bibinfo {author} {\bibfnamefont {W.~D.}\ \bibnamefont
  {Zhao}}, \bibinfo {author} {\bibfnamefont {Y.~B.}\ \bibnamefont {Yang}},
  \bibinfo {author} {\bibfnamefont {Y.}~\bibnamefont {Jiang}}, \bibinfo
  {author} {\bibfnamefont {Z.~C.}\ \bibnamefont {Mao}}, \bibinfo {author}
  {\bibfnamefont {W.~X.}\ \bibnamefont {Guo}}, \bibinfo {author} {\bibfnamefont
  {L.~Y.}\ \bibnamefont {Qiu}}, \bibinfo {author} {\bibfnamefont {G.~X.}\
  \bibnamefont {Wang}}, \bibinfo {author} {\bibfnamefont {L.}~\bibnamefont
  {Yao}}, \bibinfo {author} {\bibfnamefont {L.}~\bibnamefont {He}}, \bibinfo
  {author} {\bibfnamefont {Z.~C.}\ \bibnamefont {Zhou}}, \bibinfo {author}
  {\bibfnamefont {Y.}~\bibnamefont {Xu}}, \ and\ \bibinfo {author}
  {\bibfnamefont {L.~M.}\ \bibnamefont {Duan}},\ }\href@noop {} {\enquote
  {\bibinfo {title} {Observation of topological euler insulators with a
  trapped-ion quantum simulator},}\ } (\bibinfo {year} {2022}{\natexlab{a}}),\
  \Eprint {http://arxiv.org/abs/2201.09234} {arXiv:2201.09234 [quant-ph]}
  \BibitemShut {NoStop}%
\bibitem [{\citenamefont {Price}\ \emph {et~al.}(2015)\citenamefont {Price},
  \citenamefont {Zilberberg}, \citenamefont {Ozawa}, \citenamefont
  {Carusotto},\ and\ \citenamefont {Goldman}}]{Price}%
  \BibitemOpen
  \bibfield  {author} {\bibinfo {author} {\bibfnamefont {H.~M.}\ \bibnamefont
  {Price}}, \bibinfo {author} {\bibfnamefont {O.}~\bibnamefont {Zilberberg}},
  \bibinfo {author} {\bibfnamefont {T.}~\bibnamefont {Ozawa}}, \bibinfo
  {author} {\bibfnamefont {I.}~\bibnamefont {Carusotto}}, \ and\ \bibinfo
  {author} {\bibfnamefont {N.}~\bibnamefont {Goldman}},\ }\bibfield  {title}
  {\enquote {\bibinfo {title} {Four-dimensional quantum hall effect with
  ultracold atoms},}\ }\href {\doibase 10.1103/PhysRevLett.115.195303}
  {\bibfield  {journal} {\bibinfo  {journal} {Phys. Rev. Lett.}\ }\textbf
  {\bibinfo {volume} {115}},\ \bibinfo {pages} {195303} (\bibinfo {year}
  {2015})}\BibitemShut {NoStop}%
\bibitem [{\citenamefont {Kraus}\ \emph {et~al.}(2013)\citenamefont {Kraus},
  \citenamefont {Ringel},\ and\ \citenamefont {Zilberberg}}]{Kraus}%
  \BibitemOpen
  \bibfield  {author} {\bibinfo {author} {\bibfnamefont {Yaacov~E.}\
  \bibnamefont {Kraus}}, \bibinfo {author} {\bibfnamefont {Zohar}\ \bibnamefont
  {Ringel}}, \ and\ \bibinfo {author} {\bibfnamefont {Oded}\ \bibnamefont
  {Zilberberg}},\ }\bibfield  {title} {\enquote {\bibinfo {title}
  {Four-dimensional quantum hall effect in a two-dimensional quasicrystal},}\
  }\href {\doibase 10.1103/PhysRevLett.111.226401} {\bibfield  {journal}
  {\bibinfo  {journal} {Phys. Rev. Lett.}\ }\textbf {\bibinfo {volume} {111}},\
  \bibinfo {pages} {226401} (\bibinfo {year} {2013})}\BibitemShut {NoStop}%
\bibitem [{\citenamefont {Lohse}\ \emph {et~al.}(2018)\citenamefont {Lohse},
  \citenamefont {Schweizer}, \citenamefont {Price}, \citenamefont
  {Zilberberg},\ and\ \citenamefont {Bloch}}]{Lohse}%
  \BibitemOpen
  \bibfield  {author} {\bibinfo {author} {\bibfnamefont {Michael}\ \bibnamefont
  {Lohse}}, \bibinfo {author} {\bibfnamefont {Christian}\ \bibnamefont
  {Schweizer}}, \bibinfo {author} {\bibfnamefont {Hannah~M.}\ \bibnamefont
  {Price}}, \bibinfo {author} {\bibfnamefont {Oded}\ \bibnamefont
  {Zilberberg}}, \ and\ \bibinfo {author} {\bibfnamefont {Immanuel}\
  \bibnamefont {Bloch}},\ }\bibfield  {title} {\enquote {\bibinfo {title}
  {Exploring 4d quantum hall physics with a 2d topological charge pump},}\
  }\href {\doibase 10.1038/nature25000} {\bibfield  {journal} {\bibinfo
  {journal} {Nature}\ }\textbf {\bibinfo {volume} {553}},\ \bibinfo {pages}
  {55--58} (\bibinfo {year} {2018})}\BibitemShut {NoStop}%
\bibitem [{\citenamefont {Zilberberg}\ \emph {et~al.}(2018)\citenamefont
  {Zilberberg}, \citenamefont {Huang}, \citenamefont {Guglielmon},
  \citenamefont {Wang}, \citenamefont {Chen}, \citenamefont {Kraus},\ and\
  \citenamefont {Rechtsman}}]{Zilberberg}%
  \BibitemOpen
  \bibfield  {author} {\bibinfo {author} {\bibfnamefont {Oded}\ \bibnamefont
  {Zilberberg}}, \bibinfo {author} {\bibfnamefont {Sheng}\ \bibnamefont
  {Huang}}, \bibinfo {author} {\bibfnamefont {Jonathan}\ \bibnamefont
  {Guglielmon}}, \bibinfo {author} {\bibfnamefont {Mohan}\ \bibnamefont
  {Wang}}, \bibinfo {author} {\bibfnamefont {Kevin~P.}\ \bibnamefont {Chen}},
  \bibinfo {author} {\bibfnamefont {Yaacov~E.}\ \bibnamefont {Kraus}}, \ and\
  \bibinfo {author} {\bibfnamefont {Mikael~C.}\ \bibnamefont {Rechtsman}},\
  }\bibfield  {title} {\enquote {\bibinfo {title} {Photonic topological
  boundary pumping as a probe of 4d quantum hall physics},}\ }\href {\doibase
  10.1038/nature25011} {\bibfield  {journal} {\bibinfo  {journal} {Nature}\
  }\textbf {\bibinfo {volume} {553}},\ \bibinfo {pages} {59--62} (\bibinfo
  {year} {2018})}\BibitemShut {NoStop}%
\bibitem [{\citenamefont {Palumbo}\ and\ \citenamefont
  {Goldman}(2018)}]{Palumbo2}%
  \BibitemOpen
  \bibfield  {author} {\bibinfo {author} {\bibfnamefont {Giandomenico}\
  \bibnamefont {Palumbo}}\ and\ \bibinfo {author} {\bibfnamefont {Nathan}\
  \bibnamefont {Goldman}},\ }\bibfield  {title} {\enquote {\bibinfo {title}
  {Revealing tensor monopoles through quantum-metric measurements},}\ }\href
  {\doibase 10.1103/PhysRevLett.121.170401} {\bibfield  {journal} {\bibinfo
  {journal} {Phys. Rev. Lett.}\ }\textbf {\bibinfo {volume} {121}},\ \bibinfo
  {pages} {170401} (\bibinfo {year} {2018})}\BibitemShut {NoStop}%
\bibitem [{\citenamefont {Palumbo}\ and\ \citenamefont
  {Goldman}(2019)}]{Palumbo3}%
  \BibitemOpen
  \bibfield  {author} {\bibinfo {author} {\bibfnamefont {Giandomenico}\
  \bibnamefont {Palumbo}}\ and\ \bibinfo {author} {\bibfnamefont {Nathan}\
  \bibnamefont {Goldman}},\ }\bibfield  {title} {\enquote {\bibinfo {title}
  {Tensor berry connections and their topological invariants},}\ }\href
  {\doibase 10.1103/PhysRevB.99.045154} {\bibfield  {journal} {\bibinfo
  {journal} {Phys. Rev. B}\ }\textbf {\bibinfo {volume} {99}},\ \bibinfo
  {pages} {045154} (\bibinfo {year} {2019})}\BibitemShut {NoStop}%
\bibitem [{\citenamefont {Zhu}\ \emph {et~al.}(2020)\citenamefont {Zhu},
  \citenamefont {Goldman},\ and\ \citenamefont {Palumbo}}]{Palumbo4}%
  \BibitemOpen
  \bibfield  {author} {\bibinfo {author} {\bibfnamefont {Yan-Qing}\
  \bibnamefont {Zhu}}, \bibinfo {author} {\bibfnamefont {Nathan}\ \bibnamefont
  {Goldman}}, \ and\ \bibinfo {author} {\bibfnamefont {Giandomenico}\
  \bibnamefont {Palumbo}},\ }\bibfield  {title} {\enquote {\bibinfo {title}
  {Four-dimensional semimetals with tensor monopoles: From surface states to
  topological responses},}\ }\href {\doibase 10.1103/PhysRevB.102.081109}
  {\bibfield  {journal} {\bibinfo  {journal} {Phys. Rev. B}\ }\textbf {\bibinfo
  {volume} {102}},\ \bibinfo {pages} {081109} (\bibinfo {year}
  {2020})}\BibitemShut {NoStop}%
\bibitem [{\citenamefont {Ding}\ \emph {et~al.}(2020)\citenamefont {Ding},
  \citenamefont {Zhu}, \citenamefont {Li},\ and\ \citenamefont
  {Shao}}]{HTDing2020}%
  \BibitemOpen
  \bibfield  {author} {\bibinfo {author} {\bibfnamefont {Hai-Tao}\ \bibnamefont
  {Ding}}, \bibinfo {author} {\bibfnamefont {Yan-Qing}\ \bibnamefont {Zhu}},
  \bibinfo {author} {\bibfnamefont {Zhi}\ \bibnamefont {Li}}, \ and\ \bibinfo
  {author} {\bibfnamefont {Lubing}\ \bibnamefont {Shao}},\ }\bibfield  {title}
  {\enquote {\bibinfo {title} {Tensor monopoles and the negative
  magnetoresistance effect in optical lattices},}\ }\href {\doibase
  10.1103/PhysRevA.102.053325} {\bibfield  {journal} {\bibinfo  {journal}
  {Phys. Rev. A}\ }\textbf {\bibinfo {volume} {102}},\ \bibinfo {pages}
  {053325} (\bibinfo {year} {2020})}\BibitemShut {NoStop}%
\bibitem [{\citenamefont {Chen}\ \emph
  {et~al.}(2022{\natexlab{b}})\citenamefont {Chen}, \citenamefont {null null},
  \citenamefont {Li}, \citenamefont {Palumbo}, \citenamefont {Zhu},
  \citenamefont {Goldman},\ and\ \citenamefont {Cappellaro}}]{Cappellaro}%
  \BibitemOpen
  \bibfield  {author} {\bibinfo {author} {\bibfnamefont {Mo}~\bibnamefont
  {Chen}}, \bibinfo {author} {\bibnamefont {null null}}, \bibinfo {author}
  {\bibfnamefont {Changhao}\ \bibnamefont {Li}}, \bibinfo {author}
  {\bibfnamefont {Giandomenico}\ \bibnamefont {Palumbo}}, \bibinfo {author}
  {\bibfnamefont {Yan-Qing}\ \bibnamefont {Zhu}}, \bibinfo {author}
  {\bibfnamefont {Nathan}\ \bibnamefont {Goldman}}, \ and\ \bibinfo {author}
  {\bibfnamefont {Paola}\ \bibnamefont {Cappellaro}},\ }\bibfield  {title}
  {\enquote {\bibinfo {title} {A synthetic monopole source of kalb-ramond field
  in diamond},}\ }\href {\doibase 10.1126/science.abe6437} {\bibfield
  {journal} {\bibinfo  {journal} {Science}\ }\textbf {\bibinfo {volume}
  {375}},\ \bibinfo {pages} {1017--1020} (\bibinfo {year}
  {2022}{\natexlab{b}})},\ \Eprint
  {http://arxiv.org/abs/https://www.science.org/doi/pdf/10.1126/science.abe6437}
  {https://www.science.org/doi/pdf/10.1126/science.abe6437} \BibitemShut
  {NoStop}%
\bibitem [{\citenamefont {Tan}\ \emph {et~al.}(2021)\citenamefont {Tan},
  \citenamefont {Zhang}, \citenamefont {Zheng}, \citenamefont {Yang},
  \citenamefont {Song}, \citenamefont {Han}, \citenamefont {Dong},
  \citenamefont {Wang}, \citenamefont {Lan}, \citenamefont {Yan}, \citenamefont
  {Zhu},\ and\ \citenamefont {Yu}}]{Tan}%
  \BibitemOpen
  \bibfield  {author} {\bibinfo {author} {\bibfnamefont {Xinsheng}\
  \bibnamefont {Tan}}, \bibinfo {author} {\bibfnamefont {Dan-Wei}\ \bibnamefont
  {Zhang}}, \bibinfo {author} {\bibfnamefont {Wen}\ \bibnamefont {Zheng}},
  \bibinfo {author} {\bibfnamefont {Xiaopei}\ \bibnamefont {Yang}}, \bibinfo
  {author} {\bibfnamefont {Shuqing}\ \bibnamefont {Song}}, \bibinfo {author}
  {\bibfnamefont {Zhikun}\ \bibnamefont {Han}}, \bibinfo {author}
  {\bibfnamefont {Yuqian}\ \bibnamefont {Dong}}, \bibinfo {author}
  {\bibfnamefont {Zhimin}\ \bibnamefont {Wang}}, \bibinfo {author}
  {\bibfnamefont {Dong}\ \bibnamefont {Lan}}, \bibinfo {author} {\bibfnamefont
  {Hui}\ \bibnamefont {Yan}}, \bibinfo {author} {\bibfnamefont {Shi-Liang}\
  \bibnamefont {Zhu}}, \ and\ \bibinfo {author} {\bibfnamefont {Yang}\
  \bibnamefont {Yu}},\ }\bibfield  {title} {\enquote {\bibinfo {title}
  {Experimental observation of tensor monopoles with a superconducting
  qudit},}\ }\href {\doibase 10.1103/PhysRevLett.126.017702} {\bibfield
  {journal} {\bibinfo  {journal} {Phys. Rev. Lett.}\ }\textbf {\bibinfo
  {volume} {126}},\ \bibinfo {pages} {017702} (\bibinfo {year}
  {2021})}\BibitemShut {NoStop}%
\bibitem [{\citenamefont {Zhu}\ \emph {et~al.}(2022)\citenamefont {Zhu},
  \citenamefont {Zheng}, \citenamefont {Palumbo},\ and\ \citenamefont
  {Wang}}]{YQZhu2022}%
  \BibitemOpen
  \bibfield  {author} {\bibinfo {author} {\bibfnamefont {Yan-Qing}\
  \bibnamefont {Zhu}}, \bibinfo {author} {\bibfnamefont {Zhen}\ \bibnamefont
  {Zheng}}, \bibinfo {author} {\bibfnamefont {Giandomenico}\ \bibnamefont
  {Palumbo}}, \ and\ \bibinfo {author} {\bibfnamefont {Z.~D.}\ \bibnamefont
  {Wang}},\ }\bibfield  {title} {\enquote {\bibinfo {title} {Topological
  electromagnetic effects and higher second chern numbers in four-dimensional
  gapped phases},}\ }\href {\doibase 10.1103/PhysRevLett.129.196602} {\bibfield
   {journal} {\bibinfo  {journal} {Phys. Rev. Lett.}\ }\textbf {\bibinfo
  {volume} {129}},\ \bibinfo {pages} {196602} (\bibinfo {year}
  {2022})}\BibitemShut {NoStop}%
\bibitem [{\citenamefont {Zhao}\ and\ \citenamefont {Lu}(2017)}]{Zhao_PT}%
  \BibitemOpen
  \bibfield  {author} {\bibinfo {author} {\bibfnamefont {Y.~X.}\ \bibnamefont
  {Zhao}}\ and\ \bibinfo {author} {\bibfnamefont {Y.}~\bibnamefont {Lu}},\
  }\bibfield  {title} {\enquote {\bibinfo {title} {${P}{T}$-{S}ymmetric real
  {D}irac {F}ermions and {S}emimetals},}\ }\href {\doibase
  10.1103/PhysRevLett.118.056401} {\bibfield  {journal} {\bibinfo  {journal}
  {Phys. Rev. Lett.}\ }\textbf {\bibinfo {volume} {118}},\ \bibinfo {pages}
  {056401} (\bibinfo {year} {2017})}\BibitemShut {NoStop}%
\bibitem [{\citenamefont {Ahn}\ \emph {et~al.}(2018{\natexlab{b}})\citenamefont
  {Ahn}, \citenamefont {Kim}, \citenamefont {Kim},\ and\ \citenamefont
  {Yang}}]{BJY_linking}%
  \BibitemOpen
  \bibfield  {author} {\bibinfo {author} {\bibfnamefont {Junyeong}\
  \bibnamefont {Ahn}}, \bibinfo {author} {\bibfnamefont {Dongwook}\
  \bibnamefont {Kim}}, \bibinfo {author} {\bibfnamefont {Youngkuk}\
  \bibnamefont {Kim}}, \ and\ \bibinfo {author} {\bibfnamefont {Bohm-Jung}\
  \bibnamefont {Yang}},\ }\bibfield  {title} {\enquote {\bibinfo {title} {Band
  topology and linking structure of nodal line semimetals with ${Z}_{2}$
  monopole charges},}\ }\href {\doibase 10.1103/PhysRevLett.121.106403}
  {\bibfield  {journal} {\bibinfo  {journal} {Phys. Rev. Lett.}\ }\textbf
  {\bibinfo {volume} {121}},\ \bibinfo {pages} {106403} (\bibinfo {year}
  {2018}{\natexlab{b}})}\BibitemShut {NoStop}%
\bibitem [{Bot()}]{Bott}%
  \BibitemOpen
  \href@noop {} {\bibinfo  {journal} {By the Bott periodicity, the above
  classification of transition functions is equivalent to the homotopy
  classification of real Bloch Hamiltonians
  $\pi_4(\mathsf{Gr}_{4,\infty}^{\mathbb{R}})= \mathbb{Z}^2$ \cite{SM}}\
  }\BibitemShut {NoStop}%
\bibitem [{\citenamefont {Aharony}\ \emph {et~al.}(2013)\citenamefont
  {Aharony}, \citenamefont {Seiberg},\ and\ \citenamefont
  {Tachikawa}}]{Aharony}%
  \BibitemOpen
\bibfield  {journal} {  }\bibfield  {author} {\bibinfo {author} {\bibfnamefont
  {Ofer}\ \bibnamefont {Aharony}}, \bibinfo {author} {\bibfnamefont {Nathan}\
  \bibnamefont {Seiberg}}, \ and\ \bibinfo {author} {\bibfnamefont {Yuji}\
  \bibnamefont {Tachikawa}},\ }\bibfield  {title} {\enquote {\bibinfo {title}
  {Reading between the lines of four-dimensional gauge theories},}\ }\href
  {\doibase 10.1007/JHEP08(2013)115} {\bibfield  {journal} {\bibinfo  {journal}
  {Journal of High Energy Physics}\ }\textbf {\bibinfo {volume} {2013}},\
  \bibinfo {pages} {115} (\bibinfo {year} {2013})}\BibitemShut {NoStop}%
\bibitem [{\citenamefont {Flagga}\ and\ \citenamefont
  {Antonsen}(2004)}]{Flagga}%
  \BibitemOpen
  \bibfield  {author} {\bibinfo {author} {\bibfnamefont {Malene Steen~Nielsen}\
  \bibnamefont {Flagga}}\ and\ \bibinfo {author} {\bibfnamefont {Frank}\
  \bibnamefont {Antonsen}},\ }\bibfield  {title} {\enquote {\bibinfo {title}
  {Space-time topology (ii)---causality, the fourth stiefel--whitney class and
  space-time as a boundary},}\ }\href {\doibase
  10.1023/B:IJTP.0000049000.35083.fb} {\bibfield  {journal} {\bibinfo
  {journal} {International Journal of Theoretical Physics}\ }\textbf {\bibinfo
  {volume} {43}},\ \bibinfo {pages} {1917--1930} (\bibinfo {year}
  {2004})}\BibitemShut {NoStop}%
\bibitem [{\citenamefont {Belavin}\ \emph {et~al.}(1975)\citenamefont
  {Belavin}, \citenamefont {Polyakov}, \citenamefont {Schwartz},\ and\
  \citenamefont {Tyupkin}}]{Belavin}%
  \BibitemOpen
  \bibfield  {author} {\bibinfo {author} {\bibfnamefont {A.A.}\ \bibnamefont
  {Belavin}}, \bibinfo {author} {\bibfnamefont {A.M.}\ \bibnamefont
  {Polyakov}}, \bibinfo {author} {\bibfnamefont {A.S.}\ \bibnamefont
  {Schwartz}}, \ and\ \bibinfo {author} {\bibfnamefont {Yu.S.}\ \bibnamefont
  {Tyupkin}},\ }\bibfield  {title} {\enquote {\bibinfo {title} {Pseudoparticle
  solutions of the yang-mills equations},}\ }\href {\doibase
  https://doi.org/10.1016/0370-2693(75)90163-X} {\bibfield  {journal} {\bibinfo
   {journal} {Physics Letters B}\ }\textbf {\bibinfo {volume} {59}},\ \bibinfo
  {pages} {85--87} (\bibinfo {year} {1975})}\BibitemShut {NoStop}%
\bibitem [{\citenamefont {Avron}\ \emph {et~al.}(1988)\citenamefont {Avron},
  \citenamefont {Sadun}, \citenamefont {Segert},\ and\ \citenamefont
  {Simon}}]{Avron}%
  \BibitemOpen
  \bibfield  {author} {\bibinfo {author} {\bibfnamefont {J.~E.}\ \bibnamefont
  {Avron}}, \bibinfo {author} {\bibfnamefont {L.}~\bibnamefont {Sadun}},
  \bibinfo {author} {\bibfnamefont {J.}~\bibnamefont {Segert}}, \ and\ \bibinfo
  {author} {\bibfnamefont {B.}~\bibnamefont {Simon}},\ }\bibfield  {title}
  {\enquote {\bibinfo {title} {Topological invariants in fermi systems with
  time-reversal invariance},}\ }\href {\doibase 10.1103/PhysRevLett.61.1329}
  {\bibfield  {journal} {\bibinfo  {journal} {Phys. Rev. Lett.}\ }\textbf
  {\bibinfo {volume} {61}},\ \bibinfo {pages} {1329--1332} (\bibinfo {year}
  {1988})}\BibitemShut {NoStop}%
\bibitem [{\citenamefont {Shankar}\ and\ \citenamefont
  {Mathur}(1994)}]{Shankar}%
  \BibitemOpen
  \bibfield  {author} {\bibinfo {author} {\bibfnamefont {R.}~\bibnamefont
  {Shankar}}\ and\ \bibinfo {author} {\bibfnamefont {Harsh}\ \bibnamefont
  {Mathur}},\ }\bibfield  {title} {\enquote {\bibinfo {title} {Thomas
  precession, berry potential, and the meron},}\ }\href {\doibase
  10.1103/PhysRevLett.73.1565} {\bibfield  {journal} {\bibinfo  {journal}
  {Phys. Rev. Lett.}\ }\textbf {\bibinfo {volume} {73}},\ \bibinfo {pages}
  {1565--1569} (\bibinfo {year} {1994})}\BibitemShut {NoStop}%
\bibitem [{\citenamefont {Hawking}(1977)}]{Hawking}%
  \BibitemOpen
  \bibfield  {author} {\bibinfo {author} {\bibfnamefont {S.W.}\ \bibnamefont
  {Hawking}},\ }\bibfield  {title} {\enquote {\bibinfo {title} {Gravitational
  instantons},}\ }\href {\doibase https://doi.org/10.1016/0375-9601(77)90386-3}
  {\bibfield  {journal} {\bibinfo  {journal} {Physics Letters A}\ }\textbf
  {\bibinfo {volume} {60}},\ \bibinfo {pages} {81--83} (\bibinfo {year}
  {1977})}\BibitemShut {NoStop}%
\bibitem [{\citenamefont {Oh}\ \emph {et~al.}(2011)\citenamefont {Oh},
  \citenamefont {Park},\ and\ \citenamefont {Yang}}]{Oh}%
  \BibitemOpen
  \bibfield  {author} {\bibinfo {author} {\bibfnamefont {John~J.}\ \bibnamefont
  {Oh}}, \bibinfo {author} {\bibfnamefont {Chanyong}\ \bibnamefont {Park}}, \
  and\ \bibinfo {author} {\bibfnamefont {Hyun~Seok}\ \bibnamefont {Yang}},\
  }\bibfield  {title} {\enquote {\bibinfo {title} {Yang-mills instantons from
  gravitational instantons},}\ }\href {\doibase 10.1007/JHEP04(2011)087}
  {\bibfield  {journal} {\bibinfo  {journal} {Journal of High Energy Physics}\
  }\textbf {\bibinfo {volume} {2011}},\ \bibinfo {pages} {87} (\bibinfo {year}
  {2011})}\BibitemShut {NoStop}%
\bibitem [{Pro()}]{ProHam}%
  \BibitemOpen
  \href@noop {} {\bibinfo  {journal} {The 3D boundary Hamiltonian can be
  obtained by projecting the bulk Hamiltonian into the 3D boundary, i.e.,
  $\mathcal{H}_{BD}=P_+\mathcal{H}_0 P_+$. Here the $w$-direction projector
  $P_+=(1+i\Gamma_4\Gamma_0)/2$, the effective $4\times4$ boundary Hamiltonian
  is taken by the first non-zero block of
  $H_{eff}=\mathcal{U}\mathcal{H}_{BD}\mathcal{U}^{-1}$, where
  $\mathcal{U}=\exp[i\frac{\pi}{4} G_{232}]$ rotates $P_+$ into
  $\mathbb{1}_4\oplus \mathbb{0}_4$. Similarly, the extra perturbation terms
  that only commute with $P_+$ can survive after being projected to the
  boundary. For detailed derivations, we refer to the Refs.
  \cite{KWang2020,YQZhu2022}}\ }\BibitemShut {NoStop}%
\bibitem [{\citenamefont {Zhao}\ \emph {et~al.}(2016)\citenamefont {Zhao},
  \citenamefont {Schnyder},\ and\ \citenamefont {Wang}}]{YXZhao2016}%
  \BibitemOpen
\bibfield  {journal} {  }\bibfield  {author} {\bibinfo {author} {\bibfnamefont
  {Y.~X.}\ \bibnamefont {Zhao}}, \bibinfo {author} {\bibfnamefont {Andreas~P.}\
  \bibnamefont {Schnyder}}, \ and\ \bibinfo {author} {\bibfnamefont {Z.~D.}\
  \bibnamefont {Wang}},\ }\bibfield  {title} {\enquote {\bibinfo {title}
  {Unified theory of $pt$ and $cp$ invariant topological metals and nodal
  superconductors},}\ }\href {\doibase 10.1103/PhysRevLett.116.156402}
  {\bibfield  {journal} {\bibinfo  {journal} {Phys. Rev. Lett.}\ }\textbf
  {\bibinfo {volume} {116}},\ \bibinfo {pages} {156402} (\bibinfo {year}
  {2016})}\BibitemShut {NoStop}%
\bibitem [{\citenamefont {T\"urker}\ and\ \citenamefont
  {Moroz}(2018)}]{Turker2018}%
  \BibitemOpen
  \bibfield  {author} {\bibinfo {author} {\bibfnamefont {O\ifmmode
  \breve{g}\else~\u{g}\fi{}uz}\ \bibnamefont {T\"urker}}\ and\ \bibinfo
  {author} {\bibfnamefont {Sergej}\ \bibnamefont {Moroz}},\ }\bibfield  {title}
  {\enquote {\bibinfo {title} {Weyl nodal surfaces},}\ }\href {\doibase
  10.1103/PhysRevB.97.075120} {\bibfield  {journal} {\bibinfo  {journal} {Phys.
  Rev. B}\ }\textbf {\bibinfo {volume} {97}},\ \bibinfo {pages} {075120}
  (\bibinfo {year} {2018})}\BibitemShut {NoStop}%
\bibitem [{\citenamefont {Salerno}\ \emph {et~al.}(2020)\citenamefont
  {Salerno}, \citenamefont {Goldman},\ and\ \citenamefont {Palumbo}}]{Salerno}%
  \BibitemOpen
  \bibfield  {author} {\bibinfo {author} {\bibfnamefont {Grazia}\ \bibnamefont
  {Salerno}}, \bibinfo {author} {\bibfnamefont {Nathan}\ \bibnamefont
  {Goldman}}, \ and\ \bibinfo {author} {\bibfnamefont {Giandomenico}\
  \bibnamefont {Palumbo}},\ }\bibfield  {title} {\enquote {\bibinfo {title}
  {Floquet-engineering of nodal rings and nodal spheres and their
  characterization using the quantum metric},}\ }\href {\doibase
  10.1103/PhysRevResearch.2.013224} {\bibfield  {journal} {\bibinfo  {journal}
  {Phys. Rev. Research}\ }\textbf {\bibinfo {volume} {2}},\ \bibinfo {pages}
  {013224} (\bibinfo {year} {2020})}\BibitemShut {NoStop}%
\bibitem [{BDm()}]{BDmodes}%
  \BibitemOpen
  \href@noop {} {\bibinfo  {journal} {Notice that there is a double (pair of)
  Weyl point survives on the 3D boundary around the origin [$(\pi,\pi,\pi)$] in
  the parameter $2<m<4[-4<m<-2]$, while there are three double (pairs of) Weyl
  points with same chirality survive around three high symmetry points
  $(\pi,0,0)/(\pi,\pi,0)$, $(0,\pi,0)/(\pi,0,\pi)$, $(0,0,\pi)/(0,\pi,\pi)$
  when $0<m<2/-2<m<0$, respectively}\ }\BibitemShut {NoStop}%
\bibitem [{SM()}]{SM}%
  \BibitemOpen
\bibfield  {journal} {  }\href@noop {} {\bibinfo  {journal} {See Supplemental
  Material for details}\ }\BibitemShut {NoStop}%
\bibitem [{\citenamefont {Celi}\ \emph {et~al.}(2014)\citenamefont {Celi},
  \citenamefont {Massignan}, \citenamefont {Ruseckas}, \citenamefont {Goldman},
  \citenamefont {Spielman}, \citenamefont {Juzeli\ifmmode~\bar{u}\else
  \={u}\fi{}nas},\ and\ \citenamefont {Lewenstein}}]{Celi}%
  \BibitemOpen
\bibfield  {journal} {  }\bibfield  {author} {\bibinfo {author} {\bibfnamefont
  {A.}~\bibnamefont {Celi}}, \bibinfo {author} {\bibfnamefont {P.}~\bibnamefont
  {Massignan}}, \bibinfo {author} {\bibfnamefont {J.}~\bibnamefont {Ruseckas}},
  \bibinfo {author} {\bibfnamefont {N.}~\bibnamefont {Goldman}}, \bibinfo
  {author} {\bibfnamefont {I.~B.}\ \bibnamefont {Spielman}}, \bibinfo {author}
  {\bibfnamefont {G.}~\bibnamefont {Juzeli\ifmmode~\bar{u}\else
  \={u}\fi{}nas}}, \ and\ \bibinfo {author} {\bibfnamefont {M.}~\bibnamefont
  {Lewenstein}},\ }\bibfield  {title} {\enquote {\bibinfo {title} {Synthetic
  gauge fields in synthetic dimensions},}\ }\href {\doibase
  10.1103/PhysRevLett.112.043001} {\bibfield  {journal} {\bibinfo  {journal}
  {Phys. Rev. Lett.}\ }\textbf {\bibinfo {volume} {112}},\ \bibinfo {pages}
  {043001} (\bibinfo {year} {2014})}\BibitemShut {NoStop}%
\bibitem [{\citenamefont {Ozawa}\ and\ \citenamefont {Price}(2019)}]{Ozawa2}%
  \BibitemOpen
  \bibfield  {author} {\bibinfo {author} {\bibfnamefont {Tomoki}\ \bibnamefont
  {Ozawa}}\ and\ \bibinfo {author} {\bibfnamefont {Hannah~M.}\ \bibnamefont
  {Price}},\ }\bibfield  {title} {\enquote {\bibinfo {title} {Topological
  quantum matter in synthetic dimensions},}\ }\href {\doibase
  10.1038/s42254-019-0045-3} {\bibfield  {journal} {\bibinfo  {journal} {Nature
  Reviews Physics}\ }\textbf {\bibinfo {volume} {1}},\ \bibinfo {pages}
  {349--357} (\bibinfo {year} {2019})}\BibitemShut {NoStop}%
\bibitem [{\citenamefont {Kitagawa}\ \emph {et~al.}(2010)\citenamefont
  {Kitagawa}, \citenamefont {Berg}, \citenamefont {Rudner},\ and\ \citenamefont
  {Demler}}]{Demler}%
  \BibitemOpen
  \bibfield  {author} {\bibinfo {author} {\bibfnamefont {Takuya}\ \bibnamefont
  {Kitagawa}}, \bibinfo {author} {\bibfnamefont {Erez}\ \bibnamefont {Berg}},
  \bibinfo {author} {\bibfnamefont {Mark}\ \bibnamefont {Rudner}}, \ and\
  \bibinfo {author} {\bibfnamefont {Eugene}\ \bibnamefont {Demler}},\
  }\bibfield  {title} {\enquote {\bibinfo {title} {Topological characterization
  of periodically driven quantum systems},}\ }\href {\doibase
  10.1103/PhysRevB.82.235114} {\bibfield  {journal} {\bibinfo  {journal} {Phys.
  Rev. B}\ }\textbf {\bibinfo {volume} {82}},\ \bibinfo {pages} {235114}
  (\bibinfo {year} {2010})}\BibitemShut {NoStop}%
\bibitem [{\citenamefont {Peng}\ and\ \citenamefont {Refael}(2018)}]{Peng}%
  \BibitemOpen
  \bibfield  {author} {\bibinfo {author} {\bibfnamefont {Yang}\ \bibnamefont
  {Peng}}\ and\ \bibinfo {author} {\bibfnamefont {Gil}\ \bibnamefont
  {Refael}},\ }\bibfield  {title} {\enquote {\bibinfo {title} {Topological
  energy conversion through the bulk or the boundary of driven systems},}\
  }\href {\doibase 10.1103/PhysRevB.97.134303} {\bibfield  {journal} {\bibinfo
  {journal} {Phys. Rev. B}\ }\textbf {\bibinfo {volume} {97}},\ \bibinfo
  {pages} {134303} (\bibinfo {year} {2018})}\BibitemShut {NoStop}%
\bibitem [{\citenamefont {Chang}(2018)}]{Chang}%
  \BibitemOpen
  \bibfield  {author} {\bibinfo {author} {\bibfnamefont {Po-Yao}\ \bibnamefont
  {Chang}},\ }\bibfield  {title} {\enquote {\bibinfo {title} {Topology and
  entanglement in quench dynamics},}\ }\href {\doibase
  10.1103/PhysRevB.97.224304} {\bibfield  {journal} {\bibinfo  {journal} {Phys.
  Rev. B}\ }\textbf {\bibinfo {volume} {97}},\ \bibinfo {pages} {224304}
  (\bibinfo {year} {2018})}\BibitemShut {NoStop}%
\bibitem [{\citenamefont {Kariyado}\ and\ \citenamefont
  {Slager}(2019)}]{fluxesart}%
  \BibitemOpen
  \bibfield  {author} {\bibinfo {author} {\bibfnamefont {Toshikaze}\
  \bibnamefont {Kariyado}}\ and\ \bibinfo {author} {\bibfnamefont {Robert-Jan}\
  \bibnamefont {Slager}},\ }\bibfield  {title} {\enquote {\bibinfo {title}
  {$\ensuremath{\pi}$-fluxes, semimetals, and flat bands in artificial
  materials},}\ }\href {\doibase 10.1103/PhysRevResearch.1.032027} {\bibfield
  {journal} {\bibinfo  {journal} {Phys. Rev. Res.}\ }\textbf {\bibinfo {volume}
  {1}},\ \bibinfo {pages} {032027} (\bibinfo {year} {2019})}\BibitemShut
  {NoStop}%
\bibitem [{\citenamefont {Ezawa}(2018)}]{Ezawa2}%
  \BibitemOpen
  \bibfield  {author} {\bibinfo {author} {\bibfnamefont {Motohiko}\
  \bibnamefont {Ezawa}},\ }\bibfield  {title} {\enquote {\bibinfo {title}
  {Topological quantum quench dynamics carrying arbitrary hopf and second chern
  numbers},}\ }\href {\doibase 10.1103/PhysRevB.98.205406} {\bibfield
  {journal} {\bibinfo  {journal} {Phys. Rev. B}\ }\textbf {\bibinfo {volume}
  {98}},\ \bibinfo {pages} {205406} (\bibinfo {year} {2018})}\BibitemShut
  {NoStop}%
\bibitem [{\citenamefont {\"Unal}\ \emph {et~al.}(2019)\citenamefont {\"Unal},
  \citenamefont {Eckardt},\ and\ \citenamefont {Slager}}]{Unal2019}%
  \BibitemOpen
  \bibfield  {author} {\bibinfo {author} {\bibfnamefont {F.~Nur}\ \bibnamefont
  {\"Unal}}, \bibinfo {author} {\bibfnamefont {Andr\'e}\ \bibnamefont
  {Eckardt}}, \ and\ \bibinfo {author} {\bibfnamefont {Robert-Jan}\
  \bibnamefont {Slager}},\ }\bibfield  {title} {\enquote {\bibinfo {title}
  {Hopf characterization of two-dimensional {F}loquet topological
  insulators},}\ }\href {\doibase 10.1103/PhysRevResearch.1.022003} {\bibfield
  {journal} {\bibinfo  {journal} {Phys. Rev. Research}\ }\textbf {\bibinfo
  {volume} {1}},\ \bibinfo {pages} {022003} (\bibinfo {year}
  {2019})}\BibitemShut {NoStop}%
\bibitem [{\citenamefont {Sugawa}\ \emph {et~al.}(2018)\citenamefont {Sugawa},
  \citenamefont {Salces-Carcoba}, \citenamefont {Perry}, \citenamefont {Yue},\
  and\ \citenamefont {Spielman}}]{Spielman}%
  \BibitemOpen
  \bibfield  {author} {\bibinfo {author} {\bibfnamefont {Seiji}\ \bibnamefont
  {Sugawa}}, \bibinfo {author} {\bibfnamefont {Francisco}\ \bibnamefont
  {Salces-Carcoba}}, \bibinfo {author} {\bibfnamefont {Abigail~R.}\
  \bibnamefont {Perry}}, \bibinfo {author} {\bibfnamefont {Yuchen}\
  \bibnamefont {Yue}}, \ and\ \bibinfo {author} {\bibfnamefont {I.~B.}\
  \bibnamefont {Spielman}},\ }\bibfield  {title} {\enquote {\bibinfo {title}
  {Second chern number of a quantum-simulated non-abelian yang monopole},}\
  }\href {\doibase 10.1126/science.aam9031} {\bibfield  {journal} {\bibinfo
  {journal} {Science}\ }\textbf {\bibinfo {volume} {360}},\ \bibinfo {pages}
  {1429--1434} (\bibinfo {year} {2018})},\ \Eprint
  {http://arxiv.org/abs/https://www.science.org/doi/pdf/10.1126/science.aam9031}
  {https://www.science.org/doi/pdf/10.1126/science.aam9031} \BibitemShut
  {NoStop}%
\bibitem [{\citenamefont {Bouhiron}\ \emph {et~al.}(2022)\citenamefont
  {Bouhiron}, \citenamefont {Fabre}, \citenamefont {Liu}, \citenamefont
  {Redon}, \citenamefont {Mittal}, \citenamefont {Satoor}, \citenamefont
  {Lopes},\ and\ \citenamefont {Nascimbene}}]{Nascimbene}%
  \BibitemOpen
  \bibfield  {author} {\bibinfo {author} {\bibfnamefont {Jean-Baptiste}\
  \bibnamefont {Bouhiron}}, \bibinfo {author} {\bibfnamefont {Aurelien}\
  \bibnamefont {Fabre}}, \bibinfo {author} {\bibfnamefont {Qi}~\bibnamefont
  {Liu}}, \bibinfo {author} {\bibfnamefont {Quentin}\ \bibnamefont {Redon}},
  \bibinfo {author} {\bibfnamefont {Nehal}\ \bibnamefont {Mittal}}, \bibinfo
  {author} {\bibfnamefont {Tanish}\ \bibnamefont {Satoor}}, \bibinfo {author}
  {\bibfnamefont {Raphael}\ \bibnamefont {Lopes}}, \ and\ \bibinfo {author}
  {\bibfnamefont {Sylvain}\ \bibnamefont {Nascimbene}},\ }\href {\doibase
  10.48550/ARXIV.2210.06322} {\enquote {\bibinfo {title} {Realization of an
  atomic quantum hall system in four dimensions},}\ } (\bibinfo {year}
  {2022})\BibitemShut {NoStop}%
\bibitem [{\citenamefont {Ozawa}\ \emph {et~al.}(2016)\citenamefont {Ozawa},
  \citenamefont {Price}, \citenamefont {Goldman}, \citenamefont {Zilberberg},\
  and\ \citenamefont {Carusotto}}]{Ozawa}%
  \BibitemOpen
  \bibfield  {author} {\bibinfo {author} {\bibfnamefont {Tomoki}\ \bibnamefont
  {Ozawa}}, \bibinfo {author} {\bibfnamefont {Hannah~M.}\ \bibnamefont
  {Price}}, \bibinfo {author} {\bibfnamefont {Nathan}\ \bibnamefont {Goldman}},
  \bibinfo {author} {\bibfnamefont {Oded}\ \bibnamefont {Zilberberg}}, \ and\
  \bibinfo {author} {\bibfnamefont {Iacopo}\ \bibnamefont {Carusotto}},\
  }\bibfield  {title} {\enquote {\bibinfo {title} {Synthetic dimensions in
  integrated photonics: From optical isolation to four-dimensional quantum hall
  physics},}\ }\href {\doibase 10.1103/PhysRevA.93.043827} {\bibfield
  {journal} {\bibinfo  {journal} {Phys. Rev. A}\ }\textbf {\bibinfo {volume}
  {93}},\ \bibinfo {pages} {043827} (\bibinfo {year} {2016})}\BibitemShut
  {NoStop}%
\bibitem [{\citenamefont {Lu}\ \emph {et~al.}(2018)\citenamefont {Lu},
  \citenamefont {Gao},\ and\ \citenamefont {Wang}}]{Lu}%
  \BibitemOpen
  \bibfield  {author} {\bibinfo {author} {\bibfnamefont {Ling}\ \bibnamefont
  {Lu}}, \bibinfo {author} {\bibfnamefont {Haozhe}\ \bibnamefont {Gao}}, \ and\
  \bibinfo {author} {\bibfnamefont {Zhong}\ \bibnamefont {Wang}},\ }\bibfield
  {title} {\enquote {\bibinfo {title} {Topological one-way fiber of second
  chern number},}\ }\href {\doibase 10.1038/s41467-018-07817-3} {\bibfield
  {journal} {\bibinfo  {journal} {Nature Communications}\ }\textbf {\bibinfo
  {volume} {9}},\ \bibinfo {pages} {5384} (\bibinfo {year} {2018})}\BibitemShut
  {NoStop}%
\bibitem [{\citenamefont {Colandrea}\ \emph {et~al.}(2022)\citenamefont
  {Colandrea}, \citenamefont {Errico}, \citenamefont {Maffei}, \citenamefont
  {Price}, \citenamefont {Lewenstein}, \citenamefont {Marrucci}, \citenamefont
  {Cardano}, \citenamefont {Dauphin},\ and\ \citenamefont
  {Massignan}}]{DiColandrea}%
  \BibitemOpen
  \bibfield  {author} {\bibinfo {author} {\bibfnamefont {Francesco~Di}\
  \bibnamefont {Colandrea}}, \bibinfo {author} {\bibfnamefont {Alessio~D'}\
  \bibnamefont {Errico}}, \bibinfo {author} {\bibfnamefont {Maria}\
  \bibnamefont {Maffei}}, \bibinfo {author} {\bibfnamefont {Hannah~M}\
  \bibnamefont {Price}}, \bibinfo {author} {\bibfnamefont {Maciej}\
  \bibnamefont {Lewenstein}}, \bibinfo {author} {\bibfnamefont {Lorenzo}\
  \bibnamefont {Marrucci}}, \bibinfo {author} {\bibfnamefont {Filippo}\
  \bibnamefont {Cardano}}, \bibinfo {author} {\bibfnamefont {Alexandre}\
  \bibnamefont {Dauphin}}, \ and\ \bibinfo {author} {\bibfnamefont {Pietro}\
  \bibnamefont {Massignan}},\ }\bibfield  {title} {\enquote {\bibinfo {title}
  {Linking topological features of the hofstadter model to optical diffraction
  figures},}\ }\href {\doibase 10.1088/1367-2630/ac4126} {\bibfield  {journal}
  {\bibinfo  {journal} {New Journal of Physics}\ }\textbf {\bibinfo {volume}
  {24}},\ \bibinfo {pages} {013028} (\bibinfo {year} {2022})}\BibitemShut
  {NoStop}%
\bibitem [{\citenamefont {Cheng}\ \emph {et~al.}(2021)\citenamefont {Cheng},
  \citenamefont {Prodan},\ and\ \citenamefont {Prodan}}]{Prodan}%
  \BibitemOpen
  \bibfield  {author} {\bibinfo {author} {\bibfnamefont {Wenting}\ \bibnamefont
  {Cheng}}, \bibinfo {author} {\bibfnamefont {Emil}\ \bibnamefont {Prodan}}, \
  and\ \bibinfo {author} {\bibfnamefont {Camelia}\ \bibnamefont {Prodan}},\
  }\bibfield  {title} {\enquote {\bibinfo {title} {Revealing the boundary weyl
  physics of the four-dimensional hall effect via phason engineering in
  metamaterials},}\ }\href {\doibase 10.1103/PhysRevApplied.16.044032}
  {\bibfield  {journal} {\bibinfo  {journal} {Phys. Rev. Applied}\ }\textbf
  {\bibinfo {volume} {16}},\ \bibinfo {pages} {044032} (\bibinfo {year}
  {2021})}\BibitemShut {NoStop}%
\bibitem [{\citenamefont {Ma}\ \emph {et~al.}(2021)\citenamefont {Ma},
  \citenamefont {Bi}, \citenamefont {Guo}, \citenamefont {Yang}, \citenamefont
  {You}, \citenamefont {Feng}, \citenamefont {Sun},\ and\ \citenamefont
  {Zhang}}]{Ma}%
  \BibitemOpen
  \bibfield  {author} {\bibinfo {author} {\bibfnamefont {Shaojie}\ \bibnamefont
  {Ma}}, \bibinfo {author} {\bibfnamefont {Yangang}\ \bibnamefont {Bi}},
  \bibinfo {author} {\bibfnamefont {Qinghua}\ \bibnamefont {Guo}}, \bibinfo
  {author} {\bibfnamefont {Biao}\ \bibnamefont {Yang}}, \bibinfo {author}
  {\bibfnamefont {Oubo}\ \bibnamefont {You}}, \bibinfo {author} {\bibfnamefont
  {Jing}\ \bibnamefont {Feng}}, \bibinfo {author} {\bibfnamefont {Hong-Bo}\
  \bibnamefont {Sun}}, \ and\ \bibinfo {author} {\bibfnamefont {Shuang}\
  \bibnamefont {Zhang}},\ }\bibfield  {title} {\enquote {\bibinfo {title}
  {Linked weyl surfaces and weyl arcs in photonic metamaterials},}\ }\href
  {\doibase 10.1126/science.abi7803} {\bibfield  {journal} {\bibinfo  {journal}
  {Science}\ }\textbf {\bibinfo {volume} {373}},\ \bibinfo {pages} {572--576}
  (\bibinfo {year} {2021})},\ \Eprint
  {http://arxiv.org/abs/https://www.science.org/doi/pdf/10.1126/science.abi7803}
  {https://www.science.org/doi/pdf/10.1126/science.abi7803} \BibitemShut
  {NoStop}%
\bibitem [{\citenamefont {Chen}\ \emph
  {et~al.}(2021{\natexlab{a}})\citenamefont {Chen}, \citenamefont {Zhu},
  \citenamefont {Tan}, \citenamefont {Wang},\ and\ \citenamefont {Ma}}]{Chen}%
  \BibitemOpen
  \bibfield  {author} {\bibinfo {author} {\bibfnamefont {Ze-Guo}\ \bibnamefont
  {Chen}}, \bibinfo {author} {\bibfnamefont {Weiwei}\ \bibnamefont {Zhu}},
  \bibinfo {author} {\bibfnamefont {Yang}\ \bibnamefont {Tan}}, \bibinfo
  {author} {\bibfnamefont {Licheng}\ \bibnamefont {Wang}}, \ and\ \bibinfo
  {author} {\bibfnamefont {Guancong}\ \bibnamefont {Ma}},\ }\bibfield  {title}
  {\enquote {\bibinfo {title} {Acoustic realization of a four-dimensional
  higher-order chern insulator and boundary-modes engineering},}\ }\href
  {\doibase 10.1103/PhysRevX.11.011016} {\bibfield  {journal} {\bibinfo
  {journal} {Phys. Rev. X}\ }\textbf {\bibinfo {volume} {11}},\ \bibinfo
  {pages} {011016} (\bibinfo {year} {2021}{\natexlab{a}})}\BibitemShut
  {NoStop}%
\bibitem [{\citenamefont {Chen}\ \emph
  {et~al.}(2021{\natexlab{b}})\citenamefont {Chen}, \citenamefont {Zhang},
  \citenamefont {Wu}, \citenamefont {Huang}, \citenamefont {Nguyen},
  \citenamefont {Prodan}, \citenamefont {Zhou},\ and\ \citenamefont
  {Huang}}]{Chen2}%
  \BibitemOpen
  \bibfield  {author} {\bibinfo {author} {\bibfnamefont {Hui}\ \bibnamefont
  {Chen}}, \bibinfo {author} {\bibfnamefont {Hongkuan}\ \bibnamefont {Zhang}},
  \bibinfo {author} {\bibfnamefont {Qian}\ \bibnamefont {Wu}}, \bibinfo
  {author} {\bibfnamefont {Yu}~\bibnamefont {Huang}}, \bibinfo {author}
  {\bibfnamefont {Huy}\ \bibnamefont {Nguyen}}, \bibinfo {author}
  {\bibfnamefont {Emil}\ \bibnamefont {Prodan}}, \bibinfo {author}
  {\bibfnamefont {Xiaoming}\ \bibnamefont {Zhou}}, \ and\ \bibinfo {author}
  {\bibfnamefont {Guoliang}\ \bibnamefont {Huang}},\ }\bibfield  {title}
  {\enquote {\bibinfo {title} {Creating synthetic spaces for higher-order
  topological sound transport},}\ }\href {\doibase 10.1038/s41467-021-25305-z}
  {\bibfield  {journal} {\bibinfo  {journal} {Nature Communications}\ }\textbf
  {\bibinfo {volume} {12}},\ \bibinfo {pages} {5028} (\bibinfo {year}
  {2021}{\natexlab{b}})}\BibitemShut {NoStop}%
\bibitem [{\citenamefont {Zhao}\ \emph
  {et~al.}(2022{\natexlab{b}})\citenamefont {Zhao}, \citenamefont {Yang},
  \citenamefont {Jiang}, \citenamefont {Mao}, \citenamefont {Guo},
  \citenamefont {Qiu}, \citenamefont {Wang}, \citenamefont {Yao}, \citenamefont
  {He}, \citenamefont {Zhou}, \citenamefont {Xu},\ and\ \citenamefont
  {Duan}}]{Zhao2022}%
  \BibitemOpen
  \bibfield  {author} {\bibinfo {author} {\bibfnamefont {Wending}\ \bibnamefont
  {Zhao}}, \bibinfo {author} {\bibfnamefont {Yan-Bin}\ \bibnamefont {Yang}},
  \bibinfo {author} {\bibfnamefont {Yue}\ \bibnamefont {Jiang}}, \bibinfo
  {author} {\bibfnamefont {Zhichao}\ \bibnamefont {Mao}}, \bibinfo {author}
  {\bibfnamefont {Weixuan}\ \bibnamefont {Guo}}, \bibinfo {author}
  {\bibfnamefont {Liyuan}\ \bibnamefont {Qiu}}, \bibinfo {author}
  {\bibfnamefont {Gangxi}\ \bibnamefont {Wang}}, \bibinfo {author}
  {\bibfnamefont {Lin}\ \bibnamefont {Yao}}, \bibinfo {author} {\bibfnamefont
  {Li}~\bibnamefont {He}}, \bibinfo {author} {\bibfnamefont {Zichao}\
  \bibnamefont {Zhou}}, \bibinfo {author} {\bibfnamefont {Yong}\ \bibnamefont
  {Xu}}, \ and\ \bibinfo {author} {\bibfnamefont {Luming}\ \bibnamefont
  {Duan}},\ }\bibfield  {title} {\enquote {\bibinfo {title} {Quantum simulation
  for topological euler insulators},}\ }\href {\doibase
  10.1038/s42005-022-01001-2} {\bibfield  {journal} {\bibinfo  {journal}
  {Communications Physics}\ }\textbf {\bibinfo {volume} {5}} (\bibinfo {year}
  {2022}{\natexlab{b}}),\ 10.1038/s42005-022-01001-2}\BibitemShut {NoStop}%
\bibitem [{\citenamefont {Weisbrich}\ \emph {et~al.}(2021)\citenamefont
  {Weisbrich}, \citenamefont {Klees}, \citenamefont {Rastelli},\ and\
  \citenamefont {Belzig}}]{Rastelli}%
  \BibitemOpen
  \bibfield  {author} {\bibinfo {author} {\bibfnamefont {H.}~\bibnamefont
  {Weisbrich}}, \bibinfo {author} {\bibfnamefont {R.L.}\ \bibnamefont {Klees}},
  \bibinfo {author} {\bibfnamefont {G.}~\bibnamefont {Rastelli}}, \ and\
  \bibinfo {author} {\bibfnamefont {W.}~\bibnamefont {Belzig}},\ }\bibfield
  {title} {\enquote {\bibinfo {title} {Second chern number and non-abelian
  berry phase in topological superconducting systems},}\ }\href {\doibase
  10.1103/PRXQuantum.2.010310} {\bibfield  {journal} {\bibinfo  {journal} {PRX
  Quantum}\ }\textbf {\bibinfo {volume} {2}},\ \bibinfo {pages} {010310}
  (\bibinfo {year} {2021})}\BibitemShut {NoStop}%
\bibitem [{\citenamefont {Chan}\ and\ \citenamefont {Liu}(2017)}]{Chan}%
  \BibitemOpen
  \bibfield  {author} {\bibinfo {author} {\bibfnamefont {Cheung}\ \bibnamefont
  {Chan}}\ and\ \bibinfo {author} {\bibfnamefont {Xiong-Jun}\ \bibnamefont
  {Liu}},\ }\bibfield  {title} {\enquote {\bibinfo {title} {Non-abelian
  majorana modes protected by an emergent second chern number},}\ }\href
  {\doibase 10.1103/PhysRevLett.118.207002} {\bibfield  {journal} {\bibinfo
  {journal} {Phys. Rev. Lett.}\ }\textbf {\bibinfo {volume} {118}},\ \bibinfo
  {pages} {207002} (\bibinfo {year} {2017})}\BibitemShut {NoStop}%
\bibitem [{\citenamefont {Wang}\ \emph
  {et~al.}(2020{\natexlab{a}})\citenamefont {Wang}, \citenamefont {Price},
  \citenamefont {Zhang},\ and\ \citenamefont {Chong}}]{Wang}%
  \BibitemOpen
  \bibfield  {author} {\bibinfo {author} {\bibfnamefont {You}\ \bibnamefont
  {Wang}}, \bibinfo {author} {\bibfnamefont {Hannah~M.}\ \bibnamefont {Price}},
  \bibinfo {author} {\bibfnamefont {Baile}\ \bibnamefont {Zhang}}, \ and\
  \bibinfo {author} {\bibfnamefont {Y.~D.}\ \bibnamefont {Chong}},\ }\bibfield
  {title} {\enquote {\bibinfo {title} {Circuit implementation of a
  four-dimensional topological insulator},}\ }\href {\doibase
  10.1038/s41467-020-15940-3} {\bibfield  {journal} {\bibinfo  {journal}
  {Nature Communications}\ }\textbf {\bibinfo {volume} {11}},\ \bibinfo {pages}
  {2356} (\bibinfo {year} {2020}{\natexlab{a}})}\BibitemShut {NoStop}%
\bibitem [{\citenamefont {Yu}\ \emph {et~al.}(2020)\citenamefont {Yu},
  \citenamefont {Zhao},\ and\ \citenamefont {Schnyder}}]{Yu}%
  \BibitemOpen
  \bibfield  {author} {\bibinfo {author} {\bibfnamefont {Rui}\ \bibnamefont
  {Yu}}, \bibinfo {author} {\bibfnamefont {Y~X}\ \bibnamefont {Zhao}}, \ and\
  \bibinfo {author} {\bibfnamefont {Andreas~P}\ \bibnamefont {Schnyder}},\
  }\bibfield  {title} {\enquote {\bibinfo {title} {{4D spinless topological
  insulator in a periodic electric circuit}},}\ }\href {\doibase
  10.1093/nsr/nwaa065} {\bibfield  {journal} {\bibinfo  {journal} {National
  Science Review}\ }\textbf {\bibinfo {volume} {7}},\ \bibinfo {pages}
  {1288--1295} (\bibinfo {year} {2020})},\ \Eprint
  {http://arxiv.org/abs/https://academic.oup.com/nsr/article-pdf/7/8/1288/38882379/nwaa065.pdf}
  {https://academic.oup.com/nsr/article-pdf/7/8/1288/38882379/nwaa065.pdf}
  \BibitemShut {NoStop}%
\bibitem [{\citenamefont {Ezawa}(2019)}]{Ezawa}%
  \BibitemOpen
  \bibfield  {author} {\bibinfo {author} {\bibfnamefont {Motohiko}\
  \bibnamefont {Ezawa}},\ }\bibfield  {title} {\enquote {\bibinfo {title}
  {Electric circuit simulations of $n\mathrm{th}$-chern-number insulators in
  $2n$-dimensional space and their non-hermitian generalizations for arbitrary
  $n$},}\ }\href {\doibase 10.1103/PhysRevB.100.075423} {\bibfield  {journal}
  {\bibinfo  {journal} {Phys. Rev. B}\ }\textbf {\bibinfo {volume} {100}},\
  \bibinfo {pages} {075423} (\bibinfo {year} {2019})}\BibitemShut {NoStop}%
\bibitem [{\citenamefont {Freimuth}\ \emph {et~al.}(2013)\citenamefont
  {Freimuth}, \citenamefont {Bamler}, \citenamefont {Mokrousov},\ and\
  \citenamefont {Rosch}}]{Rosch}%
  \BibitemOpen
  \bibfield  {author} {\bibinfo {author} {\bibfnamefont {Frank}\ \bibnamefont
  {Freimuth}}, \bibinfo {author} {\bibfnamefont {Robert}\ \bibnamefont
  {Bamler}}, \bibinfo {author} {\bibfnamefont {Yuriy}\ \bibnamefont
  {Mokrousov}}, \ and\ \bibinfo {author} {\bibfnamefont {Achim}\ \bibnamefont
  {Rosch}},\ }\bibfield  {title} {\enquote {\bibinfo {title} {Phase-space berry
  phases in chiral magnets: Dzyaloshinskii-moriya interaction and the charge of
  skyrmions},}\ }\href {\doibase 10.1103/PhysRevB.88.214409} {\bibfield
  {journal} {\bibinfo  {journal} {Phys. Rev. B}\ }\textbf {\bibinfo {volume}
  {88}},\ \bibinfo {pages} {214409} (\bibinfo {year} {2013})}\BibitemShut
  {NoStop}%
\bibitem [{\citenamefont {Xiao}\ \emph {et~al.}(2009)\citenamefont {Xiao},
  \citenamefont {Shi}, \citenamefont {Clougherty},\ and\ \citenamefont
  {Niu}}]{Niu}%
  \BibitemOpen
  \bibfield  {author} {\bibinfo {author} {\bibfnamefont {Di}~\bibnamefont
  {Xiao}}, \bibinfo {author} {\bibfnamefont {Junren}\ \bibnamefont {Shi}},
  \bibinfo {author} {\bibfnamefont {Dennis~P.}\ \bibnamefont {Clougherty}}, \
  and\ \bibinfo {author} {\bibfnamefont {Qian}\ \bibnamefont {Niu}},\
  }\bibfield  {title} {\enquote {\bibinfo {title} {Polarization and adiabatic
  pumping in inhomogeneous crystals},}\ }\href {\doibase
  10.1103/PhysRevLett.102.087602} {\bibfield  {journal} {\bibinfo  {journal}
  {Phys. Rev. Lett.}\ }\textbf {\bibinfo {volume} {102}},\ \bibinfo {pages}
  {087602} (\bibinfo {year} {2009})}\BibitemShut {NoStop}%
\bibitem [{\citenamefont {Bouhon}\ and\ \citenamefont
  {Slager}(2022{\natexlab{b}})}]{Bouhon_geo2}%
  \BibitemOpen
  \bibfield  {author} {\bibinfo {author} {\bibfnamefont {Adrien}\ \bibnamefont
  {Bouhon}}\ and\ \bibinfo {author} {\bibfnamefont {Robert-Jan}\ \bibnamefont
  {Slager}},\ }\bibfield  {title} {\enquote {\bibinfo {title} {Multi-gap
  topological conversion of euler class via band-node braiding: minimal models,
  $pt$-linked nodal rings, and chiral heirs},}\ }\href {\doibase
  10.48550/arxiv.2203.16741} {\  (\bibinfo {year} {2022}{\natexlab{b}}),\
  10.48550/arxiv.2203.16741}\BibitemShut {NoStop}%
\bibitem [{\citenamefont {Alexandradinata}\ \emph {et~al.}(2014)\citenamefont
  {Alexandradinata}, \citenamefont {Dai},\ and\ \citenamefont
  {Bernevig}}]{Wi1}%
  \BibitemOpen
  \bibfield  {author} {\bibinfo {author} {\bibfnamefont {A.}~\bibnamefont
  {Alexandradinata}}, \bibinfo {author} {\bibfnamefont {Xi}~\bibnamefont
  {Dai}}, \ and\ \bibinfo {author} {\bibfnamefont {B.~Andrei}\ \bibnamefont
  {Bernevig}},\ }\bibfield  {title} {\enquote {\bibinfo {title} {Wilson-loop
  characterization of inversion-symmetric topological insulators},}\ }\href
  {\doibase 10.1103/PhysRevB.89.155114} {\bibfield  {journal} {\bibinfo
  {journal} {Phys. Rev. B}\ }\textbf {\bibinfo {volume} {89}},\ \bibinfo
  {pages} {155114} (\bibinfo {year} {2014})}\BibitemShut {NoStop}%
\bibitem [{\citenamefont {{Bzdu{\v s}ek}}\ and\ \citenamefont
  {Sigrist}(2017)}]{BzduSigristRobust}%
  \BibitemOpen
  \bibfield  {author} {\bibinfo {author} {\bibfnamefont {Tom\'{a}\v{s}}\
  \bibnamefont {{Bzdu{\v s}ek}}}\ and\ \bibinfo {author} {\bibfnamefont
  {Manfred}\ \bibnamefont {Sigrist}},\ }\bibfield  {title} {\enquote {\bibinfo
  {title} {Robust doubly charged nodal lines and nodal surfaces in
  centrosymmetric systems},}\ }\href {\doibase 10.1103/PhysRevB.96.155105}
  {\bibfield  {journal} {\bibinfo  {journal} {Phys. Rev. B}\ }\textbf {\bibinfo
  {volume} {96}},\ \bibinfo {pages} {155105} (\bibinfo {year}
  {2017})}\BibitemShut {NoStop}%
\bibitem [{\citenamefont {Hatcher}(2003)}]{Hatcher_2}%
  \BibitemOpen
  \bibfield  {author} {\bibinfo {author} {\bibfnamefont {A.}~\bibnamefont
  {Hatcher}},\ }\href@noop {} {\emph {\bibinfo {title} {{V}ector {B}undles and
  {K}-{T}heory}}}\ (\bibinfo  {publisher} {Unpublished},\ \bibinfo {year}
  {2003})\BibitemShut {NoStop}%
\bibitem [{\citenamefont {Bouhon}\ \emph {et~al.}()\citenamefont {Bouhon},
  \citenamefont {Bzdu{\v s}ek},\ and\ \citenamefont {Slager}}]{Bouhon1D_2022}%
  \BibitemOpen
  \bibfield  {author} {\bibinfo {author} {\bibfnamefont {Adrien}\ \bibnamefont
  {Bouhon}}, \bibinfo {author} {\bibfnamefont {Tom{\'a}{\v s}}\ \bibnamefont
  {Bzdu{\v s}ek}}, \ and\ \bibinfo {author} {\bibfnamefont {Robert-Jan}\
  \bibnamefont {Slager}},\ }\bibfield  {title} {\enquote {\bibinfo {title}
  {Quantization of 1d non-abelian multi-gap topological charges and the absence
  thereof},}\ }\href@noop {} {\bibinfo  {journal} {to appear}\ }\BibitemShut
  {NoStop}%
\bibitem [{\citenamefont {Wang}\ \emph
  {et~al.}(2020{\natexlab{b}})\citenamefont {Wang}, \citenamefont {Dai},
  \citenamefont {Shao}, \citenamefont {Yang},\ and\ \citenamefont
  {Zhao}}]{KWang2020}%
  \BibitemOpen
\bibfield  {journal} {  }\bibfield  {author} {\bibinfo {author} {\bibfnamefont
  {Kai}\ \bibnamefont {Wang}}, \bibinfo {author} {\bibfnamefont {Jia-Xiao}\
  \bibnamefont {Dai}}, \bibinfo {author} {\bibfnamefont {L.~B.}\ \bibnamefont
  {Shao}}, \bibinfo {author} {\bibfnamefont {Shengyuan~A.}\ \bibnamefont
  {Yang}}, \ and\ \bibinfo {author} {\bibfnamefont {Y.~X.}\ \bibnamefont
  {Zhao}},\ }\bibfield  {title} {\enquote {\bibinfo {title} {Boundary
  criticality of $\mathcal{PT}$-invariant topology and second-order nodal-line
  semimetals},}\ }\href {\doibase 10.1103/PhysRevLett.125.126403} {\bibfield
  {journal} {\bibinfo  {journal} {Phys. Rev. Lett.}\ }\textbf {\bibinfo
  {volume} {125}},\ \bibinfo {pages} {126403} (\bibinfo {year}
  {2020}{\natexlab{b}})}\BibitemShut {NoStop}%
\end{thebibliography}
%

\end{document}